\documentclass[12pt]{article} 
\usepackage{amsmath, amssymb, mathrsfs, mathtools, tensor}
\usepackage{xcolor, graphicx, subfigure}
\usepackage{caption}

\usepackage{varioref}
        \numberwithin{equation}{section}

\usepackage{enumitem, float}

\evensidemargin \oddsidemargin

\newcommand{\Romtwo}{I\(\!\)I}

\begin{document}

\begin{titlepage}

\begin{center}
\vspace{5mm}
    
{\Large \bfseries
        Effect of Impurity on
        \\[1mm]
        Inhomogeneous Vacuum and Interacting Vortices
}\\[17mm]
    
    SeungJun Jeon,
    ~~Yoonbai Kim,
    ~~Hanwool Song
    \\[4mm]
    
{\itshape
    Department of Physics, Sungkyunkwan University,
    Suwon 16419,
    Korea
\\
    sjjeon@skku.edu,~
    yoonbai@skku.edu,~
    hanwoolsong0@gmail.com
}
\end{center}
\vspace{15mm}

\begin{abstract}

We study the inhomogeneous abelian Higgs model with a magnetic impurity.
The vacuum configuration of the symmetry-broken phase is not simply the constant Higgs vacuum but is a nontrivial function of spatial coordinates, satisfying the Euler-Lagrange equations.
The vacuum of zero winding number has zero magnetic flux but its non-zero magnetic field depends on spatial coordinates.
The corresponding vacuum energy is negative for weak coupling $(\lambda < 1)$, zero for critical BPS coupling $(\lambda = 1)$, and positive for strong coupling $(\lambda > 1)$ by an over-, exact-, and under-cancellation of the huge positive impurity energy.
This distinct vacuum energies are consistent with classification of the type I and I$\!$I superconductivity in dirty conventional superconductors.
Non-BPS vortex configurations are also obtained in the presence of inhomogeneity.
Their rest energies favor energetically vortex-impurity composite in conventional type I$\!$I superconductivity, consistent with imperfect diamagnetism.
The delta function limit of Gaussian type impurity suggests the formation of vortex-lattice composite which elucidates flux-pinning in the context of inhomogeneous field theory.

\end{abstract}

\end{titlepage}

\section{Introduction}

Quantum field theories are written in terms of the fields and the coupling constants including off-shell contributions and have played a prominent role for describing the fundamental forces in microscopic level, e.g. quantum electrodynamics and unified description of electroweak interaction.
On the other side, classical field theories were introduced ahead as the effective theories derived from the Newton's second law.
Hence there does not exist any generic reason that the propagation speed and the couplings are defined as constants.
Their spatial dependence seems natural and is not excluded except for the media of constant quantities such as constant tension and mass density.
Rather, some classical phenomena are described more accurately when the couplings are allowed to be externally given functions of spatial coordinates, which reflect the environments, e.g., the description of seismic wave propagation in different media.
In the basis of these observations on emergent phenomena in diversified subjects, it is manifest and even timely that the study of classical and quantum field theories including the couplings with time and spatial dependence are necessitated.

Inhomogeneous field theories have been studied in the following directions. Janus Yang-Mills theories with a position-dependent gauge coupling have been studied with the control of holography~\cite{Bak:2003jk, Clark:2005te, DHoker:2006qeo, DHoker:2007zhm, Kim:2008dj, Kim:2009wv}.  In three and four dimensions, mass-deformed Aharony-Bergman-Jafferis-Maldacena(ABJM) and
super Yang-Mills theories allow supersymmetry-preserving inhomogeneous mass deformations which are originated by irregular form fields on the branes~ \cite{Kim:2018qle, Kim:2019kns, Arav:2020obl, Kim:2020jrs}. Field theoretic side has been examined by use of simple supersymmetric field theories in (1+1)-dimensions~\cite{Adam:2019yst, Adam:2018pvd, Adam:2018tnv, Adam:2019djg, Manton:2019xiq, Adam:2019hef, Adam:2019xuc,Kwon:2021flc}. Inhomogeneous abelian gauge theories have been studied with aiming at condensed matter systems~\cite{Hook:2013yda, Tong:2013iqa, Kim:2023abp} and recently inhomogeneous vacua of zero energy are found within the BPS limit~\cite{Kim:2024gpu,Kim:2024gfn,Jeon:2024jbs}. 

If the proposed inhomogeneous field theory is an efficient language describing physical phenomena, it must possess practicality and tractability. To be specific, validity of the proposed inhomogeneous field theory can be judged by experimental tests and, before, such testable field theoretic quantities are calculable in spite of complicated inhomogeneous part.
The preceding guidelines enforce the study of non-supersymmetric inhomogeneous field theories involving electromagnetism.
In accordance with these, we consider the abelian Higgs model with spatial inhomogeneity introduced with the magnetic impurity term. Characteristic propagation speed $v_{{\rm p}}$ is explicitly written in order to describe both nonrelativistic regime $(v_{{\rm p}}\ll c)$ and relativistic regime $(v_{{\rm p}}\sim c)$, and quartic scalar coupling is arbitrary $\lambda\ge0$.   
Though the inhomogeneous abelian Higgs model of consideration governs dynamics of a massive charged matter with another kind of the dispersion relation of acoustic phonon in superconducting state, it is unclear to describe a realistic sample of slab shape showing conventional superconductivity~\cite{tinkham2004introduction,Arovas}.
Nonetheless, this toy model shares some quantities, e.g. the London penetration depth and the correlation length, and characters such as category of type I and type I$\!$I superconductors~\cite{Abrikosov:1956sx}. It can predict the common properties of conventional superconductivity of $s$-wave order parameter. We shall discuss these aspects in the current work.

Since numerical analysis is heavily used in addition to analytic method, cylindrically symmetric Gaussian type inhomogeneity expressed by two dimensionless parameters for its size and depth is employed and the cylindrically symmetric vacuum and vortices are explored by solving the Euler-Lagrange equations for static non-BPS configurations as done in homogeneous case~\cite{Jacobs:1978ch}. The cylindrically symmetric symmetry-broken vacuum solutions are obtained as the configurations of zero vorticity without carrying U(1) charge, magnetic flux, and spin.
Their energy is negative for weak quartic scalar coupling $(\lambda<1)$, zero for critical coupling $(\lambda=1)$, and positive for strong quartic scalar coupling $(\lambda>1)$ irrespective of existence of localized inhomogeneous part. Since it is always much lower than the positive impurity energy added by inhomogeneous part, the  obtained new inhomogeneous vacuum solutions are understood as the consequence of energy-decreasing modulation through nonlinear interactions.
For various size and depth parameters, the vacuum solutions are found and their characteristic properties are computed. We derive a loose energy lower bound obeyed with arbitrary localized inhomogeneity and all the numerically obtained non-BPS inhomogeneous vacuum solutions fulfill this bound.

Subsequently, cylindrically symmetric topological vortex solutions are obtained in the inhomogeneous abelian Higgs model with arbitrary quartic scalar interaction. Its rest energy is computed for a single vortex of unit vorticity, and has, in comparison to the homogeneous case, increased value for weak quartic scalar coupling $(\lambda<1)$ and decreased value for strong quartic scalar coupling $(\lambda>1)$. This provides a plausible explanation of energetically favorable production of a vortex attached to the inhomogeneous region and its delta function limit supports a composite of vortex-lattice, which leads to imperfect diamagnetism in type I$\!$I 
superconductors~\cite{tinkham2004introduction,cohen2016fundamentals}.
For various size and depth parameters, inhomogeneous topological vortices are studied and their characteristic properties are computed. Interaction between multiple inhomogeneous topological vortices are examined in some limited cases with keeping cylindrical symmetry.

The rest of the paper is organized as follows. In section 2, we introduce the abelian Higgs model including a magnetic impurity term with arbitrary quartic scalar coupling and discuss some properties. In section 3, non-BPS inhomogeneous vacuum solution of non-zero energy is obtained for the Gaussian inhomogeneous part of various size and depth parameters including delta function limit. Section 4 devotes to finding inhomogeneous topological vortex solutions  with their rest energy and interaction.   We conclude in section 5 with discussions.


\section{Abelian Higgs Model with Inhomogeneity}

If the sample of interest is assumed to possess translational symmetry along the \(z\)-axis, the fields depend only on time \(t\) and spatially planar variables \( \boldsymbol{x} = (x_{1}, x_{2}) = (x,y) \) but are independent of \(z\) coordinate.
These systems, e.g. condensed matter samples, are also assumed to include irregular impurity and hence inhomogeneity is introduced in the vacuum expectation value \(v\).
Field theoretic description of these samples begins by defining the action
\begin{equation}
        S =
        \int dt \int d^{2} \boldsymbol{x} \int dz \,
        \Big[
                - \frac{\epsilon_{0} c^{2}}{4}
                F_{\mu\nu} F^{\mu\nu}
                - \overline{\mathcal{D}_{\mu} \phi}
                \mathcal{D}^{\mu} \phi
                - \lambda
                \frac{g^{2}}{2 \epsilon_{0} c^{2} \hbar^{2}}
                (|\phi|^{2} - v^{2} (\boldsymbol{x}))^{2}
                + s\frac{g}{\hbar}
                \sigma(\boldsymbol{x}) B
        \Big]
,\label{200}
\end{equation}
where \(\phi\) is a complex scalar field, field strength tensor \( F_{\mu\nu} \) consists of electric field \( (\boldsymbol{E})^{i} = c F_{i0} \) and magnetic field
\(      \displaystyle
        (\boldsymbol{B})^{i}
        = \epsilon^{ijk} F_{jk} /2
        \ (i = 1, 2, 3)
\)
with notation \( (\boldsymbol{B})^{3} = B \), and the gauge-covariant derivative is
\begin{equation}
        \mathcal{D}_{0} \phi
        =
        \frac{1}{v_{\text{p}}}
        \mathcal{D}_{t} \phi
        =
        \frac{1}{v_{\text{p}}}
        \Big(
                \frac{\partial}{\partial t}
                +
                i \frac{g}{\hbar}
                \frac{v_{\text{p}}}{c} \Phi
        \Big) \phi
,\qquad
        \mathcal{D}_{i} \phi
        =
        \Big(
                \frac{\partial}{\partial x^{i}}
                -
                i \frac{g}{\hbar} A^{i}
        \Big) \phi
.\label{229}
\end{equation}
Here \(v_{\text{p}}\) is the propagation speed of the complex scalar field and has mostly nonrelativistic value but can take the light speed \(v_{\text{p}} \to c\) for a relativistic motion.
Note that \( \sigma(\boldsymbol{x}) \) stands for the inhomogeneous part depending on planar coordinates due to impurities, which couple to magnetic field along the \(z\)-axis.
The magnetic impurity term is introduced first
\begin{equation}
        \Delta_{B} \mathscr{L}
        = s \frac{g}{\hbar} \sigma (\boldsymbol{x}) B
\label{210}
\end{equation}
as a supersymmetry-preserving inhomogeneous term in relativistic abelian Higgs model of critical quartic scalar coupling \(\lambda=1\) \cite{Hook:2013yda, Tong:2013iqa}, and \(s=\pm1\) is introduced for the purpose of later analysis.
Accordingly, \(v\) in the scalar potential is given by the sum of constant value \(v_0\) and inhomogeneous part
\begin{equation}
        v^{2} (\boldsymbol{x})
        = v_{0}^{2}
        + \sigma (\boldsymbol{x})
,\label{221}
\end{equation}
where \(\sigma(\boldsymbol{x})\) is assumed to vanish at spatial infinity
\begin{equation}
        \lim_{ |\boldsymbol{x}| \to \infty }
        \sigma ( \boldsymbol{x} ) = 0
.\label{212}
\end{equation}

Dynamics of the \(\mathrm{U}(1)\) gauge field \(A^\mu\) is governed by the Maxwell's equations
\begin{equation}
        \boldsymbol{\nabla} \cdot \boldsymbol{E}
        =
        \frac{\rho}{\epsilon_0}
,\label{201}
\end{equation}
\begin{equation}
        \boldsymbol{\nabla} \times \boldsymbol{B}
        - \frac{1}{c^2}
        \frac{\partial \boldsymbol{E}}{\partial t}
        =
        \frac{1}{\epsilon_{0} c^{2}}
        \Big(
                \boldsymbol{j}
                + s \frac{g}{\hbar}
                \boldsymbol{\nabla}
                \times \boldsymbol{\sigma}
        \Big)
,\label{202}
\end{equation}
where
\(
        \boldsymbol{\sigma}
        = ( 0 , 0 , \sigma )
\),
and the charge density \(\rho\) and the current density \( \boldsymbol{j} \) called the supercurrent density in condensed matter physics are assisted by spontaneous breakdown of \(\mathrm{U}(1)\) symmetry
\begin{equation}
        \rho
        = \frac{j^0}{c}
        = i \frac{g}{\hbar c v_{\text{p}}}
        \big(
                \bar{\phi} \mathcal{D}_{t} \phi
                - \overline{\mathcal{D}_{t} \phi} \phi
        \big)
,\qquad
        (\boldsymbol{j})^{i}
        = j^{i}
        = - i \frac{g}{\hbar}
        \big(
                \bar{\phi} \mathcal{D}^{i} \phi
                - \overline{\mathcal{D}^{i} \phi} \phi
        \big)
.\label{219}
\end{equation}
Euler-Lagrange equation of the complex scalar field is
\begin{equation}
        \mathcal{D}_{\mu} \mathcal{D}^{\mu} \phi
        =
        \lambda
        \frac{g^2}{ {\epsilon_0} {c^2} {\hbar^2} }
        ( |\phi|^{2} - {v}^{2} ) \phi
.\label{226}
\end{equation}

In this work, we are interested in some static neutral objects.
The Weyl gauge \( A^{0} = \Phi/c = 0 \) is naturally chosen and the corresponding electric field is zero by the Gauss' law \eqref{201}.
Translation invariance along the \(z\)-axis dictates
\(
        \partial_{3} \phi
        = \partial_{3} A^{i}
        = 0
\)
and, consistently, turning off the \(z\)-component of gauge field \( A^{3} = 0 \) leads to
\(
        \mathcal{D}_{3} \phi
        = \partial_{3} \phi
        = 0
\)
and
\(
        (\boldsymbol{B})^{1}
        = (\boldsymbol{B})^{2}
        = 0
\).
Hence spatial indices \(i, j, \cdots\) are assumed to run only planar components, i.e. \( i, j, \cdots = 1, 2 \) without \(3\) from now on.
The scalar equation \eqref{226} for static configurations becomes
\begin{equation}
        \nabla^{2} |\phi|
        =
        \frac{g^2}{\hbar^2}
        (\bar{A}^{i})^{2} |\phi|
        +
        \frac{1}{ \xi^{2} v_{0}^{2} }
        (|\phi|^{2} - v^{2} (\boldsymbol{x}))
        |\phi|
,\label{204}
\end{equation}
where the complex scalar field is written in terms of amplitude and phase as \( \phi = |\phi|     e^{ i\Omega } \), the unitary gauge also called the London gauge in superconductivity is chosen
\begin{equation}
        \bar{A}^{i}
        = A^{i} 
        - \frac{\hbar}{g} \partial_{i} \Omega
,
\end{equation}
and the correlation length is introduced
\begin{equation}
        \xi =
        \sqrt{
                \frac{1}{\lambda}
                \frac{ \epsilon_{0} c^{2} \hbar^{2} }
                { g^{2} v_{0}^{2} }
        }
.\label{213}
\end{equation}
The Amp\'ere's law \eqref{202} takes the form of a London equation for planar components \( i = 1,2 \), added by an inhomogeneous term from magnetic impurity \eqref{210}
\begin{equation}
        \epsilon^{ij} \partial_{j} B
        =
        - \frac{1}{\lambda_{\text{L}}^{2}} \bar{A}^{i}
        \bigg(
                \frac{|\phi|}{v_0}
        \bigg)^{2}
        + s \frac{\hbar}{2g v_{0}^{2}}
        \frac{1}{ \lambda_{\text{L}}^{2} }
        \epsilon^{ij} \partial_{j} \sigma
,\label{205}
\end{equation}
where the London penetration depth is introduced%
\footnote{
        According to the convention in condensed matter physics, the London penetration depth is not \(\sqrt{2}\lambda_{\text{L}}\) \eqref{214} but \(\lambda_{\text{L}}\).
        Throughout this work, we will call \(\sqrt{2} \lambda_{\text{L}}\) the London penetration depth in order to let the BPS limit as the limit of the unit Ginzburg-Landau parameter \( \sqrt{2}\lambda_{\text{L}} / \xi = 1 \).
}
\begin{equation}
        \sqrt{2} \lambda_{\text{L}}
        =
        \sqrt{
                \frac
                { \epsilon_{0} c^{2} \hbar^{2} }
                { g^{2} v_{0}^{2} }
        }
.\label{214}
\end{equation}
If the complex scalar field stands for a Cooper pair, charge is \( g=-2e \) and rest energy is
\(
        \sqrt{ 2\lambda / \epsilon_0 }
        \, g v_{0}
        = 2m_{\text{e}} c^{2}
\)
in tree level, whose value is known to be affected by band structure and quantum corrections.

If the relativistic limit is taken \( v_{\text{p}} \to c \), the model of our consideration becomes the relativistic inhomogeneous abelian Higgs model with the magnetic impurity term involving a single characteristic speed \(c\) \cite{Tong:2013iqa, Kim:2024gfn}.
In it, introduction of two length scales, the correlation length \(\xi\) \eqref{213} and the London penetration depth \( \sqrt{2} \lambda_{\text{L}} \) \eqref{214}, are analogous to existence of two mass scales, the Higgs mass \(m_{\text{H}}\) and the gauge boson mass \( m_{A_{\mu}} \), respectively, which become degenerate at the critical quartic scalar coupling \( \lambda = 1 \)
\begin{equation}
        m_{\text{H}} c^{2}
        =
        \sqrt{ \frac{ 2\lambda}{\epsilon_0} }
        g v_{0}
\quad \text{and} \quad
        m_{A_\mu} c^{2}
        =
        \sqrt{ \frac{ 2}{\epsilon_0} }
        g v_{0}
.
\end{equation}

The energy \(\mathcal{E}\) per unit length along the \(z\)-axis is read from the action \eqref{200}
\begin{align}
        \mathcal{E}
        & =
        \frac{E}{\int dz}
        = \int d^{2} \boldsymbol{x} \,
        (-T\indices{^t_t}) = \int d^{2} \boldsymbol{x} \, T_{00}
\nonumber \\
        & =
        \int d^{2} \boldsymbol{x} \,
        \bigg[
                \frac{\epsilon_0}{2}
                ( \boldsymbol{E}^{2}
                + c^{2} B^{2} )
                +
                \frac{1}{v_{\text{p}}^{2}}
                | \mathcal{D}_{t} \phi |^{2}
                +
                | \mathcal{D}_{i} \phi |^{2}
                +
                \lambda \frac{g^2}
                { 2 \epsilon_{0} c^{2} \hbar^{2} }
                ( |\phi|^{2} - v^{2} )^{2}
                -
                s \frac{g}{\hbar} \sigma B
        \bigg]
,\label{225}
\end{align}
and the energy flux density
\(
        - T\indices{^i_t}
        = c T\indices{^i_0}
\)
consists of four terms, the first Poynting vector term, the second and third terms from nonrelativistic matter of the complex scalar field, and the fourth impurity term
\begin{equation}
        - T\indices{^i_t} =
        \epsilon_{0} c^{2} \epsilon^{ij} E_{j} B
        -
        (
                \overline{\mathcal{D}_{t} \phi}
                \mathcal{D}^{i} \phi
                +
                \overline{\mathcal{D}^{i} \phi}
                \mathcal{D}_{t} \phi
        )
        + \bigg(
                1 - \frac{v_{\text{p}}}{c}
        \bigg) \Phi j^{i}
        - s \frac{g}{\hbar}
        \sigma( \boldsymbol{x} ) 
        \epsilon^{ij} E_{j} 
.\label{216}
\end{equation}
In the relativistic limit of \( v_{\text{p}} \to c \), the third term disappears and then the relativistic energy flux density is recovered.
Recall that the \(\mathrm{U}(1)\) gauge field follows the Lorentz boost transformation but the complex scalar field \(\phi\) does the Galilean boost transformation.
Since time translation symmetry is unbroken even in the presence of inhomogeneity \( \sigma(\boldsymbol{x}) \neq 0 \), the energy conservation holds trivially for static objects
\begin{equation}
        \partial_{\mu} T\indices{^\mu_t}
        = 0
.\label{215}
\end{equation}

Momentum density \( T\indices{^t_i} = T\indices{^0_i} / c \) is
\begin{equation}
        T\indices{^t_i}
        =
        \epsilon_{0} \epsilon_{ij} E^{j} B
        - \frac{1}{v_{\text{p}}^{2}}
        (
                \overline{\mathcal{D}_{t} \phi}
                \mathcal{D}_{i} \phi
                +
                \overline{\mathcal{D}_{i} \phi}
                \mathcal{D}_{t} \phi
        )
        - \bigg(
                1 - \frac{c}{v_{\text{p}}}
        \bigg) \rho A^{i}
.\label{217}
\end{equation}
Comparison of the momentum density \( T\indices{^t_i} \) \eqref{217} and the energy flux density \( T\indices{^i_t} \) \eqref{216} shows the following characters.
The first electromagnetic term is equal as a consequence of Lorentz symmetry.
The second term has the same form but different coefficient \( 1 / v_{\text{p}}^{2} \) due to Galilean boost symmetry of the matter described by a nonrelativistic complex scalar field with propagation speed \( v_{\text{p}} \).
In the relativistic limit \( v_{\text{p}} \to c \), the coefficient of the third term disappears as expected by the recovery of the Lorentz symmetry.
Though the ratio \( c / v_{\text{p}} \) becomes much larger than unity in nonrelativistic regime \( v_{\text{p}} \ll c \), the aforementioned discussion may suggest smallness of scalar amplitude which implies possible compensation of large factor \( c / v_{\text{p}} \) in nonrelativistic regime.
Generic difference surviving in the relativistic limit is given by the fourth impurity term of the energy flux density
\begin{equation}
        \lim_{ v_{\text{p}} \to c }
        \Big(
                c T \indices{^t_i}
                + \frac{1}{c} T \indices{^i_t}
        \Big)
        =
        s \frac{g}{\hbar c}
        \sigma(\boldsymbol{x})
        \epsilon_{ij} E^{j}
.
\end{equation}
Once spatial inhomogeneity \( \sigma(\boldsymbol{x}) \neq 0 \) is introduced, breakdown of spatial translation symmetry is manifest.
The stress components \(T \indices{^i_j}\) for static configurations are calculated as
\begin{equation}
        T \indices{^i_j}
        = 
        \bigg[
                \frac{\epsilon_{0} c^{2}}{2}
                B^{2}
                -
                | \mathcal{D}_{k} \phi |^{2}
                - \lambda
                \frac{g^2}
                { 2\epsilon_{0} c^{2} \hbar^{2} }
                ( |\phi|^{2} - v^{2} )^{2}
        \bigg] \delta_{ij}
        +
        \overline{\mathcal{D}_{i} \phi}
        \mathcal{D}_{j} \phi
        +
        \overline{\mathcal{D}_{j} \phi}
        \mathcal{D}_{i} \phi
,
\end{equation}
and the momentum conservation is modified as
\begin{equation}
        \partial_{\mu} T\indices{^\mu_j}
        =
        s \frac{g}{\hbar}
        \bigg[
                B + \lambda
                \frac{g}{\epsilon_{0} c^{2} \hbar}
                ( |\phi|^{2} - v^{2} )
        \bigg] \partial_{j} \sigma
.\label{227}
\end{equation}
In the presence of spatially inhomogeneous part \(\sigma\), the right hand side of \eqref{227} is nonvanishing except for static BPS solutions satisfying the Bogomolny equation as discussed in \cite{Kim:2024gfn} and thus the momentum is not conserved in general.
This non-conservation of the momentum has been expected since the translation symmetry was explicitly broken by the spatially inhomogeneous part \(\sigma(\boldsymbol{x})\) \eqref{221}--\eqref{212}.
Since we are interested in the static configurations which are electrically neutral \( \rho = 0 \) without production of the electric field \( \boldsymbol{E} = 0 \) and chose the Weyl gauge condition \( \Phi = 0 \), the momentum density is shown to be zero and any object of our interest is spinless
\begin{equation}
        J =
        \int dz \int d^{2} \boldsymbol{x} \,
        \mathcal{J}
        =
        \int dz \int d^{2} \boldsymbol{x} \,
        \epsilon^{ij} x_{i} T\indices{^t_j}
        = 0
,\label{211}
\end{equation}
where \( J \) and \( \mathcal{J} \) denote angular momentum and its density, respectively.

Let us consider a cylindrically symmetric example of inhomogeneous part \(\sigma\) as the Gaussian function which is localized and convenient for numerical studies
\begin{equation}
        \sigma (\boldsymbol{x})
        = \sigma (r)
        = -\beta v_{0}^{2}
        e^{
                - \alpha^{2}
                ( \frac{r}{ \lambda_{\text{L}} } )^{2}
        }
,\label{203}
\end{equation}
where \( r = \sqrt{x^{2} + y^{2}} \) and two dimensionless parameters \(\alpha^{-1}\) and \(\beta\) control radial size from the \(z\)-axis and depth of the Gaussian inhomogeneous part, respectively.
Note that its shape takes a Gaussian dip for \(\beta>0\) and a Gaussian bump for \(\beta<0\) with trivial homogeneous case for \(\beta=0\).
For tractability and compatibility with \(\sigma(r)\) \eqref{203}, we focus on the cylindrically symmetric configurations obtained under the ansatz
\begin{align}
        \phi
        & =
        |\phi|(r) e^{in\theta}
,\label{206} \\
        A^{i}
        & =
        - \frac{\hbar}{g}
        \epsilon^{ij} x_{j}
        \frac{A(r)}{r^{2}}
.\label{207}
\end{align}
By substitution of the ansatz, the Euler-Lagrange equations \eqref{204} and \eqref{205} are reduced to the ordinary nonlinear differential equations for scalar amplitude \(|\phi|(r)\) and angular component of the \(\mathrm{U}(1)\) gauge field \(A(r)\)
\begin{align}
        \frac{d^{2} |\phi|}{d r^{2}}
        + \frac{1}{r} \frac{d |\phi|}{dr}
        & =
        \frac{1}{r^2}
        (A - n)^{2} |\phi|
        +
        \frac{1}{ \xi^{2} v_{0}^{2} }
        ( |\phi|^{2} - v^{2} ) |\phi|
,\label{208} \\
        \frac{d^{2} A}{d r^{2}}
        - \frac{1}{r} \frac{dA}{dr}
        & =
        \frac{1}{ \lambda_{\text{L}}^{2} v_{0}^{2} }
        (A-n) |\phi|^{2}
        + s \frac{1}
        { 2 \lambda_{\text{L}}^{2} v_{0}^{2}}
        \
        r \frac{d \sigma}{dr}
.\label{209}
\end{align}
With the ansatz \eqref{207}, magnetic flux \(\Phi_B\) is expressed by the boundary values for static cylindrically symmetric configurations
\begin{align}
        \Phi_{B} & =
        \int d^{2} \boldsymbol{x} \, B
        =
        \oint_{\textrm{C}_\infty}
        d \boldsymbol{l} \cdot \boldsymbol{A}
\\      & =
        \frac{ 2\pi\hbar }{g}
        [ A(\infty) - A(0) ]
.\label{220}
\end{align}

Attractive or repulsive nature of the net interaction between two distinct objects for various values of quartic scalar coupling \(\lambda\) with turning off inhomogeneous part \(\sigma=0\) is explained by the following qualitative argument.
When the correlation length \eqref{213} is larger than the London penetration depth \eqref{214} \( \xi >\sqrt{2} \lambda_{\text{L}} \), the range of attractive scalar interaction is longer than that of repulsive electromagnetic interaction and hence the net interaction is expected to become attractive at large distance.
When the correlation length \eqref{213} is shorter than the London penetration depth \eqref{214} \( \xi < \sqrt{2} \lambda_{\text{L}} \), the range of attractive scalar interaction is shorter than that of repulsive electromagnetic interaction and hence the net interaction is expected to become repulsive at large distance.
When the correlation length \eqref{213} is equal to the London penetration depth \eqref{214} \(\xi = \sqrt{2} \lambda_{\text{L}}\), the range of attractive scalar interaction is equal to that of repulsive electromagnetic interaction and hence the net interaction vanishes due to perfect balance at the critical quartic scalar coupling \(\lambda=1\).
Phenomenological implication of the net interaction is immediate in conventional superconductivity \cite{Arovas, tinkham2004introduction}.
When correlation length is longer (shorter) than London penetration depth \( \xi > \sqrt{2} \lambda_{\text{L}} \) (\( \xi < \sqrt{2} \lambda_{\text{L}} \)) in type-I (type-\Romtwo) superconducting materials, it is theoretically classified by the range of quartic scalar coupling \(\lambda<1\) (\(1<\lambda\)), respectively.
In the case of equal length scale \( \xi = \sqrt{2} \lambda_{\text{L}} \) corresponding to the boundary between type-I and type-\Romtwo\ superconductivity \cite{Abrikosov:1956sx}, the BPS limit of \(\lambda=1\) is achieved field-theoretically.

When the quartic scalar coupling \(\lambda\) of scalar self-interaction has the critical value \(\lambda=1\), BPS limit is saturated even in the presence of inhomogeneity \cite{Tong:2013iqa}.
In addition to the expected inhomogeneous BPS vortices, the unique inhomogneous BPS vacuum of zero energy, zero magnetic flux and zero spin is found by solving the Bogomolny equations \cite{Kim:2024gfn}.
Even without cylindrical symmetry, existence and uniqueness of this inhomogeous BPS vacuum are proven under appropriate finite square-integrable impurities \cite{Jeon:2024jbs}.

In the presence of inhomogeneity with the magnetic impurity term \eqref{210}, it is intriguing to explore the existence of the inhomogeneous non-BPS vacuum with the evaluation of characteristic quantities, e.g. energy, net magnetic flux, and spin away from the critical quartic scalar coupling \( \lambda = 1 \) for BPS configurations.
In addition, the affection of this inhomogeneity to the interaction between the Abrikosov-Nielsen-Olesen vortices is also worth studying.
The so-called Bogomolny trick is devised for the analysis of some specific field-theoretic models, e.g. the abelian Higgs model with the critical quartic scalar coupling \(\lambda=1\) for derivation of energy bound \cite{Bogomolny:1975de}.
Moreover, it is also applicable for the derivation of a lower bound of the energy and even when the quartic scalar coupling is arbitrary \cite{Jacobs:1978ch}.
When the inhomogeneity is turned on in the abelian Higgs model with the critical quartic scalar coupling, the existence of a lower energy bound looks not to be guaranteed because of the magnetic impurity term \eqref{210}.
Only after the introduction of an artificial sign \(s=\pm1\) of the magnetic impurity term, the trick for BPS bound works and leads to the conclusion that the inhomogeneous abelian Higgs model in the critical quartic scalar coupling admits either vortices for \(s=+1\) or anti-vortices for \(s=-1\) but not both \cite{Kim:2024gpu, Kim:2024gfn}.
The freedom to choose \(s=+1\) accompanied with nonnegative vorticity \(n\ge0\) without loss of generality provides the lower bound of zero energy and such vacuum of zero energy is obtained as a BPS configuration.
Since we are interested in the inhomogeneous abelian Higgs model with quartic scalar self-interaction involving the BPS limit of \(\lambda=1\), the similar logic is applicable for the non-BPS cases and some useful lower energy bounds for classical configurations are achieved as follows.
When the quartic scalar coupling is not less than the critical coupling \(\lambda\ge1\), a lower bound of the energy per unit length along the \(z\)-axis is achieved
\begin{align}
        \mathcal{E}_{n} (\lambda)
        & =
        \int d^{2} \boldsymbol{x} \,
        \bigg[
                \frac{1}{2} \epsilon_{0} \boldsymbol{E}^{2}
                + 
                \frac{1}{v_{\text{p}}^2}
                | \mathcal{D}_{t} \phi |^{2}
                +
                | ( \mathcal{D}_{1} + is \mathcal{D}_{2} ) \phi|^{2}
\nonumber
\\ & \hspace{5em}
                +
                \frac{1}{2} \epsilon_{0}c^{2}
                \Big|
                        B +
                        s\frac{g}{\epsilon_{0} c^{2} \hbar}
                        ( |\phi|^{2} - v^{2}(\boldsymbol{x}) )^{2}
                \Big|^{2}
        \bigg]
\nonumber
\\ & \quad
        + (\lambda-1)
        \frac{g^2}{2 \epsilon_{0} c^{2} \hbar^{2}}
        \int d^{2} \boldsymbol{x} \,
        ( |\phi|^{2} - v^{2}(\boldsymbol{x}) )^{2}
        + s \frac{g v_{0}^{2}}{\hbar} \Phi_{B}
\nonumber
\\ & \ge
        s\frac{gv_{0}^{2}}{\hbar} \Phi_{B}
\nonumber
\\ & =
        2 \pi v_{0}^{2} |sn|
,\label{222}
\end{align}
where the last equality holds for \(n=0\) inhomogeneous vacuum or \(n\neq0\) vortices and anti-vortices, and \(s\) is chosen as \(s=+1\) (\(s=-1\)) with nonnegative (nonpositive) vorticity \(n\ge0\) (\(n\le0\)).
When it is less than the critical quartic scalar coupling \(\lambda<1\), the terms of energy are reorganized and another lower bound of the energy per unit length along the \(z\)-axis is achieved again
\begingroup \allowdisplaybreaks
\begin{align}
        & \mathcal{E}_{n} (\lambda)
\nonumber \\
        =\, &
        \lambda \int d^{2} \boldsymbol{x} \,
        \bigg[
                \frac{1}{2} \epsilon_{0} \boldsymbol{E}^{2}
                + 
                \frac{1}{v_{\text{p}}^2}
                | \mathcal{D}_{t} \phi |^{2}
                +
                | ( \mathcal{D}_{1} + is \mathcal{D}_{2} ) \phi|^{2}
                +
                \frac{1}{2} \epsilon_{0}c^{2}
                \Big|
                        B +
                        s\frac{g}{\epsilon_{0} c^{2} \hbar}
                        ( |\phi|^{2} - v^{2}(\boldsymbol{x}) )^{2}
                \Big|^{2}
        \bigg]  
\nonumber
\\ & \quad
        + (1-\lambda) \int d^{2} \boldsymbol{x} \,
        \bigg[
                \frac{1}{2} \epsilon_{0} \boldsymbol{E}^{2}
                + 
                \frac{1}{v_{\text{p}}^2}
                | \mathcal{D}_{t} \phi |^{2}
                +
                | ( \mathcal{D}_{1} + is \mathcal{D}_{2} ) \phi|^{2}
                +
                \frac{1}{2} \epsilon_{0}c^{2}
                \Big(
                        B -
                        s\frac{g}{\epsilon_{0} c^{2} \hbar}
                        \sigma(\boldsymbol{x})
                \Big)^{2}
        \bigg]
\nonumber
\\      & \quad
        + s \lambda \frac{gv_{0}^{2}}{\hbar} \Phi_{B}
        - (1-\lambda)
        \frac{g^2}{2 \epsilon_{0} c^{2} \hbar^{2}}
        \int d^{2} \boldsymbol{x} \,
        \sigma^{2} (\boldsymbol{x})
\nonumber
\\
        > \, &
        s \lambda \frac{gv_{0}^{2}}{\hbar} \Phi_{B}
        - (1-\lambda)
        \frac{g^2}{2 \epsilon_{0} c^{2} \hbar^{2}}
        \int d^{2} \boldsymbol{x} \,
        \sigma^{2} (\boldsymbol{x})
\nonumber
\\      = \, &
        2 \pi \lambda v_{0}^{2} |sn|
        - (1-\lambda) \frac{g^2}
        {2 \epsilon_{0} c^{2} \hbar^{2}}
        \int d^{2}\boldsymbol{x} \,
        \sigma^{2}(\boldsymbol{x})
.\label{223}
\end{align}
\endgroup
The last term in \eqref{223} is always negative for any inhomogeneous part \(\sigma(\boldsymbol{x})\) with \(\lambda<1\) and thus it may imply that the energy per unit length along the \(z\)-axis
\(
        \mathcal{E}_{n}
\)
for inhomogeneous vacuum can possibly be negative.
For the Gaussian inhomogeneous part \eqref{203}, the last line of \eqref{223} is computed
\begin{equation}
        \mathcal{E}_{n}
        ( \lambda, \alpha^{-1}, \beta )
        >
        2 \pi v_{0}^{2}
        \bigg[
                \lambda |sn|
                - (1-\lambda)
                \frac{\beta^2}{16 \alpha^2}
        \bigg]
.\label{224}
\end{equation}
The lower bounds \eqref{222} and \eqref{224} are given by the dashed red lines in Figure \ref{fig:303} for inhomogeneous vacuum of \(n=0\) in the next section, by those in Figure \ref{fig:403} for inhomogeneous vortex of \(n=1\) in the subsection \ref{subsec:vortex}, and by those in Figure \ref{fig:412} for two inhomogeneous vortices superimposed at the origin in the subsection \ref{subsec:vortices}.
Notice that these rigorously proven lower bounds and numerically found vacuum solutions given by the the blue solid curves match exactly in the BPS limit of the critical quartic scalar coupling \(\lambda=1\).

We conclude this section by commenting on the possible origin of the magnetic impurity term \eqref{210} in the action \eqref{200}.
Let us consider an abelian Higgs model of \( \mathrm{U}(1) \times \mathrm{U}(1) \) gauge symmetry with two complex scalar fields, \(\phi\) and \(\tilde{\phi}\), and two gauge fields, \(\hat{A}_{\mu}\) and \(\tilde{A}_{\mu}\), whose Lagrangian density is
\begin{align}
        \mathscr{L}_{\text{U1U1}}
        = &
        - \frac{\epsilon_{0} c^{2}}{4}
        \hat{F}_{\mu\nu} \hat{F}^{\mu\nu}
        - \frac{\epsilon_{0} c^{2}}{4}
        \tilde{F}_{\mu\nu} \tilde{F}^{\mu\nu}
        - \overline{ \mathcal{D}_{\mu} \phi }
        \mathcal{D}^{\mu} \phi 
        - \overline{ D_{\mu} \tilde{\phi} }
        D^{\mu} \tilde{\phi}
        - \lambda \frac{g^2}
        { 2 \epsilon_{0} c^{2} \hbar^{2} }
        ( |\phi|^{2} - v_{0}^{2} )^{2}
\nonumber
\\      &
        - \frac{g^2}
        { 2 \epsilon_{0} c^{2} \hbar^{2} }
        ( - |\phi|^{2} + |\tilde{\phi}|^{2} - \tilde{v}_{0}^{2} )^{2}
,
\end{align}
where covariant derivatives are
\begin{equation}
        \mathcal{D}_{\mu} \phi
        =
        \Big( \partial_{\mu}
        - i \frac{g}{\hbar} \hat{A}_{\mu}
        + i \frac{g}{\hbar} \tilde{A}_{\mu}
        \Big) \phi
\quad \text{and} \quad
        D_{\mu} \tilde{\phi}
        =
        \Big( \partial_{\mu}
        - i \frac{g}{\hbar} \tilde{A}_{\mu}
        \Big) \tilde{\phi}
.
\end{equation}
Notice that one scalar potential is in non-BPS case of arbitrary quartic scalar coupling \(\lambda\) and the other scalar potential is in the BPS limit.
After the energy density \( - T\indices{^t_t} \) is read for static configuration and the heavy scalar and gauge fields, \(\tilde{\phi}\) and \(\tilde{A}_{\mu}\), are integrated with the help of BPS equations as done in the BPS case \cite{Tong:2013iqa}, it becomes
\begin{align}
        - T\indices{^t_t}
        = &
        \frac{\epsilon_{0} c^{2}}{2} B^{2}
        + | \mathcal{D}_{i} \phi |^{2}
        + \lambda \frac{g^2}
        { 2\epsilon c^{2} \hbar^{2} }
        ( |\phi|^{2} - v_{0}^{2} - \sigma(\boldsymbol{x}) )^{2}
        + s \lambda
        \frac{g}{\hbar} \sigma(\boldsymbol{x}) B
\nonumber
\\      &
        + | ( D_{1}     + i s D_{2} )
                \tilde{\phi} |^{2}
        + \frac{\epsilon_{0} c^{2}}{2}
        \Big|
                \tilde{B}
                + s \frac{g}
                { \epsilon_{0} c^{2} \hbar }
                ( - |\phi|^{2} + |\tilde{\phi}|^{2} - \tilde{v}_{0}^{2} )
        \Big|^{2}
        + \cdots
,
\label{228}
\end{align}
where
\(
        \mathcal{D}_{i} \phi
        =
        ( \partial_{i} - i g A^{i} / \hbar )
        \phi
\)
\eqref{229} with the help of \( A_{\mu} = \hat{A}_{\mu} - \tilde{A}_{\mu} \), \(\cdots\) in the last line means omission of the field-independent terms, and the last two absolute square terms vanish for any static BPS solutions.
The action corresponding to the energy density \eqref{228} depicts an abelian Higgs model including a magnetic impurity term with the inhomogeneous part \(\sigma(\boldsymbol{x})\) and is the same as the action of our consideration \eqref{200} except for the coupling constant \(\lambda g\) of the magnetic impurity term instead of \( g \).
Since a plausible coupling between an externally given inhomogeneity \(\sigma(\boldsymbol{x})\) and the magnetic field \(B\) is the gauge coupling constant \(g\) independent of the quartic scalar coupling \(\lambda\), we focus on the action \eqref{200} throughout this work. This analysis in non-BPS
case also reconciles the property that the fixed sign of the magnetic impurity term $s=\pm1$ allows either
a nonnegative winding sector or a nonpositive winding sector but not both, which is
strictly proved in the BPS limit~\cite{Kim:2024gpu,Kim:2024gfn}.

\section{Inhomogeneous Vacuum of Nonzero Energy}
\label{sec:vac}

In this section, we examine the coupled equations \eqref{208}--\eqref{209} for various quartic scalar coupling \(\lambda\) and find inhomogeneous symmetry-broken vacuum solution with characteristic quantities.
According to the discussion in the previous section, we choose \(s=+1\) with nonnegative vorticity \(n\ge0\) without loss of generality.
We figure out \(\lambda\)-dependence of the energy of inhomogeneous non-BPS vacuum, particularly the sign of it.

In homogeneous limit of \( \sigma(\boldsymbol{x}) = 0 \), the constant scalar field \( \phi = v_{0} \) with zero vorticity \( n = 0 \) and zero magnetic field \( B = 0 \) is the configuration of minimum zero energy and is identified as the symmetry-broken Higgs vacuum for arbitrary positive quartic scalar coupling.
Once inhomogeneity is turned on \( \sigma(\boldsymbol{x}) \neq 0 \), this constant configuration of \(\phi = v_{0} \) and \(A^{i}=0\) can neither be a solution of the equations of motion \eqref{208}--\eqref{209} nor has zero energy.
It has rather positive impurity energy contributed purely by the inhomogeneous part \(\sigma(\boldsymbol{x})\)
\begin{equation}
        \mathcal{E}^{v_0}_{\sigma}
        =
        \frac{\lambda}
        { 4 \lambda_{\text{L}}^{2} v_{0}^{2} }
        \int d^{2} \boldsymbol{x} \,
                \sigma^{2} (\boldsymbol{x})
        > 0
.
\end{equation}
When the inhomogeneous part \(\sigma\) is Gaussian  \eqref{203}, the impurity energy is computed
\begin{equation}
        \mathcal{E}^{v_0}_{\sigma}
        = 2\pi v_{0}^{2} \lambda
        \frac{\beta^{2}}{16\alpha^{2}}
        > 0
.\label{308}
\end{equation}
On the other hand, for any nontrivial \( v(\boldsymbol{x}) \) obeying the boundary behavior at spatial infinity \eqref{212}, \( \phi = v(\boldsymbol{x}) \) with \( A^{i} = 0 \) is evidently neither a solution of the equations of motion \eqref{208}--\eqref{209} and has positive energy
\begin{equation}
        \mathcal{E}^{v(\boldsymbol{x})}_{\sigma}
        =
        \int d^{2} \boldsymbol{x} \,
        [ \partial_{i} v(\boldsymbol{x}) ]^{2}
        > 0
.
\end{equation}
For the Gaussian inhomogeneous part \eqref{203}, the computed impurity energy is independent of the size parameter \(\alpha^{-1}\) and depends only on the depth parameter \(\beta\) nontrivially
\begin{equation}
        \mathcal{E}^{v(\boldsymbol{x})}_{\sigma}
        = \pi v_{0}^{2}
        \bigg[
                - \frac{\pi^2}{6}
                - \beta
                + \frac{1}{2}
                ( \operatorname{Li}_{1} \beta )^{2}
                - \operatorname{Li}_{1} \beta
                \operatorname{Li}_{1} (1+\beta)
                + \operatorname{Li}_{2}
                \Big(
                        \frac{1}{1-\beta}
                \Big)
        \bigg]
,\label{311}
\end{equation}
where it approaches to zero minimum value at \(\beta=0\) as expected and becomes singular for \(\beta\ge1\) that implies invalidity of the formula \eqref{311} of the deep Gaussian dips.
Therefore, static minimum energy configuration should be explored by solving the coupled nonlinear Euler-Lagrange equations \eqref{208}--\eqref{209}.

To be specific, we look for the case with cylindrical symmetry for which the Gaussian type inhomogeneity \eqref{203} and the ansatz \eqref{206}--\eqref{207} are assumed.
If we have a configuration to be a candidate of the vacuum or a soliton, its energy per unit length along the \(z\)-axis of each static configuration should be taken into account and favorably has a finite value.
The energy of static neutral objects is read from the energy expression \eqref{225}
\begin{align}
        \mathcal{E}
        & =
        \int d^{2} \boldsymbol{x} \,
        \bigg[
                \frac{1}{2} \epsilon_{0} c^{2} B^{2}
                +
                | \mathcal{D}_{i} \phi |^{2}
                +
                \frac{\lambda}
                { 4 \lambda_{\text{L}}^{2} v_{0}^{2} }
                \big(
                        |\phi|^{2} - v^{2}(\boldsymbol{x})
                \big)^{2}
                -
                \frac{g}{\hbar} \sigma(\boldsymbol{x}) B
        \bigg]
\label{305} \\
        & =
        2\pi \int_{0}^{\infty} dr \,
        \bigg[
                \frac{ \lambda_{\text{L}}^{2} v_{0}^{2} }{r}
                \Big( \frac{dA}{dr} \Big)^{2}
                +
                r \Big( \frac{d|\phi|}{dr} \Big)^{2}
                +
                \frac{ (A-n)^2 }{r} |\phi|^{2}
                +
                \frac{\lambda}
                { 4 \lambda_{\text{L}}^{2} v_{0}^{2} } r
                \big(
                        |\phi|^{2} - v^{2}(\boldsymbol{x})
                \big)^{2}
\nonumber\\
                & \hspace{6em}
                - \sigma \frac{dA}{dr}
        \bigg]
,\label{309}
\end{align}
where the last equality holds for cylindrically symmetric configurations.

In order to be the vacuum solution, topologically trivial sector of zero vorticity \(n=0\) is taken into account, which guarantees minimum energy in the BPS limit \(\lambda=1\) \cite{Kim:2024gfn} and is probably consistent with minimum energy for other non-BPS case by continuity of the quartic scalar coupling \(\lambda\).
For the solution of cylindrical symmetry, finiteness of energy \(\mathcal{E}\) and regular behavior of the vacuum configuration assign the appropriate boundary conditions at spatial infinity and at the origin, respectively,
\begin{align}
        \lim_{r \to \infty} |\phi|
        = v_{0}
, & \qquad
        \lim_{r \to \infty} A
        = 0
,\label{306} \\
        \lim_{r \to 0}
        \frac{ d|\phi| }{dr}
        = 0
, & \qquad
        \lim_{r \to 0} A
        = 0
.\label{307}
\end{align}

For small \(r\), the fields are expanded up to a few terms as a part of power series
\begin{align}
        |\phi| (r)
        & \approx
        v_{0}
        \bigg\{
                \phi_{0}
                +
                \frac{1}{4} \phi_{0}
                ( \phi_{0}^{2} + \beta - 1 )
                \Big( \frac{r}{\xi} \Big)^{2}
\nonumber \\
        & \quad \qquad + 
                \frac{1}{64} \phi_{0}
                \Big[
                        ( \phi_{0}^{2} + \beta - 1 )^{2}
                        +
                        2 \phi_{0}^{2}
                        ( \phi_{0}^{2} + \beta - 1 )
                        + \frac{16 a_{0}^{2}} {\lambda^2}
                        - \frac{8 \alpha^{2} \beta}{\lambda}
                \Big]
                \Big( \frac{r}{\xi} \Big)^{4}
                + \cdots
        \bigg\}
,\label{301}
\\
        A(r) 
        & \approx
        a_{0} \Big(
                \frac{r}{ \lambda_{\text{L}} }
        \Big)^{2}
        +
        \frac{
                a_{0} \phi_{0}^{2}
                + \alpha^{2} \beta
        }{8}
        \Big( \frac{r}{ \lambda_{\text{L}} } \Big)^{4}
\nonumber \\
        & \quad
        + \frac{
                2 \lambda a_{0} \phi_{0}^{2}
                ( \phi_{0}^{2} + \beta - 1 )
                +
                \phi_{0}^{2}
                ( a_{0} \phi_{0}^{2}
                + \alpha^{2} \beta )
                - 8 \alpha^{4} \beta
        }{192}
        \Big( \frac{r}{ \lambda_{\text{L}} } \Big)^{6}
        + \cdots
,\label{302}
\end{align}
where \(\phi_0\) and \(a_0\) are two undetermined constants fixed by proper asymptotic behaviors.
Notice that the scalar amplitude \(|\phi|\) increases for \( 1 - \phi_{0}^{2} < \beta \) and decreases for \( 1 - \phi_{0}^{2} > \beta ~ (0 < \phi_{0} < 1) \) near the origin.
The quartic scalar coupling \( \lambda \) does not appear explicitly until the third order in power series expansion \eqref{301}--\eqref{302}.
Fields approach exponentially the boundary values \eqref{306} for sufficiently large \(r\)
\begin{align}
        |\phi|(r)
        & \approx
        v_{0} \bigg[
                1 -
                \phi_{\infty}
                K_{0}
                \Big( \frac{\sqrt{2}r}{\xi} \Big)
        \bigg]
,\label{303}
\\
        A(r)
        & \approx
        a_{\infty}
        \frac{r}{ \lambda_{\text{L}} }
        K_{1}
        \Big( \frac{r}{ \lambda_{\text{L}} } \Big)
,\label{304}
\end{align}
where \(\phi_\infty\) and \(a_\infty\) are undetermined constants.

\begin{figure}[H]
        \centering
        \subfigure[]{
                \includegraphics[
                        width=0.45\textwidth
                ]{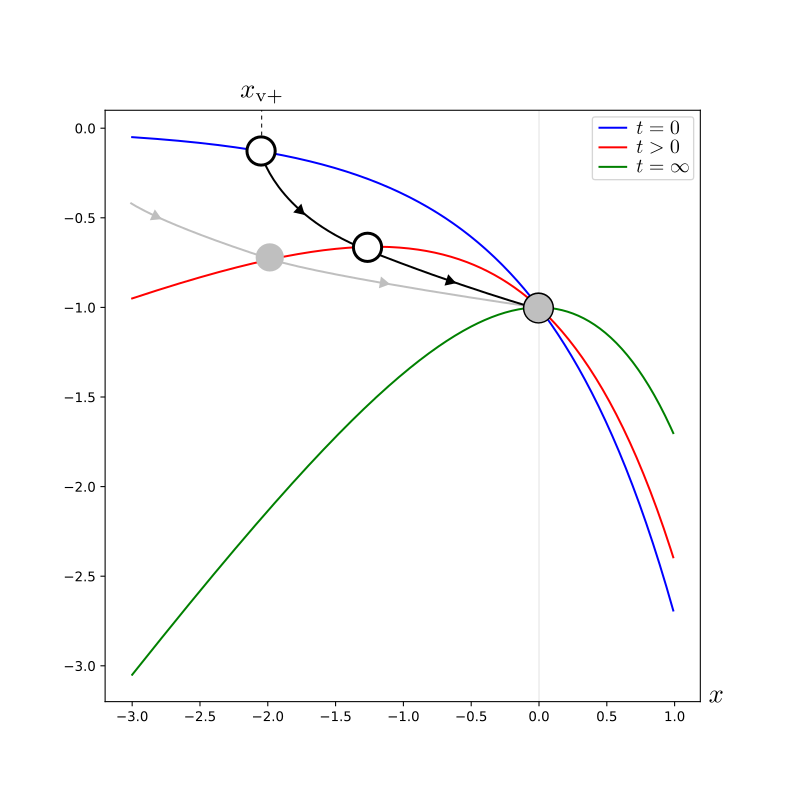}
        }
        \hfill
        \subfigure[]{
                \includegraphics[
                        width=0.45\textwidth
                ]{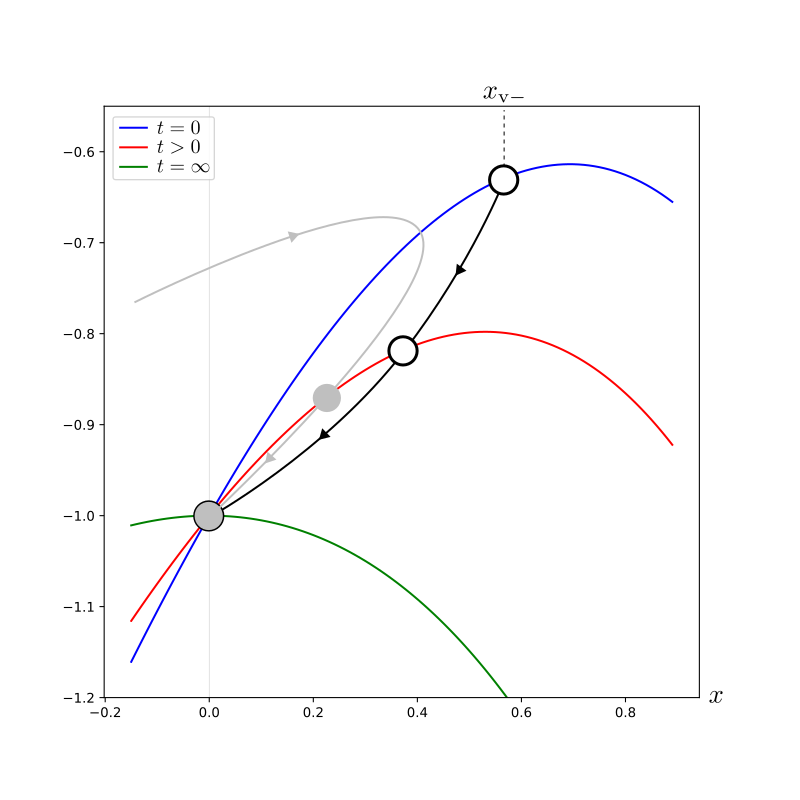}
        }
        \caption{
                The black trajectories including white balls stand for cylindrically symmetric BPS vacuum solutions and the gray trajectories including gray balls stand for cylindrially symmetric vortex solutions of nonzero vorticity \(n\):
                (a) \(\beta>0\), (b) \(\beta<0\).
        }
\label{fig:312}
\end{figure}

Profiles of the scalar amplitude \(|\phi|(r)\) are expected to connect smoothly the small \(r\) behavior \eqref{301} and the larger behavior \eqref{303}.
Since the Euler-Lagrange equations for cylindrically symmetric vacuum solutions of \(n=0\) are given by the two coupled second order nonlinear ordinary differential equations \eqref{208}--\eqref{209} and are not exactly solved in the presence of Gaussian inhomogeneous part \eqref{203} for arbitrary quartic scalar coupling, we go back to the BPS limit of critical quartic scalar coupling \( \lambda = 1 \) for tractable mathematical analysis, which lefts a Bogomolny equation \cite{Kim:2024gfn}, a single second order nonlinear equation
\begin{equation}
        \nabla^{2} \ln |\phi|^{2}
        = 
        \frac{1}{\lambda_{\rm L}^{2} v_{0}^{2}}
        [ |\phi|^{2} - v^{2}(\boldsymbol{x}) ]
.\label{312}
\end{equation}
Let us predict possible scalar profile of the vacuum solutions before numerical analysis by applying the Newtonian shooting analogy \cite{Kim:1992mm} to the equation \eqref{312}.
For cylindrically symmetric inhomogeneous vacuum solutions, the partial differential equation \eqref{312} becomes an ordinary differential equation
\begin{equation}
        \frac{d^2}{d \tilde{r}^{2}}
        \ln f(\tilde{r})
        =
        - \frac{1}{\tilde{r}}
        \frac{d}{d \tilde{r}}
        \ln f(\tilde{r})
        +
        \bigg[
                f (\tilde{r})
                -
                \frac{ v^{2} (\tilde{r}) }
                {v_{0}^{2}}
        \bigg]
,\label{313}
\end{equation}
where \(\tilde{r} = r/\lambda_{\rm L}\) and \( f(\tilde{r}) = (|\phi| / v_{0})^{2} \) are dimensionless rescaled quantities.
If we think of \(\tilde{r}\) as the time \(t\) and \(\ln f\) as the position \(x\) of a hypothetical particle, the rewritten equation
\begin{equation}
        \frac{d^{2} x}{d t^{2}}
        =
        - \frac{1}{t} \frac{dx}{dt}
        - \frac{\partial U}{\partial x}
\label{314}
\end{equation}
can be interpreted as the Newton's second law for one-dimensional motion of a particle of unit mass subject to a friction with a coefficient inversely proportional to the time \(1/t\) and the force derived from the time-dependent effective potential \(U(x, t)\)
\begin{equation}
        U(x,t)
        =
        - e^{x}
        + (1 - \beta e^{ - \alpha^{2} t^{2} } ) x
,\label{315}
\end{equation}
whose change along time is shown by the blue-, red-, and green-colored solid curves in Figure \ref{fig:312}-(a) for positive \(\beta\) and Figure \ref{fig:312}-(b) for negative \(\beta\).
Then the configuration of each vacuum solution which we are looking for can be interpreted as the description of a motion.
We introduce the energy \(\mathscr{E}\) of the particle of unit mass for convenient description of motion
\begin{equation}
        \mathscr{E}
        =
        \frac{1}{2}
        \Big( \frac{dx}{dt} \Big)^{2}
        + U(x,t)
.\label{316}
\end{equation}
For a given effective potential \(U(x, t)\) \eqref{315} with fixed size parameter \(\alpha^{-1} (\ge0)\) and depth parameter \(\beta~(\beta\neq0)\), we now argue the existence and uniqueness of a cylindrically symmetric vacuum solution satisfying the conditions
\(
        x (t=0) = x_{\text{v}}
        =
        \ln [ |\phi|(r=0) / v_{0} ]
\)
from a boundary condition \eqref{307} and \( x (t=\infty) = x_{\infty} = 0 \) from the other boundary condition \eqref{306}.

When depth parameter \(\beta\) is positive \(\beta>0\), the initial position \(x_{\text{v}+}\) should have a negative value in order to stop finally at \(x=0\).
If \(x_{\text{v}+}\) is chosen too close to zero on the effective potential curve of the blue-colored curve in Figure \ref{fig:312}-(a), say \( x_{\text{v}+}^{\infty} < x_{\text{v}+} < 0 \), the hypothetical particle passes zero at a finite time and diverges to positive infinity.
If \(x_{\text{v}+}\) is chosen too far from zero, say \( x_{\text{v}+} < x_{\text{v}+}^{-\infty} < 0 \), its energy \(\mathscr{E}\) \eqref{316} decreases too much and has a smaller value than negative unity \(\mathscr{E} < -1\) at a finite time because of the dissipation due to the friction and lowering of the time-dependent effective potential.
This particle stops at a maximum negative position at a finite time and turns back to negative infinity, and hence it can never arrive at the hilltop at the origin \(x=0\) of the effective potential at infinite time \( U(x, t=\infty) \) illustrated by the green-colored solid curve in Figure \ref{fig:312}-(a).
Thus, by continuity of the initial condition \(x_{\text{v}+}\), there is an appropriate initial position \(x_{\text{v}}\) in the region
\(
        x_{\text{v}+}^{-\infty}
        < x_{\text{v}+}
        < x_{\text{v}+}^{\infty}
        < 0
\)
between the undershoot motions obeying \( x(t=\infty) = -\infty \) and the overshoot motions obeying \( x(t=\infty) = +\infty \), for which the particle stops initially at the position \( x(t=0) = x_{\text{v}+} \), moves monotonically to the right, and stops finally at the destination \( x(t=\infty) = 0 \) on the hilltop
\( U (x=0, t=\infty) = -1 \)
of the effective potential \( U(x, t=\infty) \).
This is the cylindrically symmetric symmetry-broken vacuum solution in the BPS limit of critical quartic scalar coupling \(\lambda=1\) in the presence of a Gaussian dip \eqref{203} of positive depth parameter \(\beta>0\), whose trajectory is illustrated by the black-colored solid curve in Figure \ref{fig:312}-(a).

When depth parameter is negative \(\beta<0\), shapes of the effective potential change and its peak position always appears at positive domain \(x>0\) as shown in Figure \ref{fig:312}-(b).
Nevertheless, its shapes have a single peak and common negative infinite values at spatial infinities \(U(\pm\infty, t) = -\infty\).
Thus, according to the same argument based on the continuity, there is a motion which stops at an appropriate positive initial position \( 0 < x_{\text{v}-} = x(t=0) \) at initial time \(t=0\), moves monotonically to the left as time elapses, and stops finally at the destination \( x(t=\infty) = 0 \) at the hilltop
\( U (x=0, t=\infty) = -1 \)
of the effective potential \( U(x, t=\infty) \).
This is also the cylindrically symmetric symmetry-broken vacuum in the BPS limit of critical quartic scalar coupling \(\lambda=1\) in the presence of a Gaussian bump \eqref{203} of negative depth parameter \(\beta<0\), whose trajectory is illustrated by the black-colored solid curve in Figure \ref{fig:312}-(b).

In synthesis, for a given Gaussian inhomogeneous part of fixed size and depth parameters, there exists a unique nontrivial symmetry-broken vacuum solution in the BPS limit of critical quartic scalar coupling, which connects smoothly the boundary value of \( 0 < |\phi|(r=0) \) and \( |\phi|(r=\infty) = v_{0} \).
For a Gaussian dip of positive depth parameter, the vacuum solution increases monotonically with \( |\phi| (r=0) < v_{0} \), and, for a Gaussian bump of negative depth parameter, it decreases monotonically with \( |\phi| (r=0) > v_{0} \).
Though our proof is valid only for the solutions in the BPS limit of critical quartic scalar coupling \( \lambda = 1 \), the behavior of these vacuum solutions classified into two, one category of monotonically increasing solutions for \(\beta>0\) and the other category of monotonically decreasing solutions for \(\beta<0\), are probably kept for arbitrary quartic scalar coupling \( \lambda \). They will be confirmed by numerical works later in Figure \ref{fig:308}.
Note that the proof of existence and uniqueness of the vacuum solutions in the BPS limit is made also without cylindrical symmetry under an appropriate inequality on the amount of the inhomogeneous part \(\sigma (\boldsymbol{x})\) \cite{Jeon:2024jbs}.
To be specific, the existence and uniqueness of solution are proven with the mathematical rigor for arbitrary impurity satisfying a upper bound condition on its \(L^{2}(\mathbb{R}^{2})\)-norm
\begin{equation}
        \lVert \sigma \rVert^{2}_{2}
        =
        \int d^{2} x \,
        \sigma^{2}(\boldsymbol{x})
        <
        \frac{2 v_{0}^{4}}{\pi}
.
\end{equation}
In the aspects of various experiments of samples with diversified shape of irregular impurities, the condition implies that the vacuum solutions are always guaranteed as long as the total amount of the impurity is under control regardless of its shape.
It is imperative to produce similar existence theorems for non-BPS system, which probably require sophisticated methods even in modern mathematical theories of nonlinear partial differential equations.

\begin{figure}[H]
        \centering
        \includegraphics[
                width=0.65\textwidth
        ]{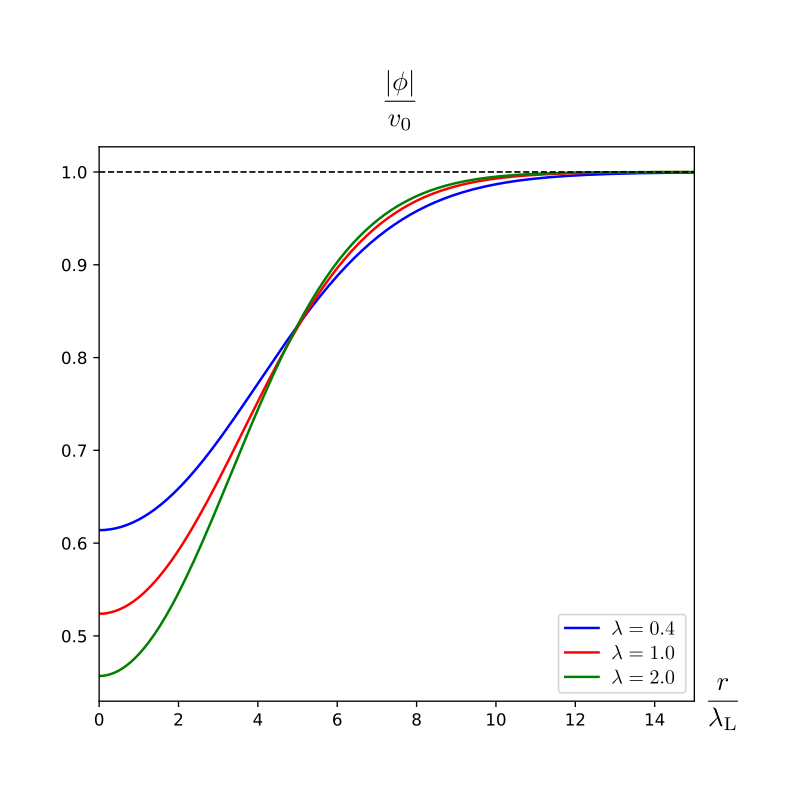}
        \caption{
                Amplitudes of the complex scalar field for inhomogeneous vacuum solution of various values of the quartic scalar coupling \(\lambda\) and for Gaussian inhomogeneous part with fixed dimensionless size parameter \( \alpha^{-1} = \sqrt{20} \approx 4.47 \) and depth parameter \(\beta = 1\).
        }
\label{fig:301}
\end{figure}

\begin{figure}[H]
        \centering
        \subfigure[]{
                \includegraphics[
                        width=0.45\textwidth
                ]{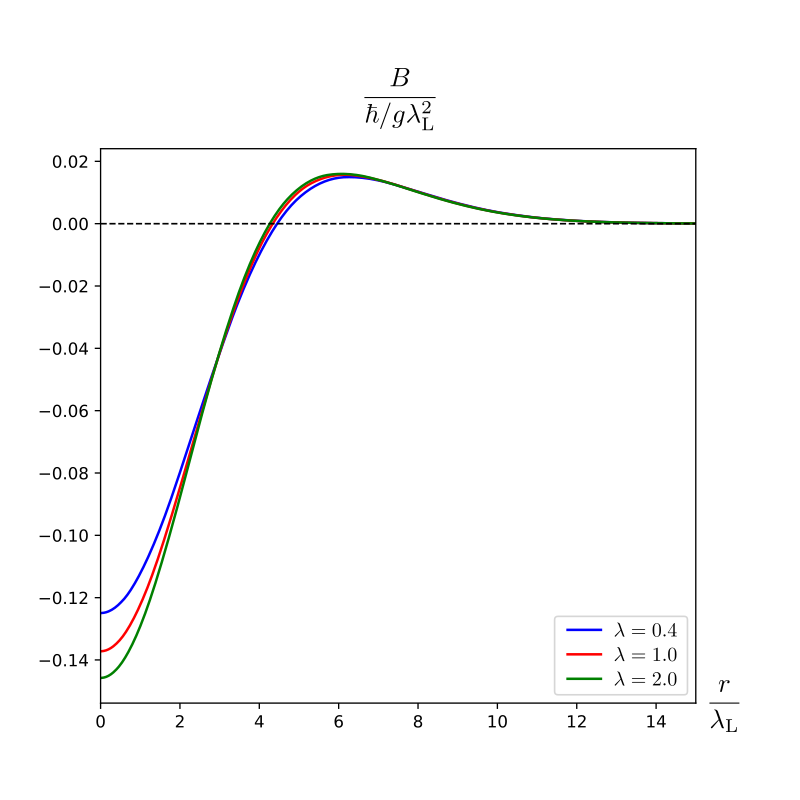}
        }
        \hfill
        \subfigure[]{
                \includegraphics[
                        width=0.45\textwidth
                ]{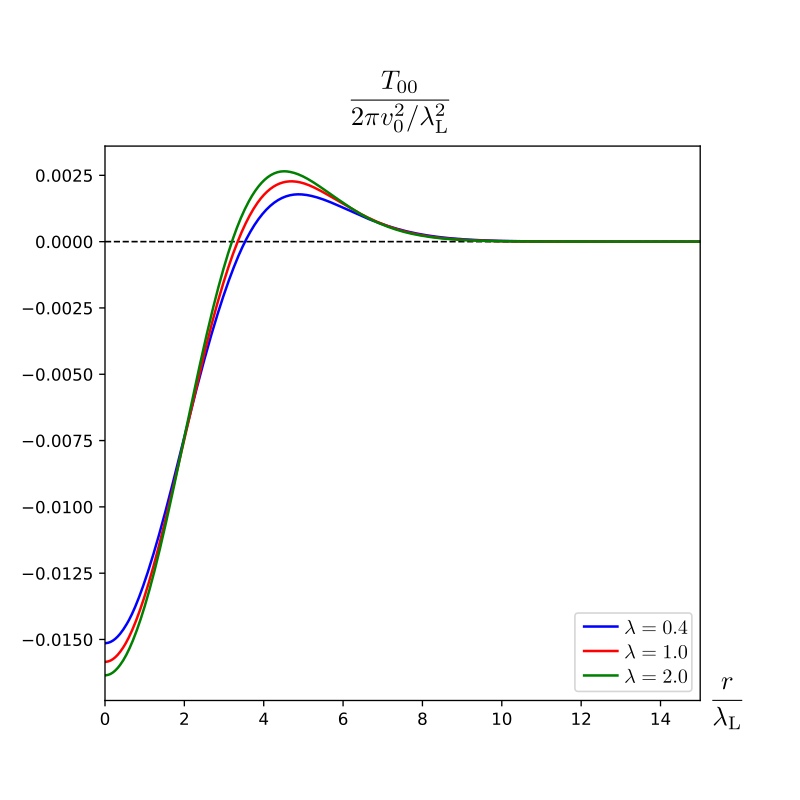}
        }
        \caption{
                (a) Magnetic fields \(B\) with
                \(
                        ( \lambda, r_{B \text{max}} /
                        \sqrt{2} \lambda_{\text{L}} )
                        = (0.4, 4.43),~(1.0, 4.36),~(2.0, 4.31)
                \)
                and (b) energy densities \(-T\indices{^t_t}\) with
                \(
                        ( \lambda, r_{ \max (-T\indices{^t_t}) } / 
                        \sqrt{2} \lambda_{\text{L}} )
                        = (0.4, 3.45),~(1.0, 3.32),~(2.0, 3.19)
                \)
                of the inhomogeneous vacuum.
                Various values of the quartic scalar coupling \(\lambda\) are taken into account for Gaussian inhomogeneous part with dimensionless size parameter \( \alpha^{-1} = \sqrt{20} \approx 4.47 \) and depth parameter \(\beta = 1\).
        }
\label{fig:302}
\end{figure}

Scalar amplitude \(|\phi| (r) \) begins from a minimum value \(v_{0} \phi_{0}\) at the origin \( r = 0 \), increases monotonically due to a Gaussian dip of depth parameter \(\beta=1>0\), and approaches rapidly the constant vacuum expectation value \(v_0\) at spatial infinity \(r\to\infty\).
As the quartic scalar coupling \(\lambda\) increases, the correlation length \(\xi\) \eqref{213} decreases and hence the three curves in Figure \ref{fig:301} show more rapid increase near the origin and more rapid approach of the boundary value \(v_0\) at large distance.
In addition, dimensionless initial value \( \phi_{0}= \phi(r=0) / v_{0} $ $(0<\phi_{0} < 1) \) of scalar field at the origin also decreases.
Those \(\lambda\)-dependent curves in inhomogeneous abelian Higgs model are different from the \(\lambda\)-independent constant vacuum \(v_0\) of the homogeneous abelian Higgs model, as shown by the three solid curves and the horizontal dashed line of \( |\phi|(r) = v_{0} \) in Figure \ref{fig:301}.
\noindent%
Every curve of both magnetic field \(B(r)\) and energy density \(-T\indices{^t_t}(r)\) begins with a negative value at the origin, increases to a positive maximum value at a finite radial variable \(r\), and decreases to zero at large distance as shown in Figure \ref{fig:302}-(a)--(b).
Thus, in \((r, \theta)\)-plane, all of them form ring-shapes with a flip from negative to positive direction.
As the quartic scalar coupling \(\lambda\) increases, the corresponding minimum value decreases but the maximum value slightly increases as shown in Figure \ref{fig:302}-(a).
Since the magnetic field \(B(r)\) is a physical quantity of electromagnetism, the observed fact that the dimensionless positions
\(      
        r_{B \text{max}} /
         \sqrt{2} \lambda_{\text{L}} 
\) of peak values
of the magnetic field are almost independent of the quartic scalar coupling \(\lambda\) is understandable.
Notice that the positions \( r_{B \text{max}} \) of peak values are located near
\(
        r\approx
        \alpha^{-1} \sqrt{2}\lambda_{\text{L}}
        \approx 4.47 \sqrt{2}\lambda_{\text{L}}
\)
as stated in the caption of Figure \ref{fig:302}-(a).
Although such a coincidence looks natural, it lacks a theoretical explanation yet.
\noindent%
For any obtained configuration, both charge density \(\rho\) \eqref{219} and angular momentum density \(\mathcal{J}\) \eqref{211} vanish everywhere by static condition without electric field and so do the electric charge
\(      \displaystyle
        \int d^{2}x \,
        \rho = 0
\)
and spin \(J=0\) \eqref{211}.
It also does not carry magnetic flux \(\Phi_{B}=0\) \eqref{220} due to boundary values \( A(0) = 0 = A(\infty) \) \eqref{306}--\eqref{307} but the magnetic field \(B\) does not vanish but rather varies spatially in order to modulate the effect of externally introduced inhomogeneity \( \sigma(\boldsymbol{x}) \neq 0 \) and to minimize its energy \(\mathcal{E}_{0}(\lambda)\) \eqref{309} as illustrated in Figure \ref{fig:302}-(a).

\begin{figure}[H]
        \centering
        \includegraphics[
                width=0.65\textwidth
        ]{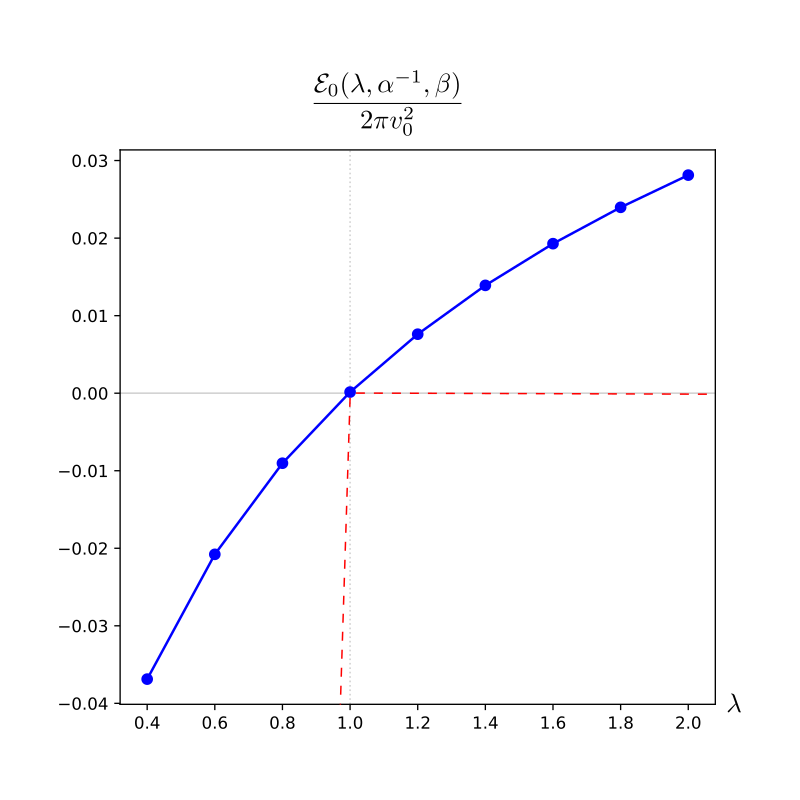}
        \caption{
                Inhomogeneous vacuum energy for various values of the quartic scalar coupling \(\lambda\) and for Gaussian inhomogeneous part with fixed size parameter \( \alpha^{-1} = \sqrt{20} \approx 4.47 \) and depth parameter \(\beta = 1\).
                Dashed lines denote lower bound of the vacuum energy \eqref{222} and \eqref{224} derived in section 2.
        }
\label{fig:303}
\end{figure}

We already know that, in the BPS limit of the critical quartic scalar coupling \(\lambda=1\) \cite{Tong:2013iqa}, the inhomogeneous BPS symmetry-broken vacuum configuration has exact zero energy consistent with the BPS relation,
\(
        \mathcal{E}(\lambda=1)
        =
        v_{0}^{2} |g \Phi_{B}|
\),
even for arbitrary BPS configurations without cylindrical symmetry \cite{Kim:2024gfn, Jeon:2024jbs}.
This exact cancellation means an amazing accurate balance between the positive impurity energy from the inhomogeneity \eqref{308} and the lowered negative energy from nontrivial nonlinear interaction.
On the other hand, similar to the magnetic field, the energy density also depends on the spatial coordinates and changes as illustrated in Figure \ref{fig:302}-(b).
As the quartic scalar coupling varies from this critical value, the BPS bound does not hold anymore and hence the energy \(\mathcal{E}_{0}\) of the minimum energy configuration need not be zero, but depends on the parameters;
one from the quartic scalar coupling \( \lambda \) of the abelian Higgs model and two more parameters from the Gaussian inhomogeneous part, the size parameter \(\alpha^{-1}\) and the depth parameter \(\beta\).
Therefore, the most intriguing question to be addressed is the energy values of the obtained inhomogeneous vacuum configurations.
Specific energy values are evaluated through numerical works and the numerical values of
\(
        \mathcal{E}_{0}
        (\lambda, \alpha^{-1}, \beta)
\)
show that the vacuum energy is not left to be zero but increases monotonically as the quartic scalar coupling \(\lambda\) increases as shown in Figure \ref{fig:303}.
Since the inhomogeneous vacuum energy is proven to be zero
\(
        \mathcal{E}_{0}
        (\lambda=1, \alpha^{-1}, \beta) = 0
\)
at the critical coupling \(\lambda=1\) for the BPS limit irrespective of arbitrary size parameter \(\alpha^{-1}\) and depth parameter \(\beta\) \cite{Kim:2024gfn}, it is always negative
\(
        \mathcal{E}_{0}
        (\lambda, \alpha^{-1}, \beta)
        < 0
\)
for \(\lambda < 1 \) and positive
\(
        \mathcal{E}_{0}
        (\lambda, \alpha^{-1}, \beta)
        > 0
\)
for \(\lambda > 1 \).
It means that the added impurity lowers the vacuum energy in weak coupling regime and grows in strong coupling regime.
This change of inhomogeneous vacuum energy
\(      \displaystyle
        \frac{
                \partial \mathcal{E}_{0}
                (\lambda, \alpha^{-1}, \beta)
        }{\partial \lambda}
        \bigg|_{\sigma(\boldsymbol{x})\neq0}
        > 0
\)
including coupling dependence is contrasted to the constancy of homogeneous vacuum energy 
\(      \displaystyle
        \frac{
                \partial \mathcal{E}_{0}
                (\lambda, \alpha^{-1}, \beta)
        }{\partial \lambda}
        \bigg|_{\sigma(\boldsymbol{x})=0}
        = 0
\)
independent of the quartic scalar coupling.

Though the inhomogeneous vacuum energy
\(
        \mathcal{E}_{0}
        (\lambda, \alpha^{-1}, \beta)
\)
is not computed analytically for arbitrary quartic scalar coupling, it can be evaluated at an extreme limit \(\lambda\to 0\) in addition to the critical value \(\lambda=1\) at the BPS limit \cite{Kim:2024gfn}.
In the limit without self-interaction \(\lambda\to 0\), the correlation length \(\xi\) \eqref{213} diverges and the last self-interaction term vanishes in the scalar equation \eqref{208}.
A reasonable approximation consistent with the numerical behavior in Figure \ref{fig:301} is \(|\phi|\approx0\) whose substitution into the gauge field equation \eqref{209} explicitly solves derivative of the \(\mathrm{U}(1)\) gauge field
\(      \displaystyle
        \frac{1}{r} \frac{dA}{dr}
        \approx
        \frac{\sigma(r)}
        {2\lambda_{\text{L}}^{2} v_{0}^{2}}
\).
Evaluation of a lower bound of the inhomogeneous vacuum energy
\(      \displaystyle
        \lim_{\lambda\to0}
        \mathcal{E}_{0}
        (\lambda, \alpha^{-1}, \beta)
\)
in the limit of no self-interaction results in a negative finite value
\begin{equation}
        \lim_{\lambda\to0}
        \mathcal{E}_{0}
        (\lambda, \alpha^{-1}, \beta)
        \approx
        - \frac{1}{4 \lambda_{\text{L}}^{2} v_{0}^{2}}
        \lim_{\lambda\to0}
        \int d^{2} \boldsymbol{x} \,
        \sigma^{2}(r)
        =
        - \frac{\beta^2}{8 \alpha^2} v_{0}^{2}
.\label{310}
\end{equation}
The obtained value diverges to negative infinity at two limits of negative infinite depth \(\beta\to\infty\) or large size \( \alpha^{-1}\to\infty \),
\(      \displaystyle
        \lim_{\substack{
                \beta \to \infty \\
                \alpha^{-1} \to \infty}}
        \mathcal{E}_{0}
        (\lambda, \alpha^{-1}, \beta)
        \to \infty
\),
and becomes trivially zero at the other two homogeneous limits of zero depth \(\beta\to0\) or zero size \( \alpha^{-1} \to 0 \) with vanishing impurity,
\(      \displaystyle
        \lim_{\substack{
                \beta \to 0 \\
                \alpha^{-1} \to 0}}
        \mathcal{E}_{0}
        (\lambda, \alpha^{-1}, \beta)
        \to 0
\),
consistent with expectation.
When \(\alpha^{-1}=\sqrt{20}\) and \(\beta=1\) for numerical works, the inhomogeneous vacuum energy in this limit has the value
\(      \displaystyle
        \lim_{\lambda\to0}
        \mathcal{E}_{0}
        (\lambda, \alpha^{-1}, \beta)
        = -2.5 v_{0}^{2}
\)
\eqref{310}.

Negativity of the inhomogeneous vacuum energy can be figured out from this attractive nature for weak quartic scalar coupling \(\lambda<1\).
Similarly, positivity of the inhomogeneous vacuum energy can be figured out from this repulsive nature for strong quartic scalar coupling \(\lambda>1\).
For the critical quartic scalar coupling \(\lambda=1\), vanishing net interaction under inhomogeneity of arbitrary shape is a characteristic of the BPS limit and exact zero energy of the inhomogeneous symmetry-broken BPS vacuum can be easily understood \cite{Kim:2024gfn, Jeon:2024jbs}.
The resultant energy
\(
        \mathcal{E}_{0}
        ( \lambda, \alpha^{-1}, \beta )
        / 2\pi v_{0}^{2}
\)
of the inhomogeneous vacuum in Figure \ref{fig:303} is compared to be much smaller than the impurity energy
\(
        \mathcal{E}_{\sigma}^{v_0}
        / 2\pi v_{0}^{2}
        =
        \beta^{2} \lambda
        / 16 \alpha^{2}
\)
\eqref{308} added by inhomogeneous part \(\sigma\) \eqref{203} in the neighborhood of the critical quartic scalar coupling for fixed size parameter \( \alpha^{-1} = \sqrt{20} \) and depth parameter \(\beta=1\)
\begin{equation}
        \frac{
                \mathcal{E}_{0}
                ( \lambda, \alpha^{-1}, \beta )
        }{ 2\pi v_{0}^{2} }
        \ll
        \frac{ \mathcal{E}_{\sigma}^{v_0} }
        { 2\pi v_{0}^{2} }
        =
        \frac{5}{4} \lambda
.
\end{equation}
The slope of the energy curve in Figure \ref{fig:303} is much smaller than \(5/4\) but becomes steeper as the quartic scalar coupling decreases, that seems compatible with the upper bound \(5/4\). 

In the inhomogeneous U(1) gauge theory away from the BPS limit, negativity of the vacuum energy in weak coupling regime $\lambda<1$ and positivity of the vacuum energy in strong coupling regime $\lambda>1$ are interpreted as an over-cancellation and a under-cancellation of the added positive impurity energy due to Gaussian inhomogeneous part \eqref{308}, respectively. The non-zero energy of inhomogeneous vacuum relative to the zero energy of constant Higgs vacuum is a measurable energy gap which is detected by an appropriately designed experiment. It suggests that the types I and I$\!$I of conventional superconductors can be distinguished at the level of vacuum energies by the addition of impurities in some local regions.%

\begin{figure}[H]
        \centering
        \subfigure[]{
                \includegraphics[
                        width=0.45\textwidth
                ]{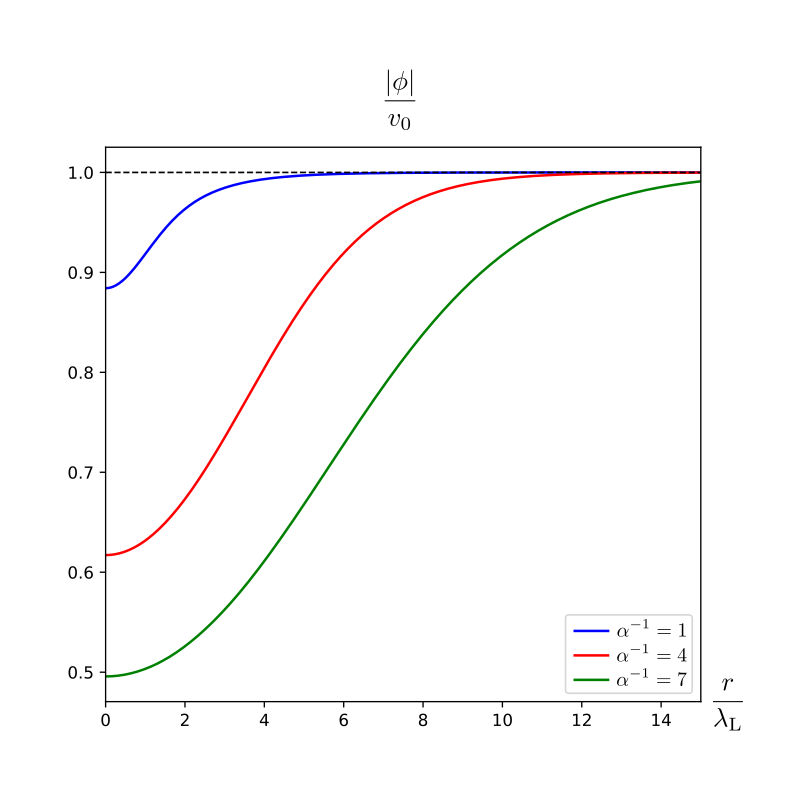}
        }
        \hfill
        \subfigure[]{
                \includegraphics[
                        width=0.45\textwidth
                ]{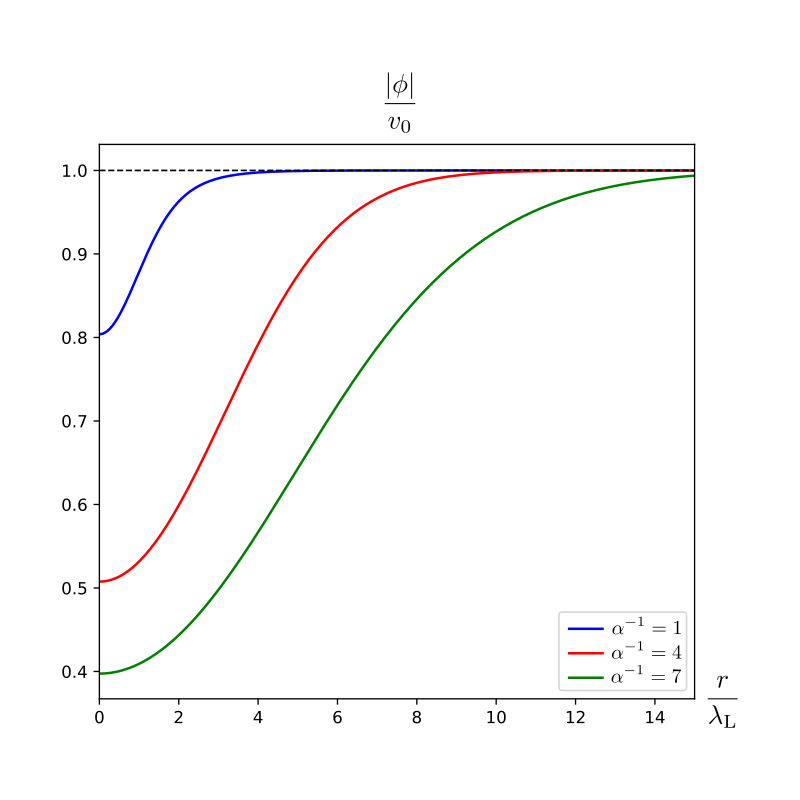}
        }
        \caption{
                Amplitudes of the complex scalar field for quartic scalar coupling (a) \(\lambda=0.5\) and (b) \(\lambda=1.5\) for Gaussian inhomogeneous part with various values of the size parameter \( \alpha^{-1}\) and fixed depth parameter \(\beta = 1\).
        }
\label{fig:304}
\end{figure}

As the dimensionless size parameter \(\alpha^{-1}\) in the Gaussian inhomogeneous part \(\sigma(r)\) \eqref{203} increases with a fixed dimensionless depth parameter \(\beta=1\), the minimum value of the scalar amplitude at the origin decreases and the size of \(r\)-dependent region becomes larger independent of the quartic scalar coupling \(\lambda\) as shown in Figure \ref{fig:304}-(a)--(b)
(see also Figure 4-(a) in \cite{Kim:2024gfn} for the critical quartic scalar coupling \(\lambda=1\)).

\begin{figure}[H]
        \centering
        \subfigure[]{
                \includegraphics[
                        width=0.45\textwidth
                ]{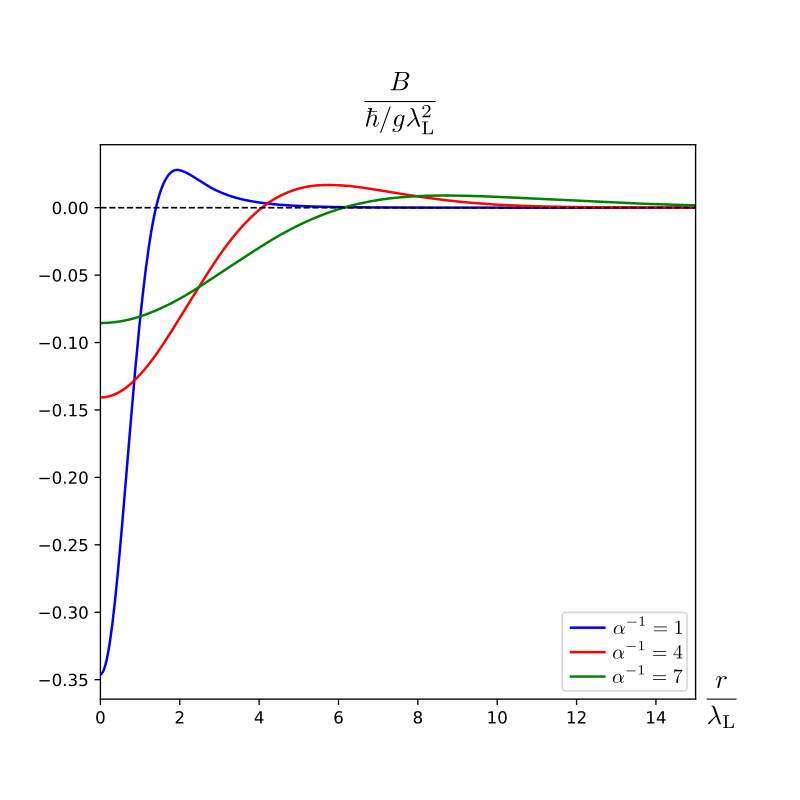}
        }
        \hfill
        \subfigure[]{
                \includegraphics[
                        width=0.45\textwidth
                ]{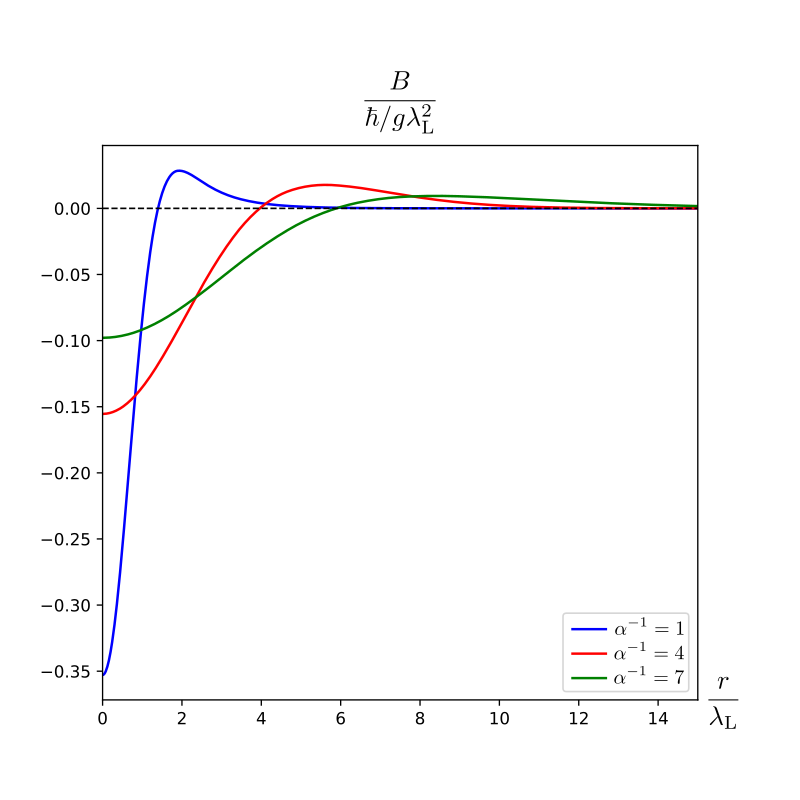}
        }
        \caption{
                Magnetic fields \(B\) of the inhomogeneous vacuum for quartic scalar coupling (a) \(\lambda=0.5\) with
                \(
                        ( \alpha^{-1}, r_{B \text{max}} / 
                        \sqrt{2}\lambda_{\text{L}} )
                        = (1, 1.37),~(4, 4.07),~(7, 6.13)
                \)
                and (b) \(\lambda=1.5\) with
                \(
                        ( \alpha^{-1}, r_{B \text{max}} / 
                        \sqrt{2} \lambda_{\text{L}} )
                        = (1, 1.36),~(4, 3.98),~(7, 5.96)
                \)
                for Gaussian inhomogeneous part with various values of the size parameter \( \alpha^{-1}\) and fixed depth parameter \(\beta = 1\).
        }
\label{fig:305}
\end{figure}

\begin{figure}[H]
        \centering
        \subfigure[]{
                \includegraphics[
                        width=0.45\textwidth
                ]{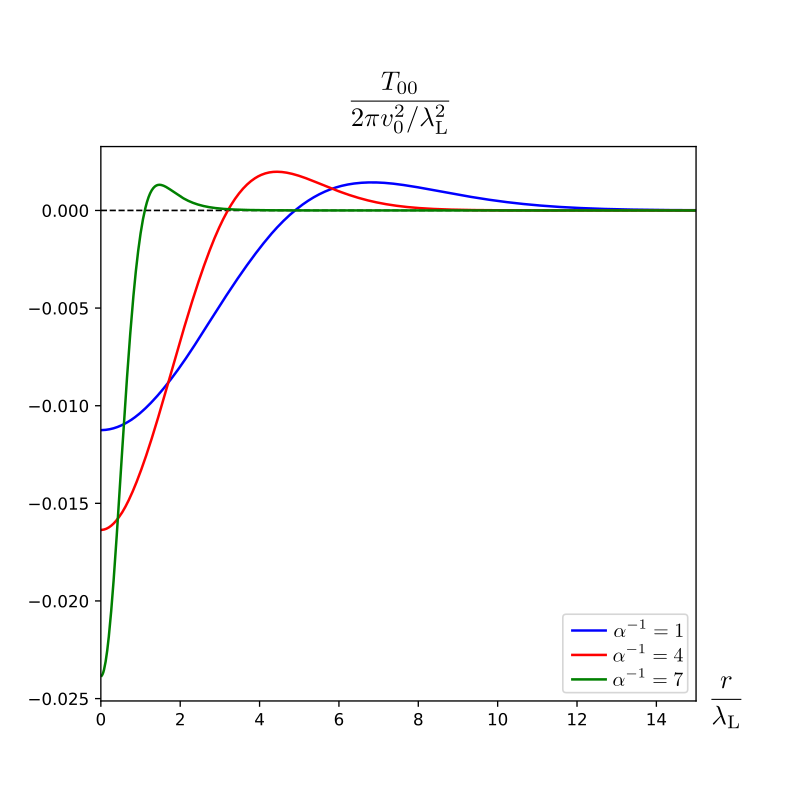}
        }
        \hfill
        \subfigure[]{
                \includegraphics[
                        width=0.45\textwidth
                ]{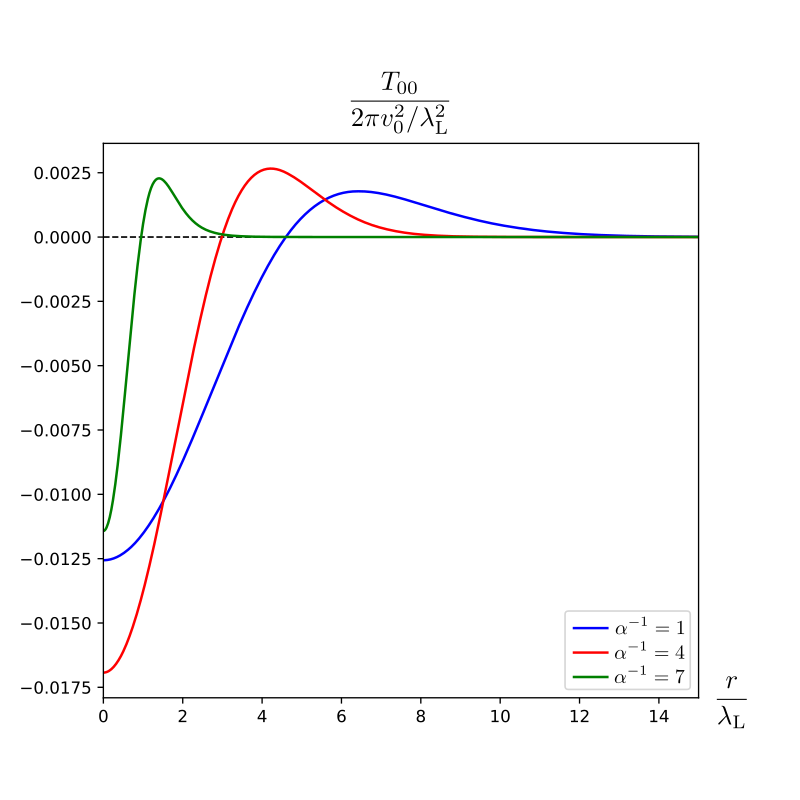}
        }
        \caption{
                Energy densities \(-T\indices{^t_t}\) of the inhomogeneous vacuum for quartic scalar coupling (a) \(\lambda=0.5\) with
                \(
                        ( \alpha^{-1}, r_{\max( -T\indices{^t_t} )} / 
                        \sqrt{2}\lambda_{\text{L}} )
                        = (1, 1.05),~(4, 3.14),~(7, 4.83)
                \)
                and (b) \(\lambda=1.5\) with
                \(
                        ( \alpha^{-1}, r_{\max( -T\indices{^t_t} )} / 
                        \sqrt{2} \lambda_{\text{L}} )
                        = (1, 0.99),~(4, 2.99),~(7, 4.55)
                \)
                for Gaussian inhomogeneous part with various values of the size parameter \( \alpha^{-1}\) and fixed depth parameter \(\beta = 1\).
        }
\label{fig:306}
\end{figure}

\noindent%
The ring-shaped magnetic field \(B(r)\) is peaked at a finite radial position.
The peak positions of magnetic field obtained through numerical analysis seem to be almost independent of the quartic scalar coupling \(\lambda\) but seems to be almost linearly related with size parameter \(\alpha^{-1}\) of the inhomogeneous part \(\sigma(r)\) \eqref{203} in unit of the London penetration depth \( \sqrt{2}\lambda_{\text{L}} \) \eqref{214} as given in the caption of Figure \ref{fig:305}.
Comparison between the data of \(\lambda=0.5\) and \(\lambda=1.5\) with fixed size and depth parameters \(\alpha^{-1}\) and \(\beta\) show that the peak positions are different and seem to depend slightly on the correlation length \(\xi\) \eqref{213}.
\noindent%
Similar to the magnetic field, the energy density \(-T\indices{^t_t}\) involves an energy dip about the origin and a positive ring-shaped bump irrespective of the values of size parameter \(\alpha^{-1}\) as in Figure \ref{fig:306}-(a)--(b).
The radial positions of positive energy peaks are almost proportional to the size parameter \(\alpha^{-1}\).

\begin{figure}[H]
        \centering
        \subfigure[]{
                \includegraphics[
                        width=0.45\textwidth
                ]{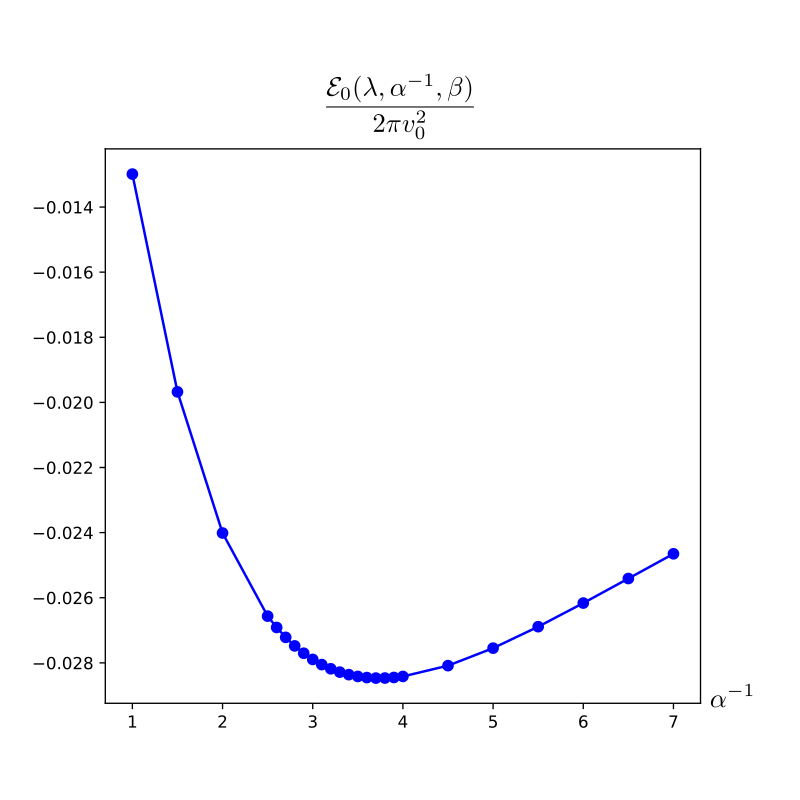}
        }
        \hfill
        \subfigure[]{
                \includegraphics[
                        width=0.45\textwidth
                ]{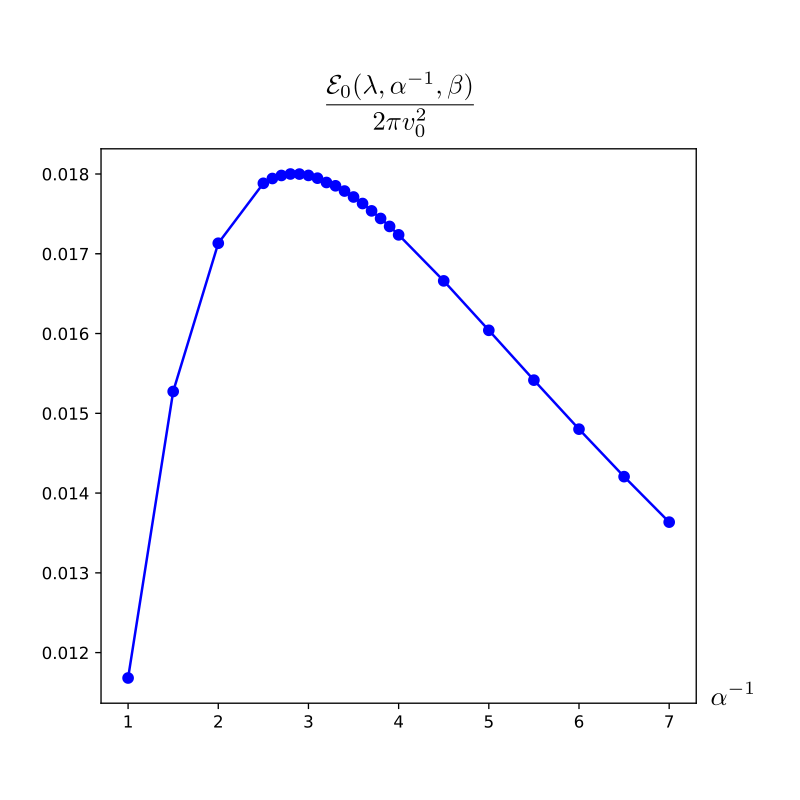}
        }
        \caption{
                Inhomogeneous vacuum energy for quartic scalar coupling (a) \(\lambda=0.5\) and (b) \(\lambda=1.5\) for Gaussian inhomogeneous part with various values of the size parameter \( \alpha^{-1}\) and a fixed depth parameter \(\beta = 1\).
        }
\label{fig:307}
\end{figure}

\begin{figure}[H]
        \centering
        \subfigure[]{
                \includegraphics[
                        width=0.45\textwidth
                ]{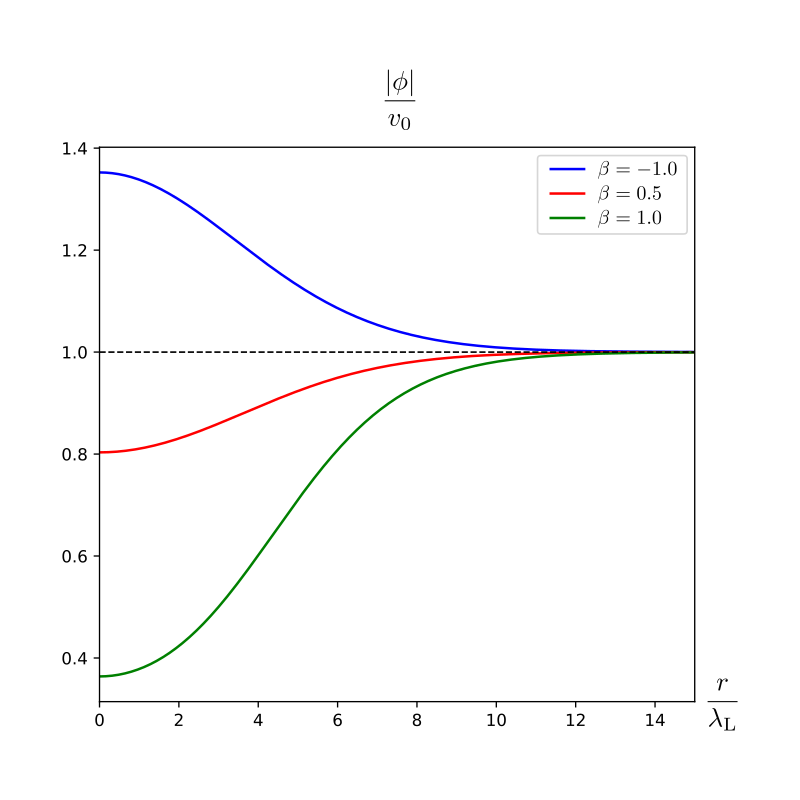}
        }
        \hfill
        \subfigure[]{
                \includegraphics[
                        width=0.45\textwidth
                ]{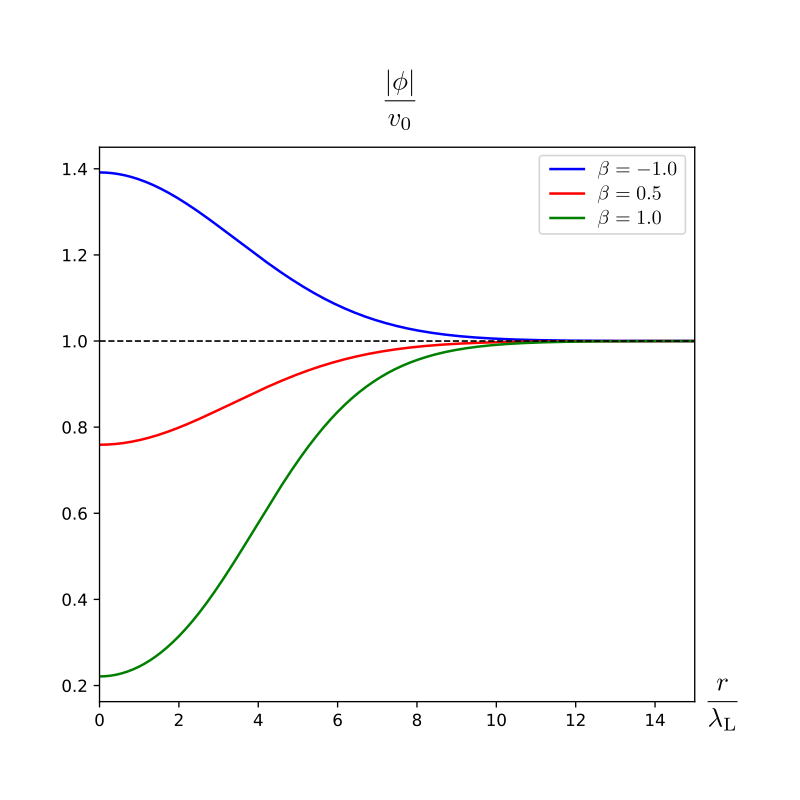}
        }
        \caption{
                Amplitudes of the complex scalar field for quartic scalar coupling (a) \(\lambda=0.5\) and (b) \(\lambda=1.5\) for Gaussian inhomogeneous part with fixed size parameter \( \alpha^{-1} = \sqrt{20} \approx 4.47 \) and various values of the depth parameter \(\beta\).
        }
\label{fig:308}
\end{figure}

When the quartic scalar coupling has the critical value \(\lambda=1\), the BPS bound is saturated \cite{Tong:2013iqa} and thus the energy of the inhomogeneous symmetry-broken vacuum has the minimum energy whose value is strictly zero 
\(
        \mathcal{E}_{0}
        (\lambda=1, \alpha^{-1}, \beta)
        = 0
\)
independent of the size parameter \(\alpha^{-1}\) \cite{Kim:2024gfn}.
When it is smaller than the critical value \(\lambda<1\), e.g. \(\lambda=0.5\), the inhomogenous vacuum energy is likely to be always negative 
\(
        \mathcal{E}_{0}
        (\lambda < 1, \alpha^{-1}, \beta)
        < 0
\)
for every size parameter \(\alpha^{-1}\) and has a minimum value as shown in Figure \ref{fig:307}-(a).
Since the scale of net interaction determined by the correlation length \(\xi\) relative to the London penetration depth \( \sqrt{2}\lambda_{\text{L}} \), the effect of such net interaction can possibly be maximized in the vicinity of the correlation length
\(
        \xi     \approx
        \alpha^{-1} \sqrt{2}\lambda_{\text{L}}
\).
Therefore, net attractive interaction for \(\lambda<1\) lowers maximally the inhomogenous vacuum energy about
\(
        \xi \approx
        \alpha^{-1} \sqrt{2}\lambda_{\text{L}}
\)
to make it have the minimum negative value, e.g. minimum value
\(
        \mathcal{E}_{0}
        (0.5, 3.6 < \alpha^{-1} < 3.7, 1 )
        / 2\pi v_{0}^{2}
        \approx -0.029
\)
as shown in Figure \ref{fig:307}-(a) and net repulsive interaction for \(\lambda>1\) raises maximally the inhomogenous vacuum energy about
\(
        \xi \approx
        \alpha^{-1} \sqrt{2} \lambda_{\text{L}}
\)
to make it have the maximum positive value, e.g. maximum value
\(
        \mathcal{E}_{0}
        (1.5, 2.8 < \alpha^{-1} < 2.9, 1 )
        / 2\pi v_{0}^{2}
        \approx 0.018
\)
as shown in Figure \ref{fig:307}-(b).

It is the turn to evaluate the energy of inhomogeneous vacuum configuration in both limits of zero and infinite size parameter.
In the limit of zero size parameter \(\alpha^{-1}\to 0\), the Gaussian type inhomogeneous part \(\sigma(r)\)  \eqref{203} becomes zero everywhere
\(      \displaystyle
        \lim_{\alpha^{-1}\to0}
        \sigma(r) = 0
\)
and this disappearance of inhomogeneity lets the magnetic impurity term \eqref{210} vanish.
Thus the trivial solution of constant symmetry-broken vacuum \(|\phi| = v_{0}\) and \(A(r)=0\) leads to null result
\(      \displaystyle
        \lim_{\alpha^{-1}\to0}
        \mathcal{E}_{0}
        (\lambda, \alpha^{-1}, \beta)
        = 0
\).
In the limit of infinite size parameter \(\alpha^{-1} \to \infty\), the Gaussian type inhomogeneous part \(\sigma(r)\) \eqref{203} becomes a new constant
\(      \displaystyle
        \lim_{\alpha^{-1}\to\infty}
        \sigma(r) = - \beta
\)
and subsequently the configuration of minimum zero energy is nothing but the new constant vacuum of \(|\phi|(r) = v_{0}^{2} (1-\beta)\) and \(A(r)=0\).
The corresponding energy including the constant magnetic impurity term proportional to magnetic flux \(\Phi_{B}\) is computed to be zero
\(      \displaystyle
        \lim_{\alpha^{-1}\to\infty}
        \mathcal{E}_{0}(\lambda, \alpha^{-1}, \beta)
        = 0
\)
again.
In both graphs (a) and (b) of Figure \ref{fig:307}, each curve of inhomogeneous vacuum energy 
\(
        \mathcal{E}_{0}
        (\lambda, \alpha^{-1}, \beta)
\)
approaches zero at both ends of \(\alpha^{-1}=0\) and \(\alpha^{-1}=\infty\).

Typical profiles of the scalar amplitude \(|\phi|(r)\) show monotonic behavior of rapid approach of a boundary value at infinity \(|\phi|(\infty) = v_{0}\) \eqref{306} in Figure \ref{fig:308} for various values of depth parameter \(\beta\) with a fixed size parameter \(\alpha^{-1}\).
The second term in power series expansion for small \(r\) \eqref{301} decides increasing (decreasing) behavior for \( \phi_{0}^{2} + \beta - 1 > 0 \) (\( \phi_{0}^{2} + \beta - 1 < 0 \)) near the origin, irrespective of the quartic scalar coupling \(\lambda\) as in Figure \ref{fig:308}-(a) for \(\lambda=0.5<1\) and (b) \(\lambda=1.5>1\)
(see also Figure 1 for the critical quartic scalar coupling \(\lambda=1\) in \cite{Kim:2024gfn}).
Thus the initial value \(|\phi|(0)\) decreases as positive depth parameter \(\beta\) increases but should be larger than unity \( |\phi|(0) = v_{0}\phi_{0} > 1 \) for negative \(\beta\) smaller than critical value \(\beta < 1 - \phi_{0}^{2}\) irrespective of the quartic scalar coupling \(\lambda\) either.
These two categories, one of monotonically increasing vacuum solutions from the values between zero and \(v_{0}\), \(0<|\phi|(r=0)<v_{0}\), and the other of monotonically decreasing vacuum solutions of the values larger than constant vacuum expectation value, \(v_{0}<|\phi|(r=0)\), are consistent with the expectation made by use of the Newtonian shooting analogy stated after Figure \ref{fig:301} for the BPS case of critical quartic scalar coupling \(\lambda=1\), and are valid even for arbitrary quartic scalar coupling \(\lambda\).
When the depth parameter of Gaussian bump \eqref{203} decreases to negative infinity \(\beta\to-\infty\), the initial value can increase to positive infinity.
Meanwhile, as the depth parameter of Gaussian dip \eqref{203} increases to positive infinity \(\beta\to\infty\), the decrease of the initial value is limited by its nonnegativity \( 0\le \phi_{0} \).

\begin{figure}[H]
        \centering
        \subfigure[]{
                \includegraphics[
                        width=0.45\textwidth
                ]{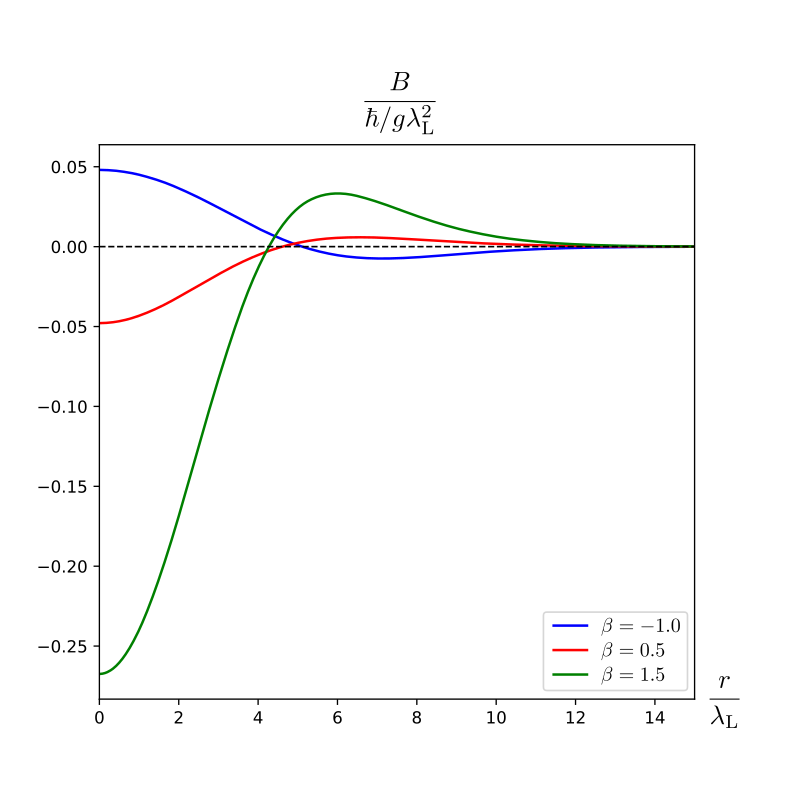}
        }
        \hfill
        \subfigure[]{
                \includegraphics[
                        width=0.45\textwidth
                ]{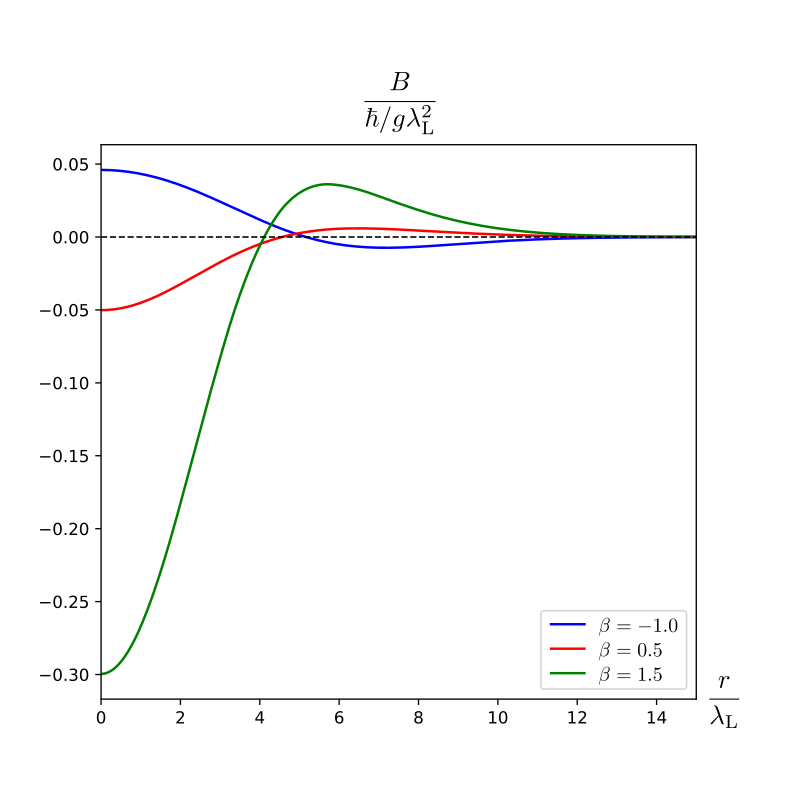}
        }
        \caption{
                Magnetic fields \(B\) of the inhomogeneous vacuum for quartic scalar coupling (a) \(\lambda=0.5\) and (b) \(\lambda=1.5\) for Gaussian inhomogeneous part with fixed size parameter \( \alpha^{-1} = \sqrt{20} \approx 4.47 \) and various values of the depth parameter \(\beta\).
        }
\label{fig:309}
\end{figure}

\begin{figure}[H]
        \centering
        \subfigure[]{
                \includegraphics[
                        width=0.45\textwidth
                ]{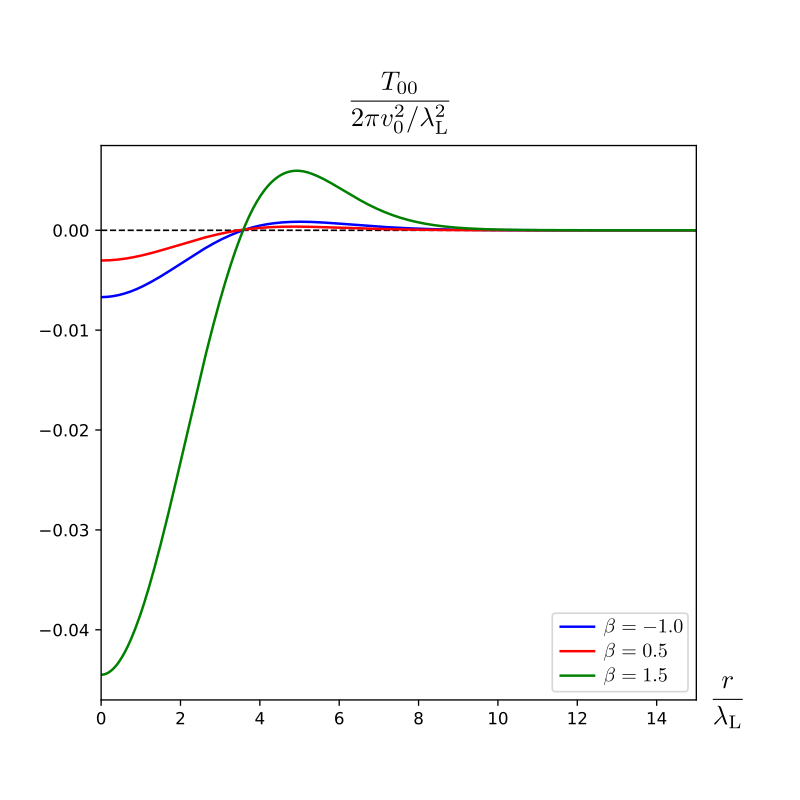}
        }
        \hfill
        \subfigure[]{
                \includegraphics[
                        width=0.45\textwidth
                ]{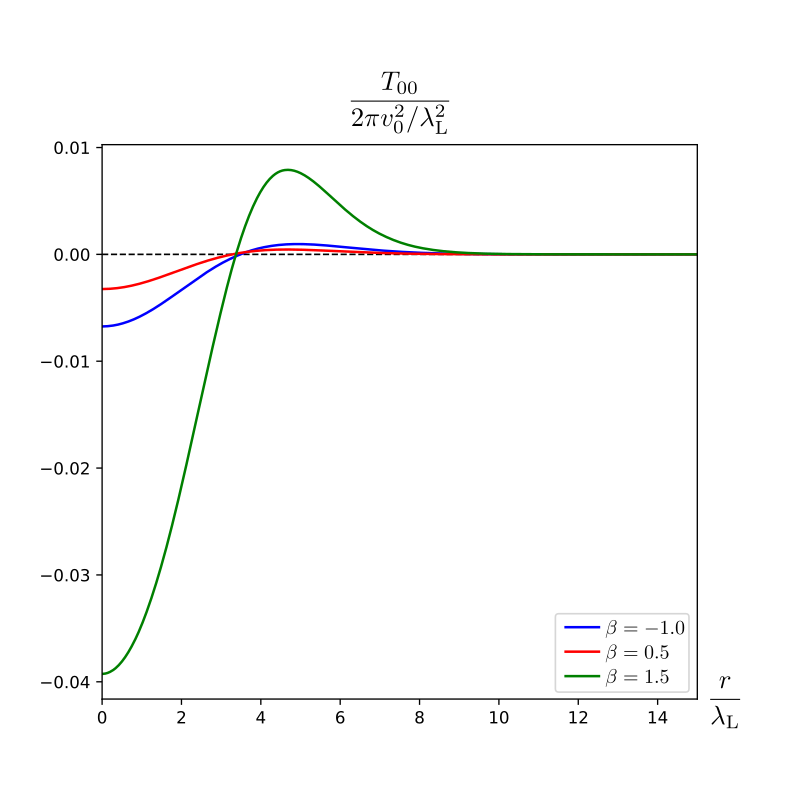}
        }
        \caption{
                Energy densities \(-T\indices{^t_t}\) of the inhomogeneous vacuum for quartic scalar coupling (a) \(\lambda=0.5\) and (b) \(\lambda=1.5\) for Gaussian inhomogeneous part with fixed size parameter \( \alpha^{-1} = \sqrt{20} \approx 4.47 \) and various values of the depth parameter \(\beta\).
        }
\label{fig:310}
\end{figure}

Accordingly magnetic field \(B(r)\) also exhibits two qualitatively different behaviors.
When negative \(\beta\) is smaller than a critical value obeying \( \phi_{0}^{2} + \beta - 1 < 0 \), magnetic field begins with a positive maximum value at the origin \(B(0) = 2a_{0}/\lambda_{ \text{L} }^{2} >0 ~(a_{0} > 0)\), decreases to a negative minimum value by negative coefficient of the quartic term \( a_{0} \phi_{0}^{2} < - \alpha^{2} \beta \) in \eqref{302}, and approaches rapidly zero from below zero as shown by the blue-colored solid curves of \(\beta = -1\) in Figure \ref{fig:309}-(a)--(b).
When \(\beta\) is larger than a critical value, magnetic field begins with a negative minimum value at the origin \(B(0) = 2a_{0}/\lambda_{ \text{L} }^{2}<0~(a_{0}<0)\), increases to a positive maximum value by positive coefficient of the quartic term \( - a_{0} \phi_{0}^{2} < \alpha^{2} \beta \) in \eqref{302}, and approaches rapidly zero from above zero as shown by the red- and green-colored solid curves of \(\beta = 0.5\) and \(1.5\) in Figure \ref{fig:309}-(a)--(b).
Nonetheless, magnetic flux \(\Phi_{B}\) \eqref{220} is left to be zero for any inhomogenous symmetry-broken vacuum of any depth parameter \(\beta\).
\noindent%
Different from the profiles of magnetic field, energy density \(-T\indices{^t_t}\) always has negative minimum value of a dip around the origin and forms a ring-shaped bump of positive energy density at a finite radius.
Its depth and height deepen and become larger, respectively, as depth parameter \(\beta\) increases.
See Figure \ref{fig:310}-(a)--(b)
(and also \(-T\indices{^t_t}\) profiles for the critical quartic scalar coupling \(\lambda = 1\) in Figure 2 of \cite{Kim:2024gfn}).
Hence the aforementioned behavior of energy density of the inhomogeneous vacuum is common irrespective of the quartic scalar coupling \(\lambda\).

\begin{figure}[H]
        \centering
        \subfigure[]{
                \includegraphics[
                        width=0.45\textwidth
                ]{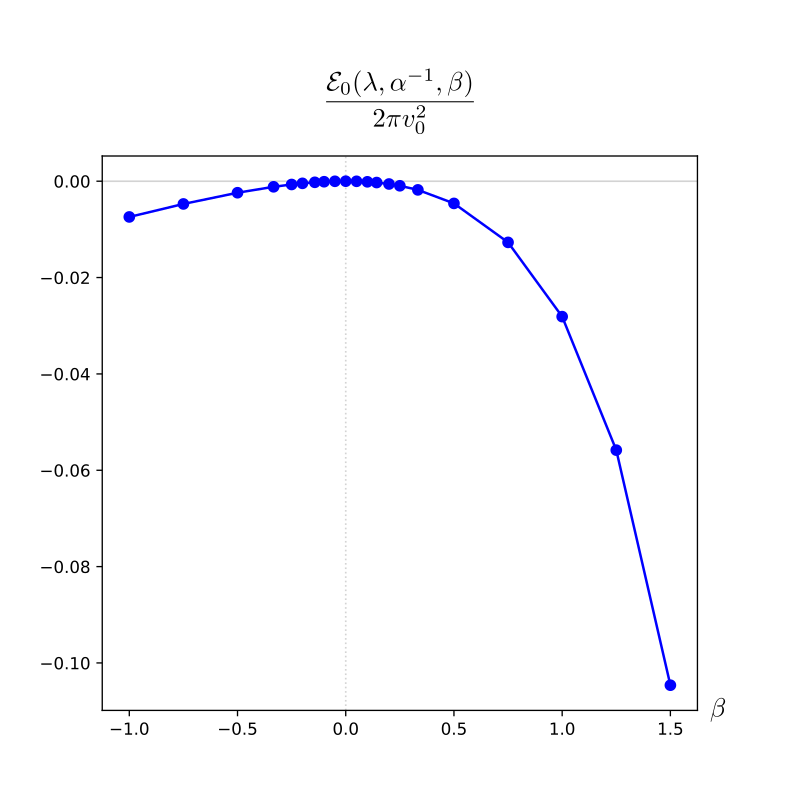}
        }
        \hfill
        \subfigure[]{
                \includegraphics[
                        width=0.45\textwidth
                ]{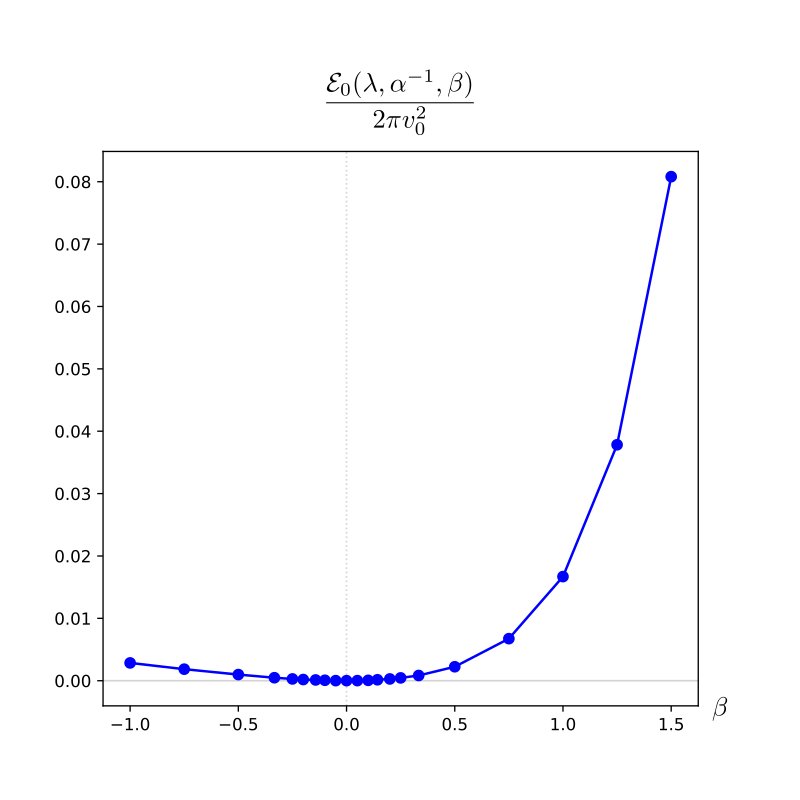}
        }
        \caption{
                Inhomogeneous vacuum energy for quartic scalar coupling (a) \(\lambda=0.5\) and (b) \(\lambda=1.5\) for Gaussian inhomogeneous part with fixed radius parameter \( \alpha^{-1} = \sqrt{20} \approx 4.47 \) and various values of the depth parameter \(\beta\).
        }
\label{fig:311}
\end{figure}

If the depth parameter \(\beta\) varies with keeping the size parameter \(\alpha^{-1}\) a finite fixed value, the inhomogeneous vacuum energy gradually changes as long as the quartic scalar coupling is not critical \(\lambda\neq1\).
When it is smaller than the critical value \(\lambda<1\), inhomogeneous vacuum energy has maximum zero energy
\(
        \mathcal{E}_{0}
        (\lambda < 1, \alpha^{-1}, \beta = 0)
        =0
\)
in the homogeneous limit and is always negative
\(
        \mathcal{E}_{0}
        (\lambda < 1, \alpha^{-1}, \beta \neq 0)
        < 0
\)
that is consistent with attractive nature of net interaction.
As the depth parameter goes away further from the homogeneous limit of \( \beta = 0 \), the negative inhomogenous vacuum energy gets smaller as shown in Figure \ref{fig:311}-(a) of \(\lambda=0.5\).
When it is greater than the critical value \(\lambda>1\), inhomogeneous vacuum energy has minimum zero energy
\(
        \mathcal{E}_{0}
        (\lambda>1, \alpha^{-1}, \beta = 0)
        =0
\)
in the homogeneous limit and is always positive
\(
        \mathcal{E}_{0}
        (\lambda>1, \alpha^{-1}, \beta \neq 0)
        > 0
\)
that is consistent with repulsive nature of net interaction.
As the depth parameter goes away further from the homogeneous limit of \(\beta = 0\), the positive inhomogenous vacuum energy gets larger as shown in Figure \ref{fig:311}-(b) of \(\lambda=1.5\).
This growing behavior of large absolute energy
\(
        | \mathcal{E}_{0}
        ( \lambda, \alpha^{-1}, \beta ) |
\)
due to increasing absolute depth parameter \(|\beta|\) seems to be related to the impurity energy \(\mathcal{E}_{\sigma}^{v_0}\) \eqref{308} induced by the Gaussian inhomogeneous part \eqref{203}, which is quadratically proportional to the depth parameter \(\beta\), \( \mathcal{E}_{\sigma}^{v_{0}} \propto \beta^{2} \).
The inhomogeneous vacuum energy \( \mathcal{E}_{0} (\lambda,\alpha^{-1},\beta) \) is lowered or grown rapidly for Gaussian dips of \(\beta>0\) (the right side of the blue-colored solid curves) in comparison with Gaussian bumps of \(\beta<0\) (the left side of the blue-colored solid curves).
This left-right asymmetry of both curves in Figure \ref{fig:311}-(a)--(b), smaller absolute energy for \(\beta<0\) and larger absolute energy for \(\beta>0\), implies that reduction of the impurity energy \(\mathcal{E}_{\sigma}^{v_0}\) \eqref{308} added by the inhomogeneous part \(\sigma\) \eqref{203} is less excessive for Gaussian bump of negative \(\beta\) (\(\beta<0\)) through limitless growth of the scalar amplitude
\(      \displaystyle
        \lim_{\beta\to -\infty}
        |\phi| (0) =
        \lim_{\beta\to -\infty}
        v_{0} \phi_{0}
        \to + \infty
        ,
\)
and much more excessive for Gaussian dip of positive \(\beta\) (\(\beta>0\)) because of the lower bound of the initial value of scalar amplitude
\(
        |\phi|(0) = v_{0} \phi_{0} \ge 0
\).
In infinite strength limits \( \beta \to \mp \infty \) of the inhomogeneity of Gaussian bump (\(\beta<0\)) and dip (\(\beta>0\)) with keeping the size parameter \(\alpha^{-1}\) finite, the inhomogeneous vacuum energy is expected to diverge to positive or negative infinite however physical relevance of these infinite energies seem is unclear.

In this section, we have examined the abelian Higgs model of arbitrary quartic scalar interaction in the presence of inhomogeneity of a Gaussian function and have found inhomogeneous symmetry-broken vacuum as the family of three parameters:
One is the quartic scalar coupling \(\lambda\) (\(0\le \lambda\)) and two from the Gaussian inhomogeneity consist of a depth parameter \(\beta\) (\(-\infty<\beta<\infty\)) and a size parameter \(\alpha^{-1}\) (\(0\le\alpha^{-1}\)).
The inhomogeneous vacuum energy
\(
        \mathcal{E}_{0}
        ( \lambda, \alpha^{-1}, \beta )
\)
is exactly zero 
\(
        \mathcal{E}_{0}
        ( 1, \alpha^{-1}, \beta )
        = 0
\)
in the BPS limit of critical quartic scalar coupling \(\lambda=1\) for arbitrary \(\alpha^{-1}\) and \(\beta\), negative
\(
        \mathcal{E}_{0}
        ( \lambda, \alpha^{-1}, \beta )
        < 0
\)
in weak coupling regime \( 0\le\lambda<1 \) of attractive net interaction, and positive
\(
        \mathcal{E}_{0}
        ( \lambda, \alpha^{-1}, \beta )
        > 0
\)
in strong coupling regime \( \lambda>1 \) of repulsive net interaction, as shown in Figure \ref{fig:303}.
With fixed quartic scalar coupling \(\lambda\neq1\) and depth parameter \(\beta\), the inhomogeneous vacuum energy has an extremum value about the correlation length
\(
        \xi =
        \alpha^{-1}\sqrt{2}\lambda_{\text{L}}
\),
a minimum value 
\(      \displaystyle
        \mathcal{E}_{0}
        ( \lambda, \frac{\xi}{
                \sqrt{2}\lambda_{\text{L}}
        }, \beta )
\)
in weak coupling regime \( 0\le\lambda<1 \) and a maximum value 
\(      \displaystyle
        \mathcal{E}_{0}
        ( \lambda, \frac{\xi}{
                \sqrt{2}\lambda_{\text{L}}
        }, \beta )
\)
in strong coupling regime \( \lambda > 1 \) as shown in Figure \ref{fig:307}.
In both limits of zero and infinite size parameter \( \alpha^{-1} \to 0 \) and \( \alpha^{-1} \to \infty \), the inhomogeneous vacuum energies approach zero
\(      \displaystyle
        \lim_{\alpha^{-1}\to 0}
        \mathcal{E}_{0}
        (\lambda, \alpha^{-1}, \beta)
        = 0 =
        \lim_{\alpha^{-1}\to \infty}
        \mathcal{E}_{0}
        (\lambda, \alpha^{-1}, \beta)
\)
due to the revival of homogeneous vacuum.
With any fixed quartic scalar coupling except unity \(\lambda\neq1\) and a finite size parameter \(\alpha^{-1}\), absolute value of the inhomogeneous vacuum energy
\(
        | \mathcal{E}_{0}
        ( \lambda, \alpha^{-1}, \beta ) |
\)
has an extremum value for the case without inhomogeneity of zero depth parameter \(\beta=0\), and increases as absolute value of the depth parameter \(|\beta|\) increases as shown in Figure \ref{fig:311}-(a)--(b).

\begin{figure}[H]
	\centering
	\subfigure[]{
		\includegraphics[
		width=0.45\textwidth
		]{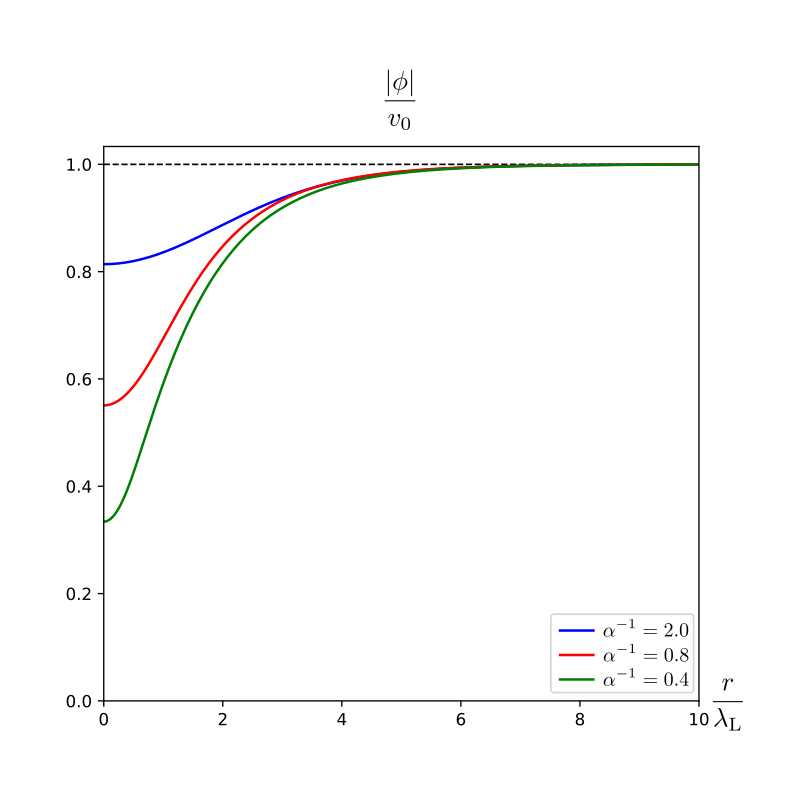}
	}
	\hfill
	\subfigure[]{
		\includegraphics[
		width=0.45\textwidth
		]{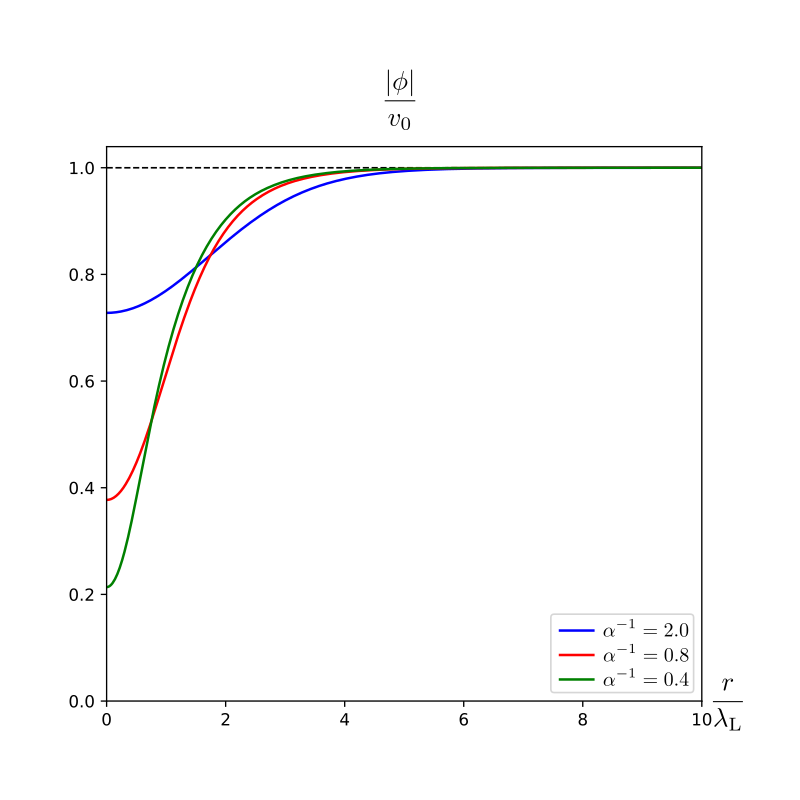}
	}
	\\
	\subfigure[]{
		\includegraphics[
		width=0.45\textwidth
		]{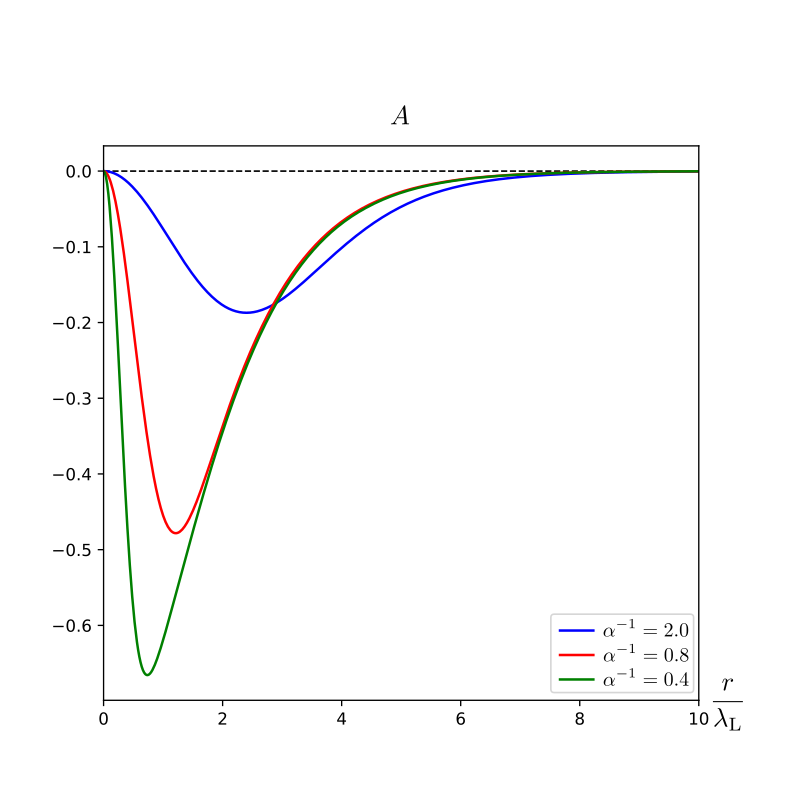}
	}
	\hfill
	\subfigure[]{
		\includegraphics[
		width=0.45\textwidth
		]{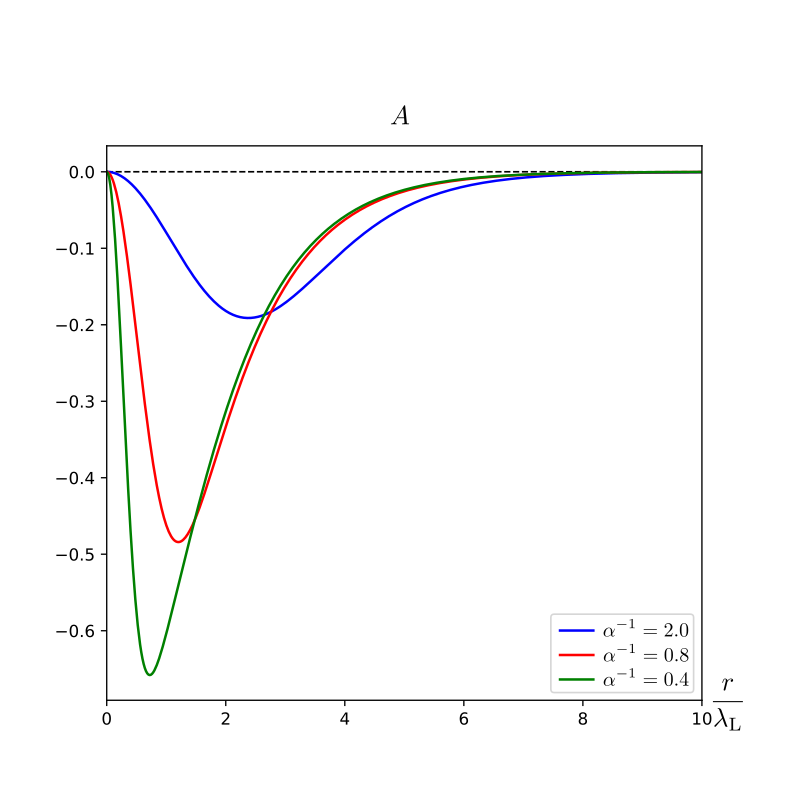}
	}
	\caption{
		Scalar amplitude \(|\phi|\) for two quartic scalar couplings (a) \(\lambda = 0.5 < 1 \) and (b) \(\lambda = 1.5 >1 \), and radial part of the \( \mathrm{U}(1) \) gauge field for two quartic scalar couplings (c) \(\lambda = 0.5 < 1\) and (d) \(\lambda = 1.5 >1 \) with a fixed ratio of size and depth parameter \( \eta = 0.8 \) and various values of the size parameter \( \alpha^{-1}=0.4,~0.8,~2.0 \).
	}
	\label{fig:313}
\end{figure}

An intriguing limit of Gaussian type inhomogeneous part \eqref{203} is the delta function limit which is taken by the zero size limit \(\alpha^{-1}\to 0\) with keeping the ratio of size and depth parameters constant \(\eta = \beta/4\alpha^{2}\) as
\begin{equation}
	\sigma (r)
	=
	- 4 \eta \alpha^{2} v_{0}^{2}
	e^ {-\alpha^{2} ( \frac{r}{\lambda_{\text{L}}} )^{2}}
	\equiv
	- 4\pi\eta \lambda_{\rm L}^{2} v_{0}^{2} \delta_{\alpha} (r)
	.\label{317}
\end{equation}
The limit \( \displaystyle \lim_{\alpha^{-1} \to 0} \delta_{\alpha} (r) \) results in the two-dimensional delta function \(\delta^{(2)} (\boldsymbol{x})\) whose expression is obtained from the two-dimensional Gauss' law for a point charge at the origin
\begin{equation}
	\nabla^{2} \ln |\boldsymbol{x}|
	= 2\pi \delta^{(2)} (\boldsymbol{x})
	.\label{318}
\end{equation}
When the Euler-Lagrange equations \eqref{208}--\eqref{209} are reexamined by use of the formula \eqref{318} for the delta function impurity, the field behaviors near the origin vary significantly from the leading order in comparison to both homogeneous and smooth inhomogeneous Gaussian case
\begin{align}
	|\phi|(r) =&\, f_{0} v_{0} \Big( \frac{r}{\lambda_{ \text{L} }} \Big)^{\lambda\eta} \bigg[ 1 - \frac{\lambda}{8(\lambda\eta+1)} \Big( \frac{r}{\lambda_{ \text{L} }} \Big)^{2} + \cdots \bigg] ,\\
	A(r) =&\, - \lambda\eta - \frac{f_{0}^{2}}{4(\lambda\eta+1)} \Big( \frac{r}{\lambda_{\rm L}} \Big)^{2(\lambda\eta+1)} + \cdots ,\label{322}
\end{align}
where \( \lambda\eta > 0 \) is chosen.\footnote{In the viewpoint of consistency between the coupled Euler-Lagrange equations \eqref{208}--\eqref{209} in this delta function limit, the coupling of magnetic impurity term seems to favor $\lambda g$ in \eqref{228} derived from the U(1)$\times$U(1) gauge theory instead of our choice $g$ in \eqref{210}. }
%
The profile of scalar amplitude \(|\phi|\) mimics that of a cylindrically symmetric vortex with vorticity \(\lambda\eta\) as in \eqref{402} and the \(\mathrm{U}(1)\) gauge field \eqref{207} is singular for nonzero \(\lambda\eta\) at the origin. Numerical analysis is performed for two quartic scalar couplings, e.g. $\lambda=0.5$ and $\lambda=1.5$. If the size parameter $\alpha^{-1}$ decreases to zero with keeping the ratio $\eta=\beta/4\alpha^{2}$ a constant value $\eta=0.8$, the minimum value of scalar amplitude at the origin $|\phi|(0)$ decreases and approaches zero as shown in Figure~\ref{fig:313}$-$(a)--(b) and the minimum value of $A(r_{{\rm min}})$ at a radial distance $r_{{\rm min}}$ approaches the destined value $-\lambda\eta$ in \eqref{322}, e.g. $\lambda\eta=0.4$ and $\lambda\eta=1.2$, with zero radial distance limit $r_{{\rm min}}\rightarrow0$ as shown in Figure~\ref{fig:313}$-$(c)--(d).

\begin{figure}[H]
	\centering
	\subfigure[]{
		\includegraphics[
		width=0.45\textwidth
		]{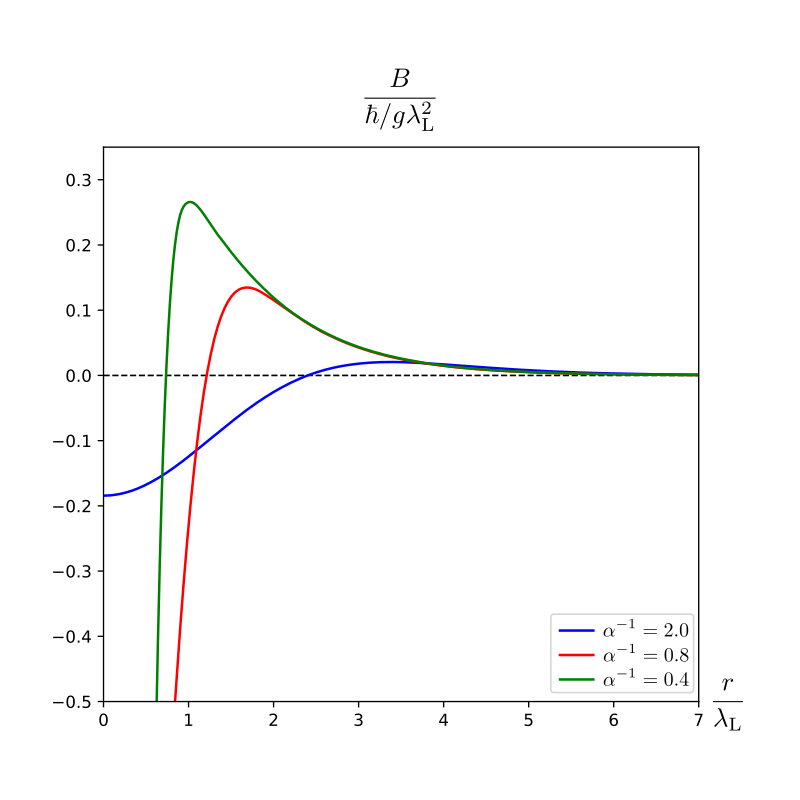}
	}
	\hfill
	\subfigure[]{
		\includegraphics[
		width=0.45\textwidth
		]{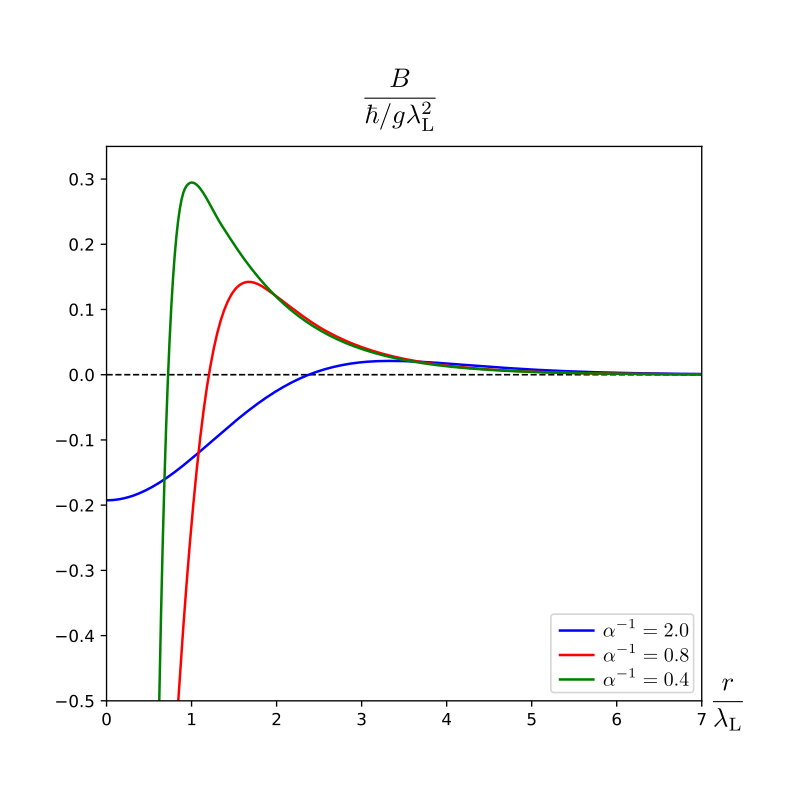}
	}
	\caption{
		Magnetic field for quartic scalar coupling (a) \( \lambda = 0.5 \) and (b) \( \lambda = 1.5 \) for a fixed ratio of size and depth parameter \( \eta = 0.8 \) and various values of the size parameter \( \alpha^{-1} \).
	}
	\label{fig:314}
\end{figure}

Magnetic field \( B \) near the origin is concentrated at the site \( r=0 \) of a delta function type impurity, e.g. lattice,
\begin{equation}
	B(r) \approx
	\frac{\hbar}{g}
	\bigg[ \lambda\eta 
	\frac{\delta (r)}{r}
	- \frac{f_{0}^{2}}{2\lambda_{ \text{L} }^{2}}
	\Big( \frac{r}{\lambda_{ \text{L} }} \Big)^{2\lambda\eta}
	+ \cdots
	\bigg]
	,
	\label{321}
\end{equation}
in addition to continuous distribution consisting of infinitely many expanded terms, which satisfies the boundary condition $A(0)=0$ \eqref{307} and $A(\infty)=0$ \eqref{306}. A procedure to singular limit is illustrated through numerical data. As the size parameter $\alpha^{-1}$ decreases as 0.4, 0.8, 2.0 with constant $\eta=0.8$, the minimum value of magnetic field at the origin $B(0)$ decrease rapidly to negative infinity for both $\lambda$'s: $(\alpha^{-1},B(0))=(2.0,-0.18)$, $(0.8,-2.07)$, $(0.4,-9.49)$ for $\lambda=0.5$ and $(\alpha^{-1},B(0))=(2.0,-0.19)$, $(0.8,-2.11)$, $(2.0,-9.48)$ for $\lambda=1.5$ as shown in Figure~\ref{fig:314}$-$(a)--(b).    
The corresponding magnetic flux \eqref{220} is nonzero and finite
\begin{equation}
	\Phi_{B} = \frac{2\pi\hbar}{g} \lambda\eta
	,\label{320}
\end{equation}
and $\lambda\eta$ plays effectively a role of continuous vorticity in a dirty superconducting sample. Since the first delta function term in magnetic field \eqref{321} results in the exact magnetic flux \eqref{320}, spatial integration of all the other infinitely many expanded terms leads to null net contribution to the magnetic flux through complete cancellation.  
Since the magnetic field at the each location of impurity site is infinite value but has a finite magnetic flux as described in \eqref{320},
this penetration of the magnetic field at each delta function type impurity site, e.g. lattice, can provide a field theoretic description of imperfect diamagnetism in dirty superconductor~\cite{tinkham2004introduction}.

\begin{figure}[H]
	\centering
	\subfigure[]{
		\includegraphics[
		width=0.45\textwidth
		]{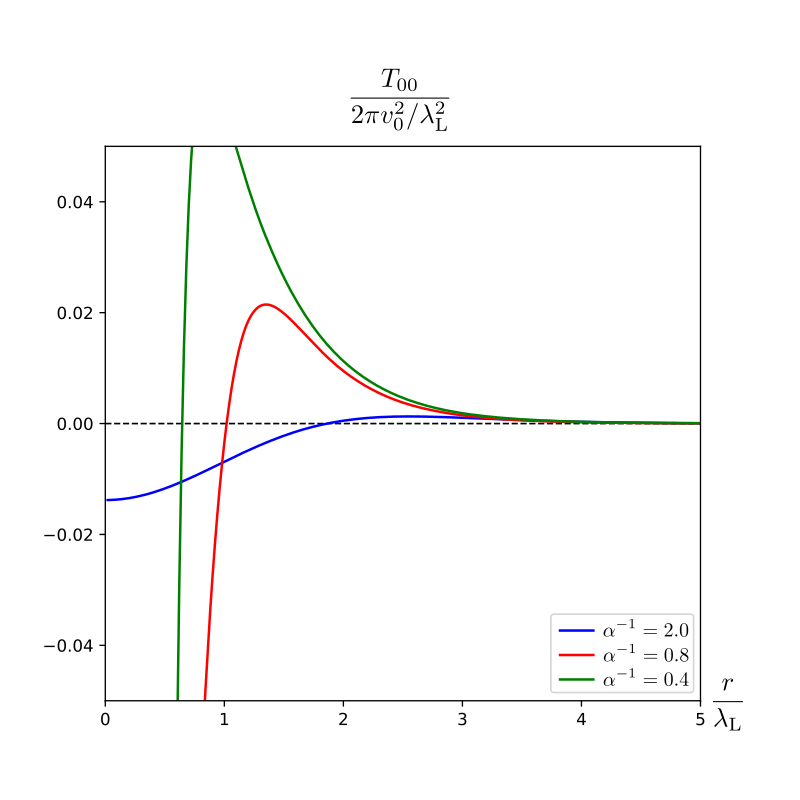}
	}
	\hfill
	\subfigure[]{
		\includegraphics[
		width=0.45\textwidth
		]{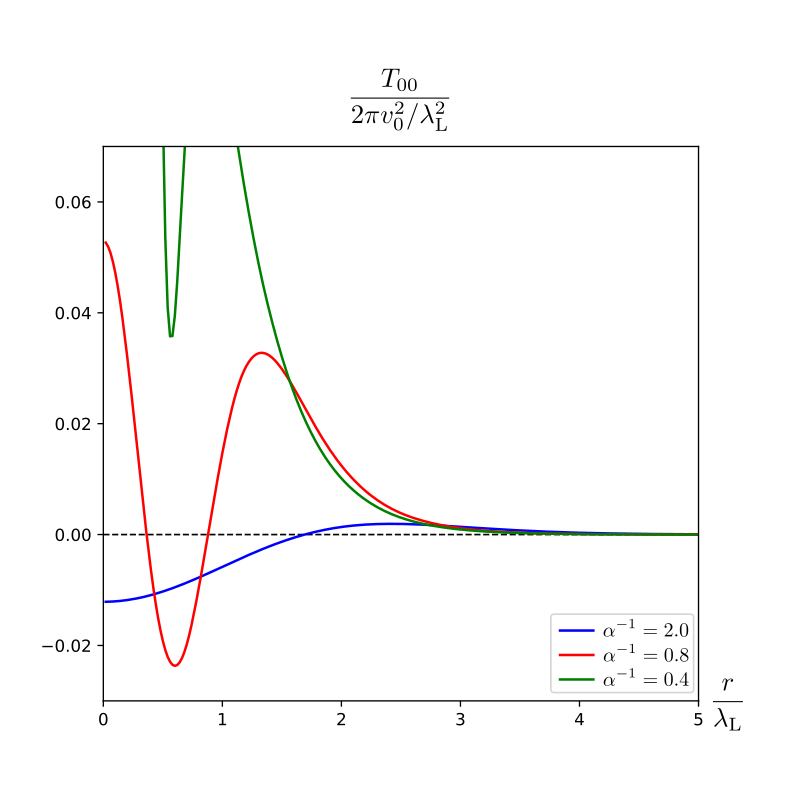}
	}
	\subfigure[]{
		\includegraphics[
		width=0.45\textwidth
		]{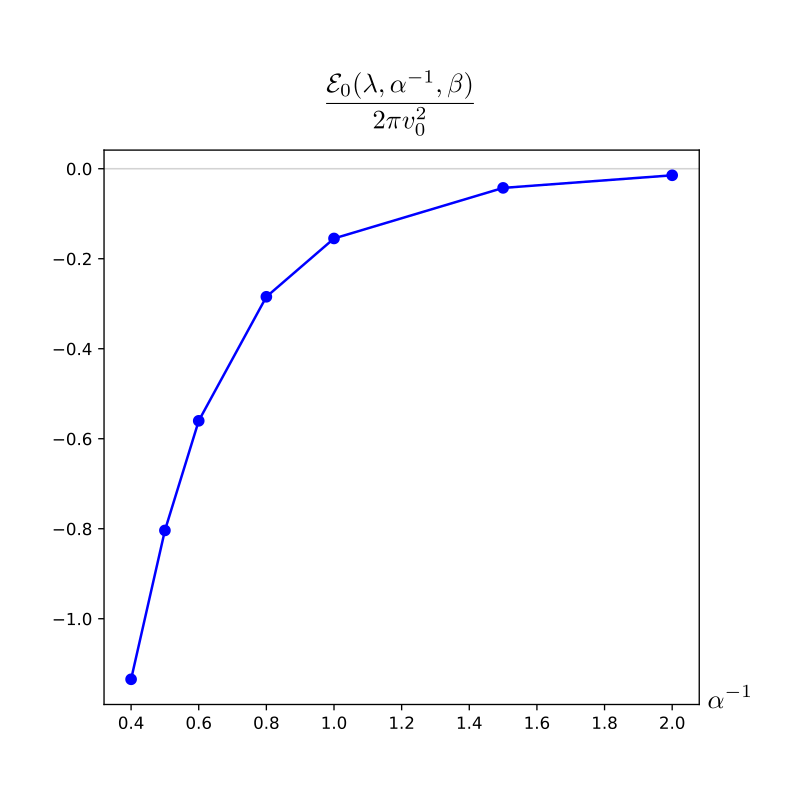}
	}
	\hfill
	\subfigure[]{
		\includegraphics[
		width=0.45\textwidth
		]{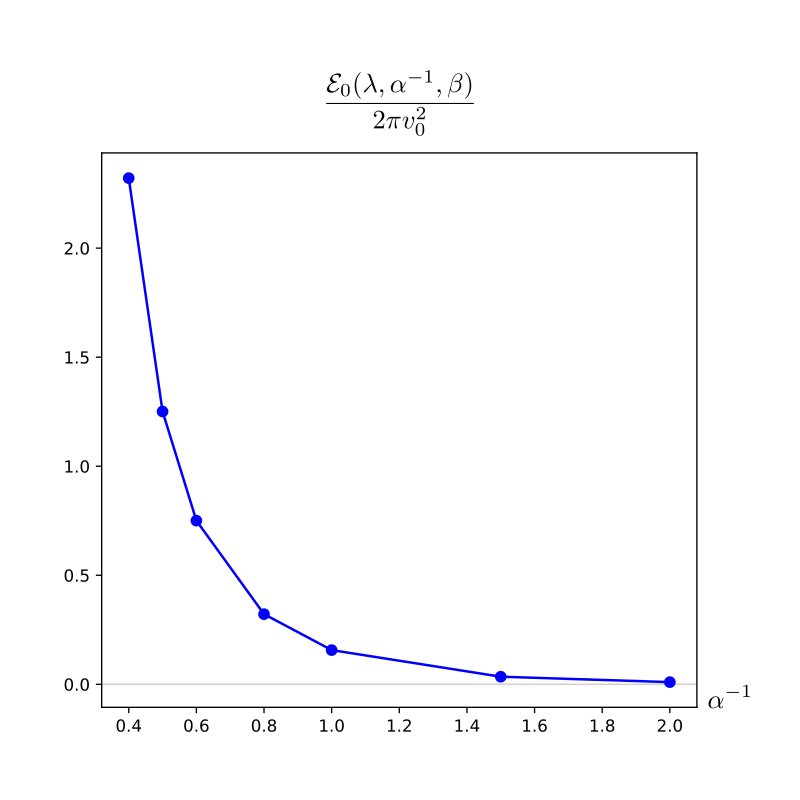}
	}
	\caption{
		Energy density $-T^{t}_{~t}$ for quartic scalar coupling (a) \( \lambda = 0.5  \) and (b) \( \lambda = 1.5  \) for a fixed ratio of size and depth parameter \( \eta = 0.8 \) and various values of the size parameter \( \alpha^{-1} \). Values of energy $\mathcal{E}_{0}(\lambda,\alpha^{-1},\eta=0.8)$  for quartic scalar coupling, (c) \( \lambda = 0.5  \) and (d) \( \lambda = 1.5  \), and various values of the size parameter \( \alpha^{-1} \).    
	}
	\label{fig:315}
\end{figure}

Energy density \( - T\indices{^t_t} \) near the origin also consists of singular delta function pieces and continuous nonsingular part
\begin{align}
	-T\indices{^t_t} = v_{0}^{2} \bigg[ (\lambda-1) \eta^{2} \lambda_{ \text{L} }^{2} \Big( \frac{\delta(r)}{r} \Big)^{2} - \lambda\eta \frac{\delta(r)}{r} + \frac{\lambda}{4\lambda_{ \text{L} }^{2}} + 
	\frac{2 f_{0}^{2} \lambda^{2} \eta^{2}}{\lambda_{ \text{L} }^{2}} \Big(\frac{r}{\lambda_{\rm L}}\Big)^{2(\lambda\eta-1)} + \cdots \bigg]
	.\label{319}
\end{align}
Numerical data exhibits a procedure to singular limit. As the size parameter $\alpha^{-1}$ decreases to zero with constant $\eta=0.8$, the graphs in 
Figure~\ref{fig:315}$-$(a)--(b) show the development of a singular dip of energy density at the origin as $(\alpha^{-1},-T^{t}_{~t})=(2.0,-0.01)$, $(0.8,-0.60)$, $(0.4,-8.57)$ for $\lambda=0.5$ and a singular bump of energy density at the origin as $(\alpha^{-1},-T^{t}_{~t})=(2.0,-0.01)$, $(0.8,0.05)$, $(0.4,5.74)$ for $\lambda=1.5$. Hence the first term proportional to $(\delta(r)/r)^{2}$ in the energy density \eqref{319} contributes dominantly to energy which is leading divergence. This infinite energy is easily expected since the infinite impurity energy $\mathcal{E}_{\sigma}^{v_{0}}$ \eqref{308} and $\mathcal{E}_{\sigma}^{v(\boldsymbol{x})}$ \eqref{311} is peaked at the origin, is proportional to square of two-dimensional delta function, and thus can not easily be tamed to be a finite value by local nonlinear interactions. Notice that the sign of energy is determined only by the quartic scalar coupling, specifically by the sign of $\lambda-1$ in the first energy term in \eqref{319}, in the delta function limit of our consideration: Energy of this vacuum configuration of positive nonzero $\lambda\eta$ is negative in weak coupling regime $(\lambda<1)$ and positive in strong coupling $(\lambda>1)$, that matches perfectly with numerical data throughout this section and Figure~\ref{fig:315}$-$(c)--(d). At the critical quartic scalar coupling $\lambda=1$, from the energy density \eqref{319}, the first leading divergent term vanishes and the second delta function term gives a negative energy $-2\pi v_{0}^{2}\eta$ related to an effective vorticity $\lambda\eta=\eta$. Suppose that the inhomogeneous vacuum is expected to have zero energy according to the BPS bound even in the delta function limit. Then the net contribution from spatial integration of all the other infinitely many expanded terms in \eqref{319} must produce the positive energy $2\pi v_{0}^{2}\eta$ for exact cancellation.

If a delta function bump of negative depth parameter \(\beta<0\) is considered, \(\lambda\eta\) becomes negative.
Hence the scalar field diverges at the origin as \(|\phi| \sim r^{-|\lambda\eta|}\) and any regular solution is forbidden.

All the non-BPS vacuum solutions of various values of the quartic scalar couplings \(\lambda\), obtained in this section, are connected continuously the BPS vacuum solutions of critical quartic scalar coupling.
Therefore, as long as the lower bounds \eqref{222}--\eqref{223} are satisfied, nonexistence of such classical lower energy configuration disconnected from the BPS vacuum solution is not guaranteed though unlikely.
Nevertheless this nonexistence issue is extremely difficult and far beyond the scope of the current work.

\section{Vortices in the Presence of Inhomogeneity}

Even in the inhomogeneous nonrelativisic abelian Higgs model, the manifold of Higgs vacua keeps to be a circle \( \mathrm{S}^1 \) as long as the localized inhomogeneous part \(\sigma\) with
\(      \displaystyle
        \lim_{ |\boldsymbol{x}| \to\infty }
        \sigma = 0
\)
\eqref{212} is assumed.
The winding between the circle of the \( \text{U}(1) \) Higgs vacua and the circle \( \partial \mathbb{R}^{2} = \text{S}^{1} \) at spatial infinity \( |\boldsymbol{x}|\to\infty \) is kept to be topological with the first homotopy
\(
        \Pi_{1}
        (\mathrm{S}^{1})
        = \mathbb{Z}
\),
and thus topologically stable gauged vortices of finite energy are expected to be supported even in the presence of inhomogeneity.

In this section, we turn on nonzero vorticity \( n = 1, 2, 3, \cdots \), and consider \(n\) inhomogenous topological vortices superimposed at the center of cylindrically symmetric inhomogeneous part \eqref{203}.
Hence the ansatz \eqref{206}--\eqref{207} is valid and so are the equations \eqref{208}--\eqref{209}.
Nonsingular behavior of the fields at the origin requires
\begin{equation}
        \lim_{r \to 0}
        |\phi|
        = 0
, \qquad
        \lim_{r \to 0} A
        = 0
,\label{400}
\end{equation}
and finiteness of the energy fixes the boundary conditions at spatial infinity
\begin{equation}
        \lim_{r \to \infty} |\phi|
        = v_{0}
, \qquad
        \lim_{r \to \infty} A
        = n
.\label{401}
\end{equation}

Power series expansion of the fields is attempted near the origin
\begingroup \allowdisplaybreaks
\begin{align}
        |\phi|(r) & \approx
        v_{0} \phi_{n0} r^{n}
        \bigg\{
                1 -
                \frac
                { 4na_{n0} + \lambda(1-\beta) }
                {4 \lambda (n+1)}
                \Big( \frac{r}{\xi} \Big)^{2}
\nonumber \\ & \quad
                + \frac
                { [ 4n a_{n0} + \lambda (1-\beta) ]^{2}
                        + 16 (n+1) a_{n0}^{2}
                        - 4 (n+1) (n+2\lambda)
                        ( \alpha^{2}\beta
                        - \phi_{n0}^{2} \delta_{n1} )
                } { 32 \lambda^{2} (n+1)(n+2) }
                \Big( \frac{r}{\xi} \Big)^{4}
\nonumber \\ & \quad
                + \cdots
        \bigg\}
,\label{402}
\\
        A(r) & \approx
        a_{n0}
        \Big(
                \frac{r}{\lambda_{\text{L}}}
        \Big)^{2}
        +
        \frac{
                \alpha^{2}\beta
                - 
                \phi_{n0}^{2}
                \delta_{n1}
        }{8}
        \Big(
                \frac{r}{\lambda_{\text{L}}}
        \Big)^{4}
\nonumber \\ & \quad
        -
        \frac{
                8 \alpha^{4}\beta
                -
                \phi_{n0}^{2}
                [
                        12 a_{n0}
                        + \lambda(1-\beta)
                ]
                \delta_{n1}
                +
                16 \phi_{n0}^{2}
                \delta_{n2}
        }{192}
        \Big(
                \frac{r}{\lambda_{\text{L}}}
        \Big)^{6}
        + \cdots
,\label{403}
\end{align}
\endgroup
where the two undetermined constants \( \phi_{n0} \) and \( a_{n0} \) have different values for different vorticity \( n \), quartic scalar coupling \(\lambda\), and parameters \(\alpha^{-1}\) and \(\beta\) of Gaussian inhomogeneous part \(\sigma(r)\) \eqref{203}.
The asymptotic solution of fields approaches the boundary value \eqref{401} with the same exponential behavior for the scalar amplitude \eqref{303} and the gauge field
\begin{equation}
        A(r) \approx
        n - a_{n\infty}
        \frac{r}{ \lambda_{\text{L}} }
        K_{1}
        \Big(
        \frac{r}{ \lambda_{\text{L}} }
        \Big)
,\label{404}
\end{equation}
where the constants \(\phi_\infty\) in \eqref{303} and \(a_{n\infty}\) in \eqref{404} are related to the constants \(\phi_{n0}\) and \(a_{n0}\) near the origin for smooth connection of the fields.

Once a vorticity \(n\) also called the winding number is given, the magnetic flux is quantized as
\(      \displaystyle
    \Phi_{B} = 
    \frac{2\pi\hbar}{g} n
\)
for any inhomogeneous topological vortex solution as the first Chern number independent of the quartic scalar coupling \(\lambda\) and irrespective of the shape of inhomogeneous part \( \sigma (\boldsymbol{x}) \) \eqref{221}--\eqref{212}.
The inhomogeneous topological vortices are  also spinless \eqref{211} as long as any static configurations without carrying electric field are concerned.
Contrary to the aforementioned two quantities, energy per unit length along the \(z\)-axis \(\mathcal{E}\) \eqref{305} is not calculated analytically except for the BPS case of critical coupling \(\lambda=1\).
When the quartic scalar coupling has the critical value \(\lambda = 1\), the correlation length and the London penetration depth are equal \(\xi = \sqrt{2} \lambda_{\text{L}}\).
Non-perturbatively, the BPS bound is saturated
\( \displaystyle
        \mathcal{E}_{n}^{\lambda=1}
        =
        n \mathcal{E}_{1}^{\lambda=1}
        =
        2\pi v_{0}^{2} n
\)
which suggests non-interacting vortices at least in tree level because of exact cancellation of attraction by the Higgs and repulsion by the gauge field. In addition, this character is not spoiled by introduction of spatial inhomogeneity \cite{Tong:2013iqa}.
On the other hand, once it has arbitrary value \(\lambda \neq 1\), all convenient BPS properties disappear in the calculation of energy of both homogeneous and inhomogeneous vacuum configuration.
Hence analysis relies on numerical studies \cite{Jacobs:1978ch} and the introduced inhomogeneity with magnetic impurity term \eqref{210} may induce the problem more complicated.

Before numerical analysis with arbitrary quartic scalar coupling, again we consider the BPS limit of critical quartic scalar coupling and examine the Bogomolny equation for \(n\) separated vortices at positions
\(
        \{ \boldsymbol{x}_{a}
        \mid a = 1, 2, \cdots, n \}
\)
\cite{Kim:2024gfn}
\begin{equation}
        \nabla^{2} \ln
        \frac{ |\phi|^{2} }
        { \displaystyle \prod_{a=1}^{n}
          | \boldsymbol{x}
          - \boldsymbol{x}_{a} |^{2} }
        =
        \frac{1}{ \lambda_{\text{L}}^{2} v_{0}^{2} }
        [ |\phi|^{2} - v^{2}(\boldsymbol{x}) ]
.\label{407}
\end{equation}
Application of the Newtonian shooting analogy, introduced for mathematical analysis of existence and uniqueness of cylindrically symmetric inhomogeneous vacuum solution in the section \ref{sec:vac}, is also relevant for cylindrically symmetric inhomogeneous vortex solutions of vorticity \(n~(n\ge1)\).
With the help of the two-dimensional Gauss' law \eqref{318} for delta function, the Bogomolny equation \eqref{407} is rewritten as the Newton's second law for one-dimensional motion of a hypothetical particle of unit mass subject to an impact at initial time \( t = r/\lambda_{\text{L}} = 0 \), a friction with a coefficient inversely proportional to the time \(1/t\), and the force derived from the time-dependent effective potential \(U(x, t)\) \eqref{315}
\begin{equation}
        \frac{d^{2} x}{d t^{2}}
        =
        2n \frac{ \delta(t) }{t}
        - \frac{1}{t} \frac{dx}{dt}
        - \frac{\partial U}{\partial x}
.\label{408}
\end{equation}
Once the impact of infinite strength \(\delta(0) = +\infty\) is exerted to the particle at initial position at negative infinity \( x(0) = \ln [|\phi|(r=0) / v_{0}] = -\infty \), this nonrelativistic particle moves infinite distance during finite elapsed time \(t_{0} ~ (t_{0} > 0)\) and reaches an appropriate negative finite position \( x_{0} = x(t_{0}) < 0 \) with an arbitrary positive velocity \( v_{0}^{\rm in} = dx(t_{0}) / dt > 0 \).
Disappearance of the delta function term for the impact reduces the equation \eqref{408} to the previous equation \eqref{314} for the motion between time \(t=t_{0}>0\) and \(t=+\infty\).

Consider first positive depth parameter \(\beta>0\) as illustrated by the three colored solid curves in Figure \ref{fig:312}-(a).
If the velocity \(v_{0}^{\rm in}\) is chosen to be too large, the hypothetical particle passes zero at a finite time and diverges to positive infinity.
If \(v_{0}^{\rm in}\) is chosen too small, its energy \(\mathscr{E}\) \eqref{316} decreases too much and has a smaller value than negative unity \(\mathscr{E}<-1\) at a finite time because of the dissipation due to the friction and lowering of the time-dependent effective potential \( U(x,t) \). Hence it cannot arrive at the hilltop \( U(x=0, t=\infty) = -1 \) at the origin \(x=0\) of the effective potential at infinite time \(U(x, t=\infty)\) \eqref{315}.
Thus, by the continuity of the velocity \(v_{0}^{\rm in}\) at time \(t_0\), there is a motion between these overshoot and undershoot motions, which starts at negative infinity at initial value \(x(0)=-\infty\), increases monotonically with slowing velocity, and arrives at the destination at infinite time \(x(\infty) = 0\) as illustrated by the gray-colored solid curve in Figure \ref{fig:312}-(a).
This unique motion corresponds to the cylindrically symmetric vortex solution of vorticity \(n\) whose monotonically increasing profile of scalar amplitude connects from \(|\phi|(0)=0\) to \(|\phi|(\infty)=v_{0}\).

Consider second negative depth parameter \(\beta<0\) as illustrated by the three colored solid curves in Figure \ref{fig:312}-(b).
If the velocity \(v_{0}^{\rm in}\) at time $t_{0}$ is chosen to be larger than a critical value \(v_{0}^{\text{top}}\), the particle passes the hilltop of the effective potential at a finite time and diverges to positive infinity at infinite time \(x(+\infty) = +\infty\).
Since the motion with \(x(t_{0}) = x_{0}\) and \(v_{0}^{\rm in} = v_{0}^{\text{top}}\) corresponds to the motion which stops at the hilltop \(x(t_{1})=x_{\rm top}\) of the time-dependent effective potential at a finite time \(t_{1} (> t_{0})\), all such motions with \(x(t_{0}) = x_{0}\) and \(v_{0}^{\rm in} < v_{0}^{\text{top}}\) proceed to positive infinity at infinite time \(\displaystyle \lim_{t\to\infty} x(t) = +\infty\) along the right decreasing side of the effective potential whose hilltop gradually moves to the left.
When the velocity \(v_{0}^{\rm in}\) is smaller than but close to \(v_{0}^{\rm top}\),
the particle turns near the hilltop of the effective potential at a finite time.
Even if the turning point \(x_{\rm turn}\) on the left increasing side of the effective potential \(U(x,t)\) \eqref{316} is located to close to the hilltop, the particle stays near \(x_{\rm turn}\) for sufficiently large time to make the time-dependent friction with coefficient \(1/t\) weak enough.
Meanwhile the hilltop of the time-dependent effective potential moves to the left faster than the motion of the turned particle. Then the particle turns again to the right and moves to positive infinity at infinite time \( \displaystyle \lim_{t\to\infty} x(t) = + \infty \) along the decreasing right side of the time-dependent effective potential \( U(x,t) \).
If the velocity \(v_{0}^{\rm in}\) is smaller than another appropriately chosen critical value \(v_{0}^{\text{bottom}}\) smaller than \(v_{0}^{\rm top}\),
the particle turns at a negative position on the left side of the effective potential at a finite time \(x_{\rm turn} < 0\) and goes back to negative infinity at infinite time \(\displaystyle \lim_{t\to\infty} x(t) = - \infty\) along the decreasing left side of the time-dependent effective potential \( U(x,t) \).
Thus by the hilltop \( U(x=0, t=\infty) = -1 \) position of the time-dependent effective potential at infinite time \(x_{\rm top} (t=\infty) = 0\) and continuity of the velocity \(v_{0}^{\rm in}\) at time \(t_{0}\), there exists an appropriate velocity \(v_{0}^{\rm in}\) in the region \(v_{0}^{\rm bottom} < v_{0}^{\rm in} < v_{0}^{\rm top}\) between the discussed overshoot and undershoot motions.
This corresponds to the motion that the particle starts at negative infinity at initial time \(x(0) = -\infty\), increases monotonically to the positive turning point at a finite time
\(t_{\rm turn}\), \(0 < x(t_{\rm turn}) = x_{\rm turn} < x_{\rm top}\),
moves back along the left decreasing side of the effective potential \(U(x,t)\) \eqref{315}, and stops finally at the hilltop \(x(t=\infty) = 0\) of the effective potential at infinite time \(U(x,t=\infty)\).
This motion is energetically obtained only under the condition that particle energy \(\mathscr{E} = U(x_{\rm turn}, t_{\rm turn})\) for the particle stopped at the turning position \(x_{\rm turn}\) at time \(t_{\rm turn}\) is exactly equal to the total power loss by the friction \(\displaystyle \int_{t_{\rm turn}}^{\infty} dt \, \frac{1}{t} \frac{dx}{dt}\) and potential change
\begin{align}
\int_{t_{\rm turn}}^{\infty} dt \frac{1}{t} \frac{dx}{dt} = U(x_{\rm turn}, t_{\rm turn}) - U(0,+\infty) = 1 - e^{x_{\rm turn}} + (1-\beta e^{-\alpha^{2} t_{\rm turn}^{2}})x_{\rm turn} .
\end{align}
This unique motion illustrated by the gray-colored solid curve in Figure \ref{fig:312}-(b) corresponds to the cylindrically symmetric vortex solution of vorticity \(n\) in the BPS limit of \(\lambda=1\) in the presence of a Gaussian bump \eqref{203} of negative depth parameter \(\beta<0\).

In the following two subsections, numerical works show that all the obtained cylindrically symmetric vortex solutions are classified into two categories of positive and negative depth parameter \(\beta\) irrespective of the size parameter \(\alpha^{-1}\) and quartic scalar coupling \(\lambda\) including its critical value \(\lambda = 1\).

\subsection{Rest energy of inhomogeneous vortex}
\label{subsec:vortex}

\begin{figure}[H]
        \centering
        \includegraphics[
                width=0.65\textwidth
        ]{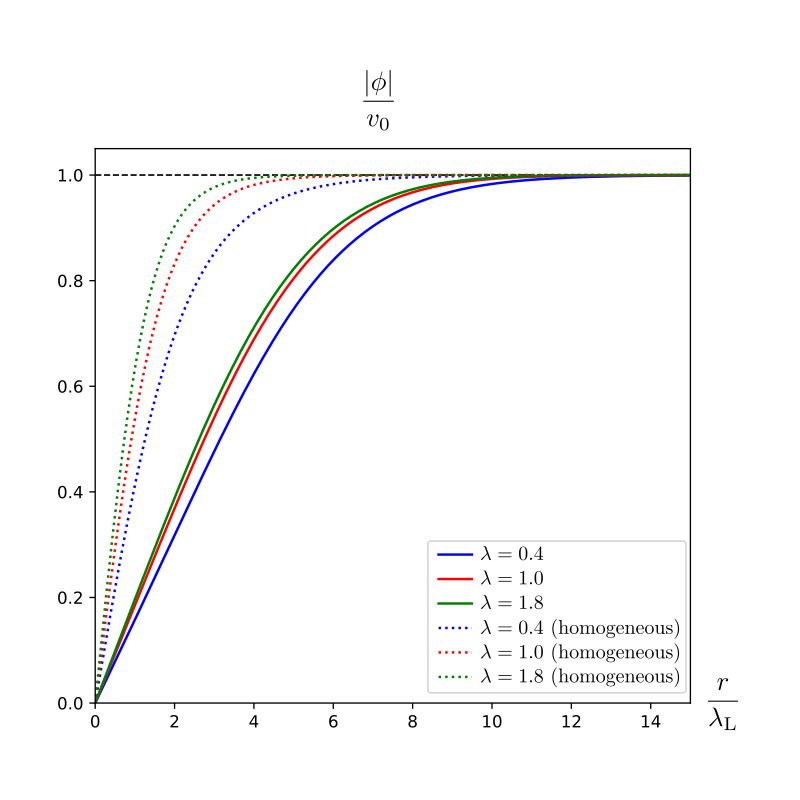}
        \caption{
                Amplitudes of the complex scalar field of an \(n=1\) vortex for different values of the quartic scalar coupling \(\lambda = 0.4, 1.0, 1.8 \) and for Gaussian inhomogeneous part with fixed dimensionless size parameter \( \alpha^{-1} = \sqrt{20} \approx 4.47 \) and depth parameter \(\beta=1\).
                The homogeneous counterparts with zero depth parameter \(\beta=0\) are given by three dotted curves for comparison.
        }
\label{fig:401}
\end{figure}

\begin{figure}[H]
        \centering
        \subfigure[]{
                \includegraphics[
                        width=0.45\textwidth
                ]{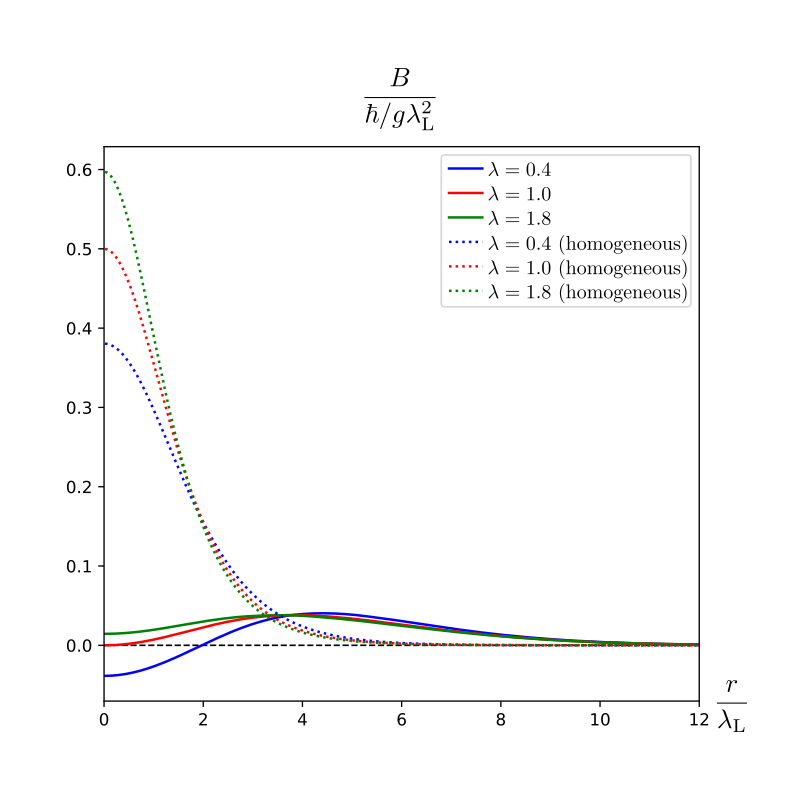}
        }
        \hfill
        \subfigure[]{
                \includegraphics[
                        width=0.45\textwidth
                ]{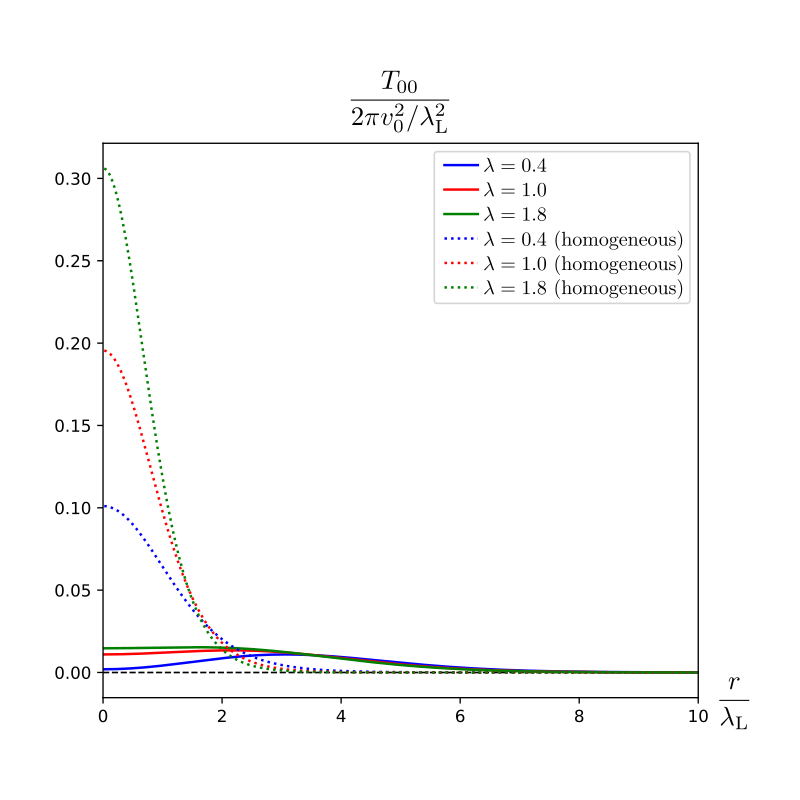}
        }
        \caption{
                (a) Magnetic fields \(B\) and (b) energy densities \(-T\indices{^t_t}\) of an \(n=1\) vortex for different values of the quartic scalar coupling \(\lambda = 0.4, 1.0, 1.8 \) and for Gaussian inhomogeneous part with fixed dimensionless size parameter \( \alpha^{-1} = \sqrt{20} \approx 4.47 \) and depth parameter \(\beta=1\).
                The homogeneous counterparts with zero depth parameter \(\beta=0\) are plotted as dotted curves for comparison.
        }
\label{fig:402}
\end{figure}

First, topological vortex solutions of unit vorticity \(n=1\) are considered with the Gaussian type inhomogeneous part \eqref{203}.
Numerical works show change of the scalar amplitude profiles connecting the boundary values at the origin \eqref{400} and at spatial infinity \eqref{401} as in Figure \ref{fig:401}.
Specifically, with a chosen set of size and depth parameters \( \alpha^{-1} = \sqrt{20} \approx 4.47 \) and \(\beta = 1 \), the region of varying scalar amplitude is characterized not only by the length scale \(\xi\) for homogeneous vortices but also  by the competition between two length scale \(\xi\) and \( \alpha^{-1} \lambda_{\text{L}} \) for inhomogeneous vortices.
As the quartic scalar coupling \(\lambda\) increases, the correlation length \(\xi\) \eqref{213} decreases and hence both the three solid and dotted curves in Figure \ref{fig:401} show more rapid increase from zero scalar amplitude near the origin and approach exponentially the boundary value \(v_0\) at large distance irrespective of the presence of inhomogeneity.
\noindent
The corresponding curves of magnetic field \(B(r)\) and energy density \(-T\indices{^t_t}(r)\) are shown in Figure \ref{fig:402}-(a)--(b).
Inhomogeneity of the Gaussian bump of depth parameter \(\beta=1\) \eqref{203} repels both magnetic field \(B\) and energy density \(-T\indices{^t_t}\) of larger distance, of which shapes are drastically changed in comparison with those of \(n=1\) vortices in the homogeneous limit.
Thus they are distributed in wide region and every peak in these ring-shaped distributions is located about the radius \( \alpha^{-1} \lambda_{\text{L}} \).
As the quartic scalar coupling \(\lambda\) decreases, the correlation length \(\xi\) \eqref{213} increases and the distributions become growing and their peaks move gradually outside.

\begin{figure}[H]
        \centering
        \includegraphics[
                width=0.65\textwidth
        ]{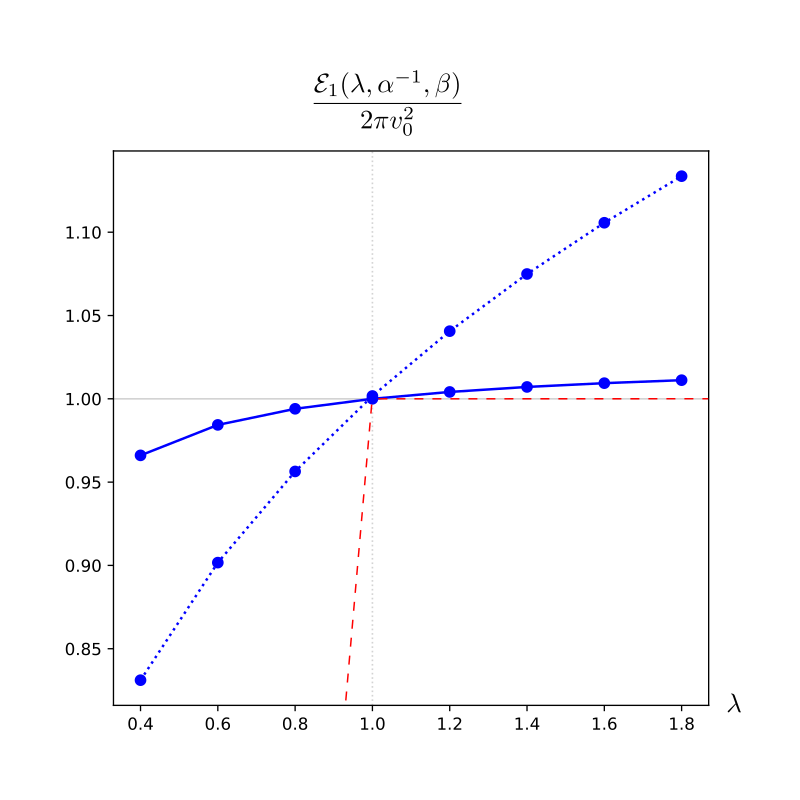}
        \caption{
                The energy of an \(n=1\) vortex for various values of the quartic scalar coupling \(\lambda\) and for Gaussian inhomogeneous part with fixed size parameter \( \alpha^{-1} = \sqrt{20} \approx 4.47 \) and depth parameter \(\beta = 1\) is plotted by a solid curve in the unit of \(n=1\) BPS vortex
                \(
                        \mathcal{E}_{1}
                        ( \lambda=1, \alpha^{-1}, \beta)
                        = 2\pi v_{0}^{2}
                \).
                The homogeneous counterpart with zero depth parameter \(\beta = 0\) is plotted by a dotted curve for comparison.
        }
\label{fig:403}
\end{figure}

It is known that, in the BPS limit of critical quartic scalar coupling \(\lambda=1\) \cite{Tong:2013iqa, Kim:2024gfn}, the rest energy of an isolated vortex of unit vorticity \(n=1\) is unaffected for the variety of inhomogeneous part \(\sigma(\boldsymbol{x})\) \eqref{221}--\eqref{212}.
The rest energy
\(
        \mathcal{E}_{1}
        ( \lambda, \sqrt{20}, 1)
\)
of a vortex is computed by numerical works for various \(\lambda\) with a fixed Gaussian inhomogeneous part \eqref{203}.
When \(\lambda\) increases, the rest energy of a single vortex increases and reproduces the rest energy of a single BPS vortex
\(
        \mathcal{E}_{1}
        ( \lambda=1, \alpha^{-1}, \beta)
        = 2\pi v_{0}^{2}
\)
at the critical quartic scalar coupling as shown by the solid curve in Figure \ref{fig:403}.
Effect of Gaussian inhomogeneous part \(\sigma(r)\) \eqref{203} to the rest energy of a single vortex is to reduce the amount of \(\lambda\)-dependence, i.e., the slope of increasing rest energy of an inhomogeneous vortex of unit vorticity drawn by a solid curve is smaller than that of a vortex of unit vorticity in homogeneous limit \(\sigma=0\) drawn by a dotted curve.
Thus the presence of inhomogeneous part \(\sigma(\boldsymbol{x})\) absorbs some residual interaction due to the quartic scalar coupling away from the critical quartic scalar coupling \(\lambda\) and makes additional or reduced energy to create a single vortex of unit vorticity less.
The data obtained in Figure \ref{fig:403} explains the following properties in the context of field theory:
When the quartic scalar coupling is less than the critical value \(\lambda < 1\) corresponding to type I superconductivity, the significantly increased rest energy of single vortex means the energetic disfavor of vortex-impurity composite, the production of a vortex at the impurity site, in a dirty type I superconducting sample.
This tendency is consistent with perfect diamagnetism in conventional type I superconductivity.
When the quartic scalar coupling is larger than the critical value \(\lambda > 1\) corresponding to type \Romtwo\ superconductivity, the significantly decreased rest energy of a single vortex means the energetic favor of vortex-impurity composite, the production of a vortex at the impurity site, in a dirty type \Romtwo\ superconducting sample.
This tendency is consistent with imperfect diamagnetism in conventional type \Romtwo\ superconductivity.
Obviously, the boundary between two type I and \Romtwo\ criteria is given by the BPS limit of nonrelativistic inhomogeneous abelian Higgs model with critical quartic scalar coupling \(\lambda = 1\) in which the rest energy of each vortex is unaltered by the inhomogeneous part and even the interaction between two separated BPS vortices remains to be zero \cite{Kim:2024gfn}.

\begin{figure}[H]
        \centering
        \subfigure[]{
                \includegraphics[
                        width=0.45\textwidth
                ]{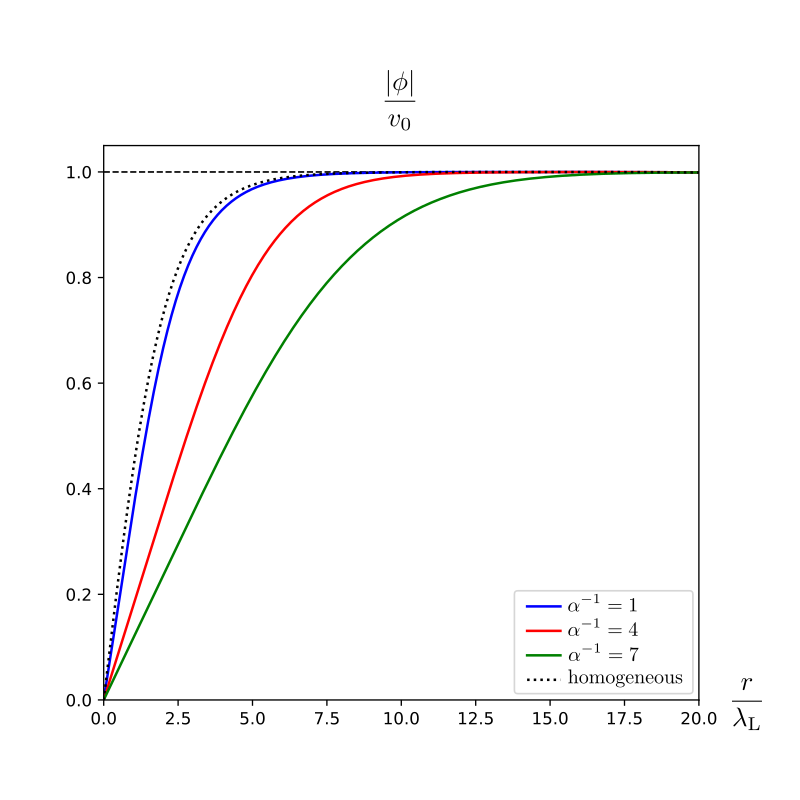}
        }
        \hfill
        \subfigure[]{
                \includegraphics[
                        width=0.45\textwidth
                ]{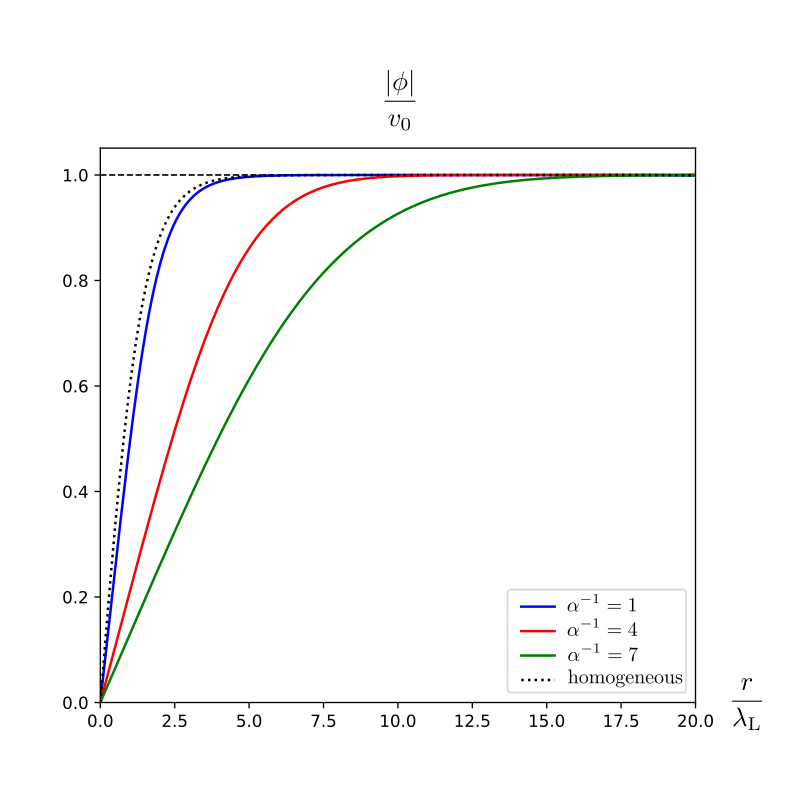}
        }
        \caption{
                Amplitudes of the complex scalar field for two quartic scalar couplings, (a) \(\lambda=0.5\) and (b) \(\lambda=1.5\) for Gaussian inhomogeneous part with various values of the size parameter \( \alpha^{-1}\) and fixed depth parameter \(\beta = 1\).
                The homogeneous counterparts with zero depth parameter \(\beta=0\) are plotted as dotted curves for comparison.
        }
\label{fig:404}
\end{figure}

\begin{figure}[H]
        \centering
        \subfigure[]{
                \includegraphics[
                        width=0.45\textwidth
                ]{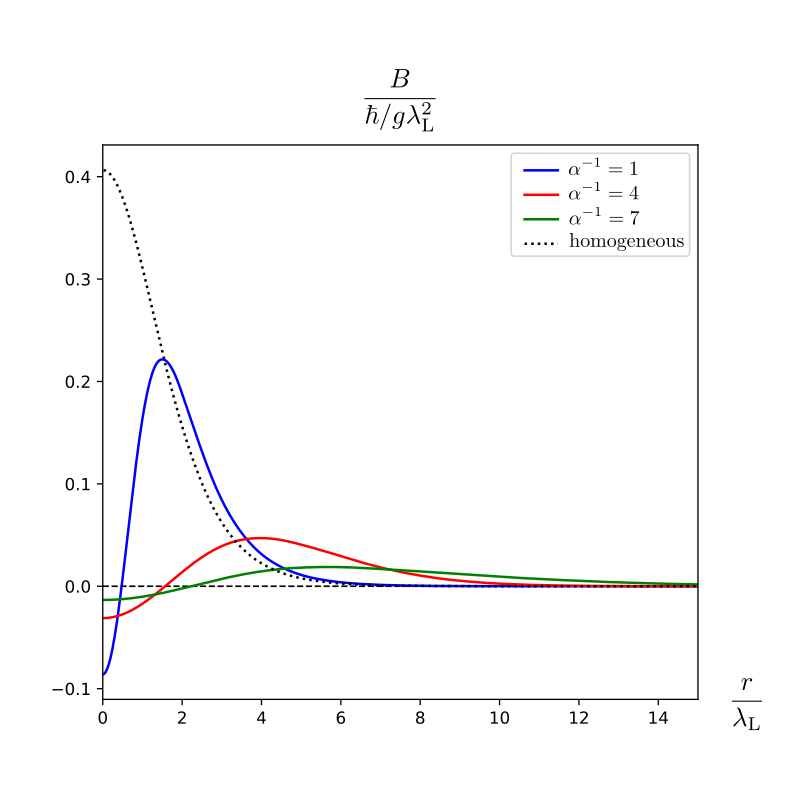}
        }
        \hfill
        \subfigure[]{
                \includegraphics[
                        width=0.45\textwidth
                ]{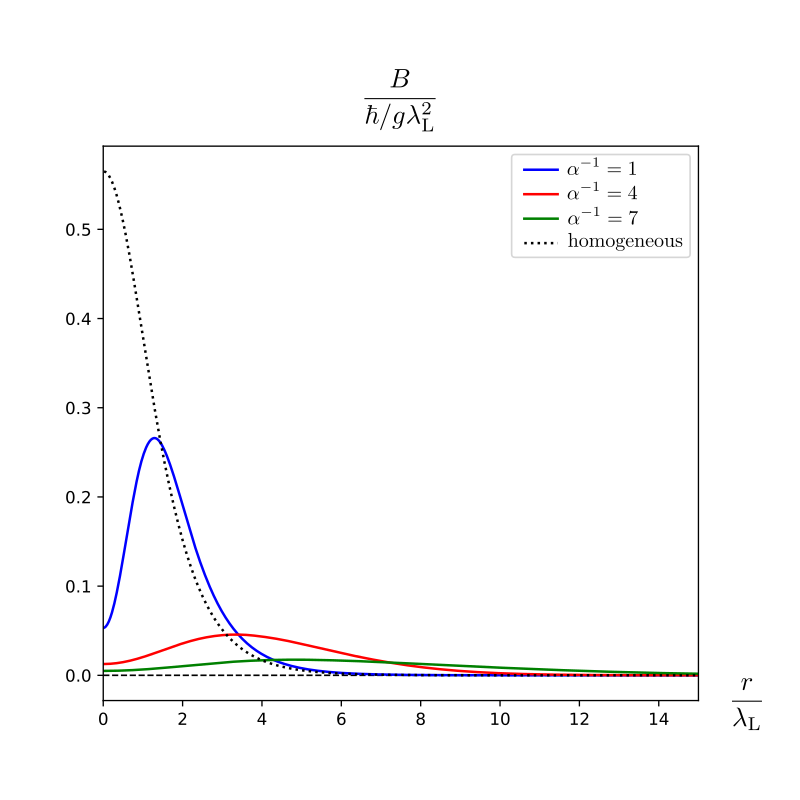}
        }
        \\
        \subfigure[]{
                \includegraphics[
                        width=0.45\textwidth
                ]{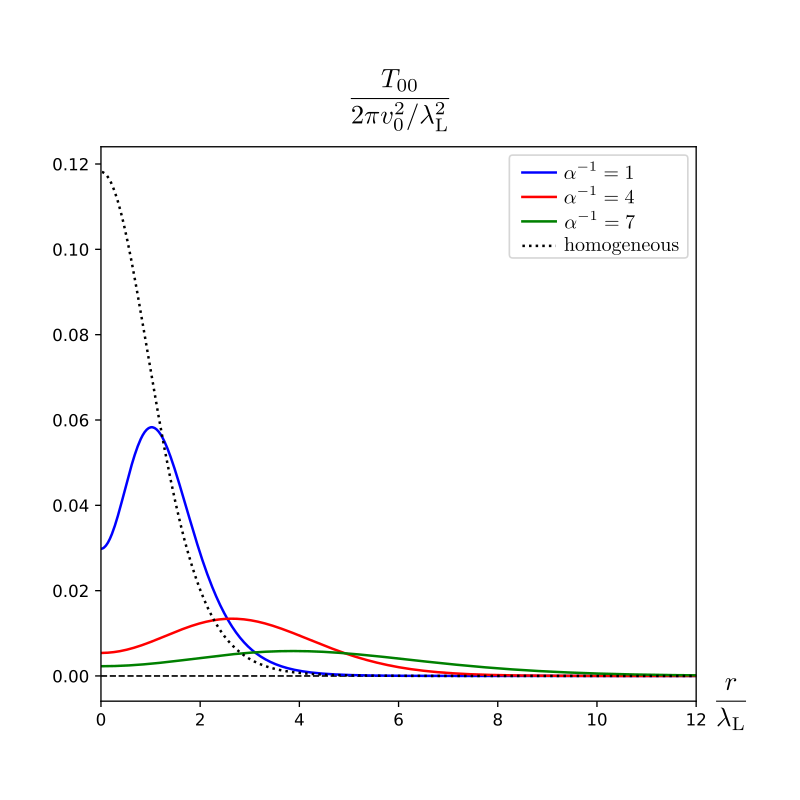}
        }
        \hfill
        \subfigure[]{
                \includegraphics[
                        width=0.45\textwidth
                ]{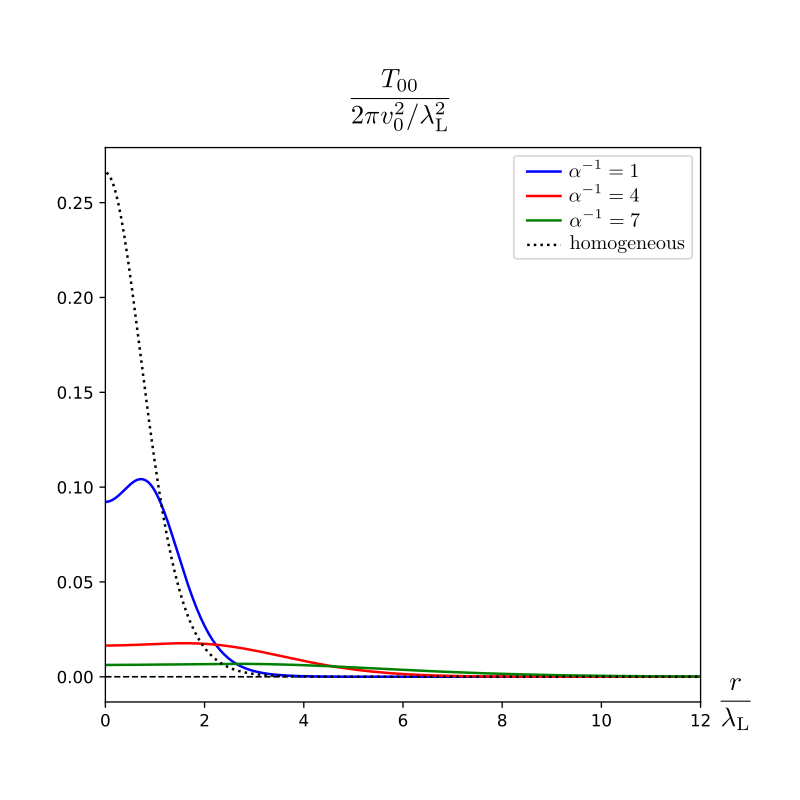}
        }
        \caption{
                Magnetic fields \(B\) for two quartic scalar couplings (a) \(\lambda=0.5\) and (b) \(\lambda=1.5\) and energy density \(-T\indices{^t_t}\) for two quartic scalar couplings (c) \(\lambda=0.5\) and (d) \(\lambda=1.5\) for Gaussian inhomogeneous part with various values of the size parameter \( \alpha^{-1}\) and fixed depth parameter \(\beta = 1\).
                The homogeneous counterparts of magnetic field and energy density with zero depth parameter \(\beta=0\) are given by dotted curves for comparison.
        }
\label{fig:405}
\end{figure}

For two chosen values of quartic scalar coupling \(\lambda\), one less than unity \(\lambda = 0.5 < 1\) and the other larger than unity \(\lambda = 1.5 > 1\), numerical data are computed for various size parameters \(\alpha^{-1} = 1, 4, 7\) and the obtained results are summarized in the following graphs.
Profiles of scalar amplitude \(|\phi|(r)\) of an \(n=1\) topological vortex connect smoothly the small \(r\) behavior \eqref{402}--\eqref{403} and the large \(r\) behavior \eqref{303} and \eqref{404}, as shown in Figure \ref{fig:404}.
Change of the size parameter \(\alpha^{-1}\), \(\alpha^{-1} = 1, 4, 7\) with keeping quartic scalar coupling \(\lambda\) and depth parameter  \(\beta=1\) induces shape change of inhomogeneous vortex solutions of unit winding number \(n=1\), i.e., the radial distances of \(r\)-dependent region increase almost proportional to the size parameter \(\alpha^{-1}\).
When the quartic scalar coupling \(\lambda\) varies, (a) \(\lambda = 0.5\) and (b) \(\lambda = 1.5\), the correlation length \(\xi\) \eqref{213} is inversely proportional to its square root however the \(\alpha\)-dependent behavior is sustained.
\noindent
Both the magnetic fields and energy densities of the \(n=1\) vortex solutions are peaked at the origin in homogeneous limit of \(\beta=0\) as shown by the dotted curves in Figure \ref{fig:405} and become ring-shaped with a peak of radius about \(\alpha^{-1}\lambda_{\text{L}}\) in the presence of the inhomogeneous part \(\sigma(r)\) with depth parameter \(\beta=1\) as shown by the solid curves in Figure \ref{fig:405}.
On dependence of the quartic scalar coupling \( \lambda \), its value of the magnetic field at the origin \(B(0)\) is negative \(B(0)<0\) for \(\lambda=0.5\) and positive \(B(0)>0\) for \(\lambda=1.5\) for depth parameter \(\beta=1\) irrespective of size parameter \(\alpha^{-1}\).
In the BPS limit of critical quartic scalar coupling \(\lambda=1\), the magnetic fields begins with zero \(B(0)=0\) at the origin for the depth parameter \(\beta=1\)
(see the expansion of the gauge field (4.52) near the origin and Figure 6 in Ref. \cite{Kim:2024gfn}).
Therefore, with unit depth parameter \(\beta=1\), the value of magnetic field at the origin \(B(0)\) is always negative for all \(\lambda<1\) and positive for all \(\lambda>1\) irrespective of size parameter \(\alpha^{-1}\).
Since the expansion of \(A(r)\) near the origin \eqref{403} gives the magnetic field at the origin
\(
        B(0) =
        \dfrac{2\hbar}
        {g\lambda_{\text{L}}^{2}}
        a_{n0}
\)
and thus the sign of the coefficient \(a_{n0}\) is also negative for all \(\lambda<1\) and positive for all \(\lambda>1\) for depth parameter \(\beta=1\) and arbitrary size parameter \(\alpha^{-1}\).

\begin{figure}[H]
        \centering
        \subfigure[]{
                \includegraphics[
                        width=0.45\textwidth
                ]{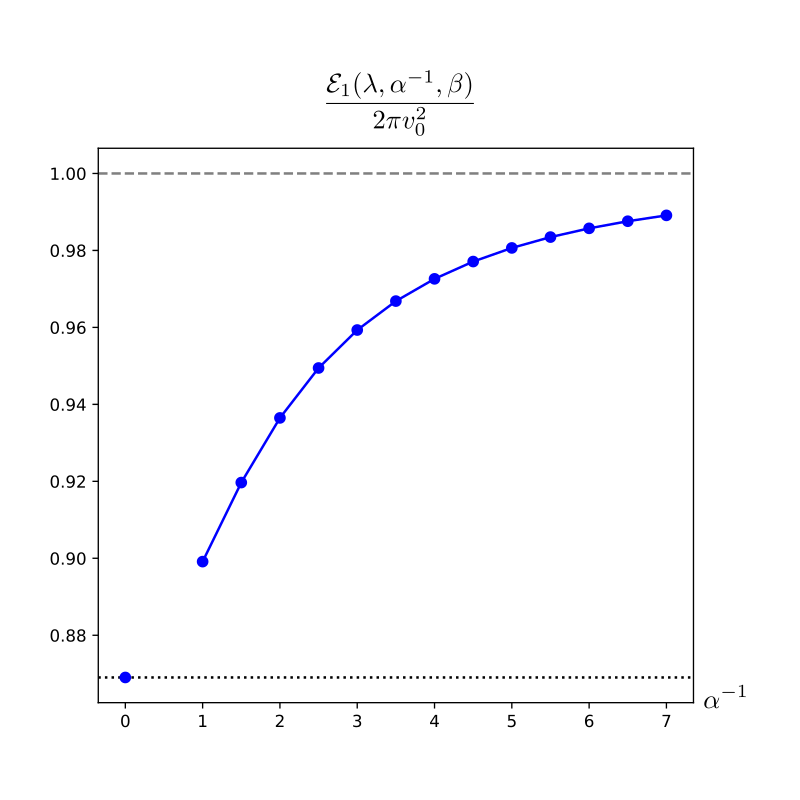}
        }
        \hfill
        \subfigure[]{
                \includegraphics[
                        width=0.45\textwidth
                ]{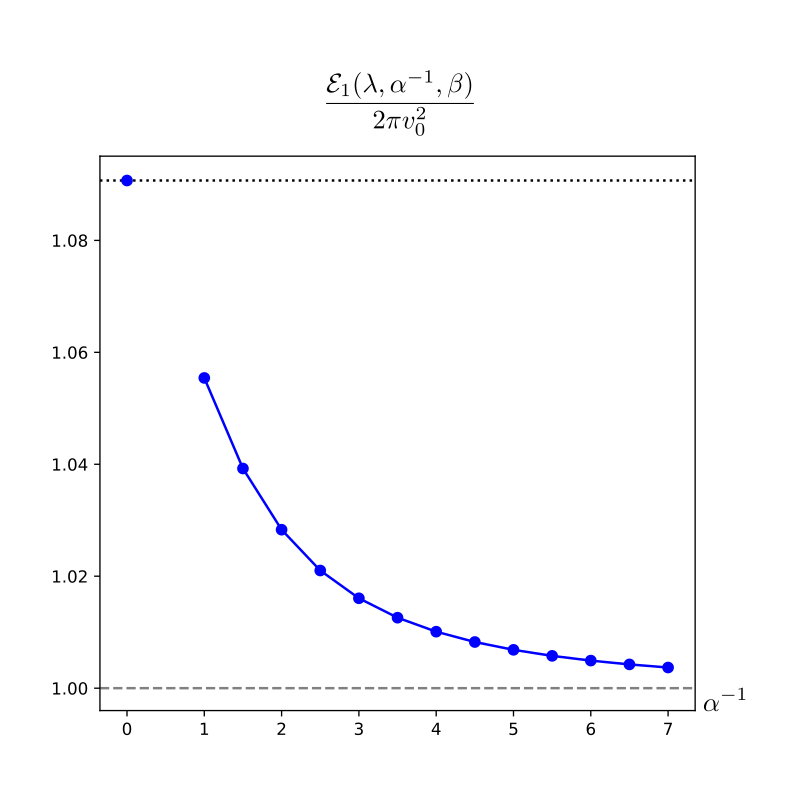}
        }
        \caption{
                The energy of an \(n=1\) vortex \(\mathcal{E}_{1}(\lambda, \alpha^{-1}, 1)\) for various values of the size parameter \(\alpha^{-1}\) of Gaussian inhomogeneous part with fixed depth parameter \(\beta=1\) for two quartic scalar couplings \(\lambda = 0.5 < 1\) and \(\lambda = 1.5 > 1\).
                The homogeneous counterpart with zero depth parameter \(\beta=0\) is plotted by a dotted horizontal line and the energy \( \mathcal{E}_{1} ( \lambda=1, \alpha^{-1}, \beta ) = 2 \pi v_{0}^{2} \) of a BPS vortex is given by a dashed line for comparison.
        }
\label{fig:406}
\end{figure}

When the quartic scalar coupling is less than the critical value \(\lambda<1\), e.g. \(\lambda=0.5\) in Figure \ref{fig:406}-(a), the rest energy of an \(n=1\) topological vortex is always smaller than that of a unit BPS vortex
\(
        \mathcal{E}_{1}
        ( \lambda<1, \alpha^{-1}, \beta )
        <
        \mathcal{E}_{1}
        ( \lambda=1, \alpha^{-1}, \beta )
\)
but the energy gap between the two decreases as the inhomogeneous region increase.
When the quartic scalar coupling is greater than the critical value \(\lambda>1\), e.g. \(\lambda=1.5\) in Figure \ref{fig:406}-(b), the rest energy of an \(n=1\) topological vortex is always larger than that of a unit BPS vortex
\(
        \mathcal{E}_{1}
        (\lambda= 1, \alpha^{-1}, \beta )
        <
        \mathcal{E}_{1}
        ( \lambda>1, \alpha^{-1}, \beta )
\)
but the energy gap between the two decreases as the inhomogeneous region increase.
It implies that the change of rest energy expressed by the energy gap
\(
        \mathcal{E}_{1}
        ( \lambda \neq 1, \alpha^{-1}, \beta )
        -
        \mathcal{E}_{1}
        ( 1, \alpha^{-1}, \beta )
\)
is induced by the effect of nonzero net interaction and is reduced by the inhomogeneity whose size is parameterized in terms of \(\alpha^{-1}\).
The larger the contamination by impurity becomes in a sample, the more the cancellation of net interaction occurs for the rest energy.

\begin{figure}[H]
        \centering
        \subfigure[]{
                \includegraphics[
                        width=0.45\textwidth
                ]{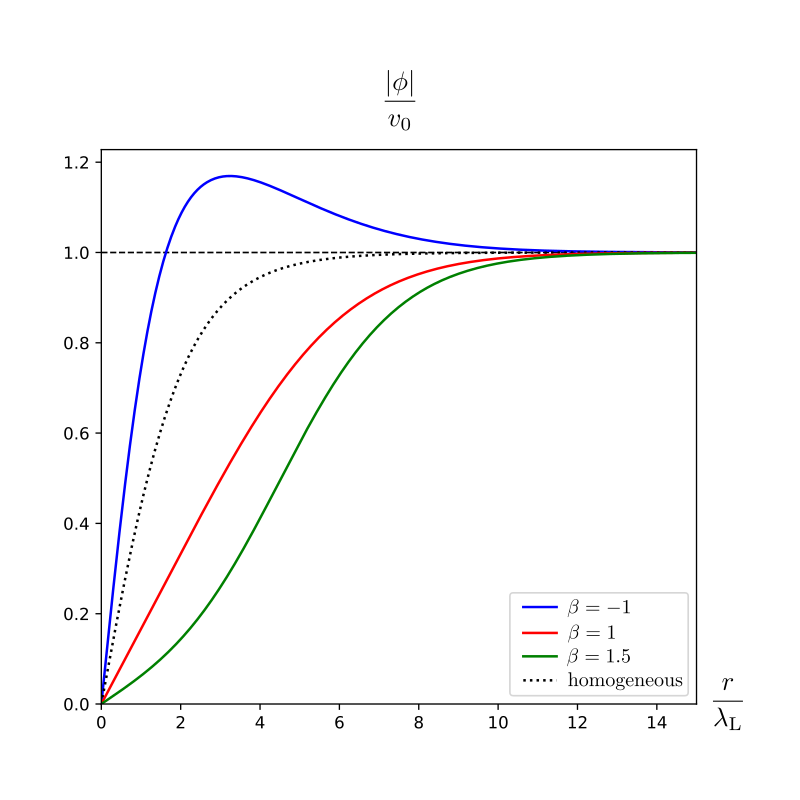}
        }
        \hfill
        \subfigure[]{
                \includegraphics[
                        width=0.45\textwidth
                ]{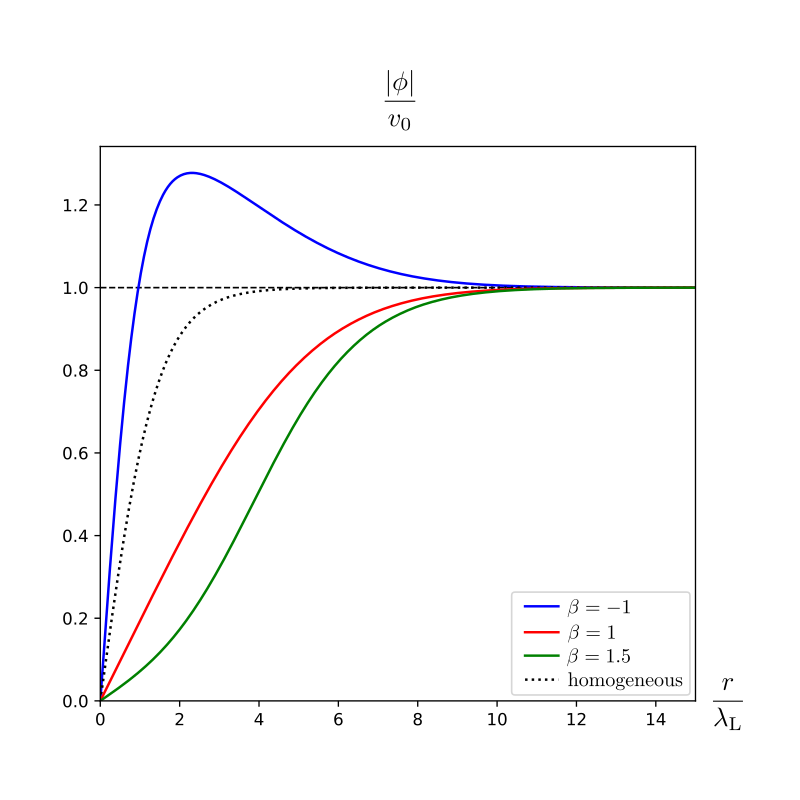}
        }
        \caption{
                Amplitudes of the complex scalar field for two quartic scalar couplings (a) \(\lambda=0.5\) and (b) \(\lambda=1.5\) for Gaussian inhomogeneous part with various values of the depth parameter \(\beta\) and fixed size parameter \( \alpha^{-1} = \sqrt{20} \approx 4.47 \).
                The homogeneous counterpart with zero depth parameter \(\beta=0\) is plotted as a dotted curve for comparison.
        }
\label{fig:407}
\end{figure}

For two chosen values of \(\lambda\), one less than unity \(\lambda = 0.5 < 1\) and the other larger than unity \(\lambda = 1.5 > 1\), numerical works of the \(n=1\) vortex solutions are performed for various depth parameters \(\beta\) and the obtained results are summarized in the following graphs.
When depth parameter \(\beta\) is larger than a critical negative value, scalar amplitude of each \(n=1\) vortex is monotonically increasing from one boundary value \(|\phi|(0)=0\) \eqref{400} to the other boundary value \(|\phi|(\infty) = v_{0}\) irrespective of the value of the quartic scalar coupling as illustrated by the dotted, red-colored and green-colored solid curves in Figure \ref{fig:407}-(a)--(b).
Meanwhile negative depth parameter \(\beta\) of the Gaussian dip is smaller than a critical value, the slope parameter \(\phi_{n0}\) for the scalar amplitude \eqref{402} expanded near the origin becomes large enough and passes \(v_{0}\) at a finite distance, reaches a maximum value \(\max(|\phi|) > v_{0}\), and then decreases to the boundary value \(|\phi|(\infty) = v_{0}\) \eqref{401} irrespective of the value of quartic scalar coupling as illustrated by the blue-colored solid curves in the Figure \ref{fig:407}-(a)--(b).

\begin{figure}[H]
        \centering
        \subfigure[]{
                \includegraphics[
                        width=0.45\textwidth
                ]{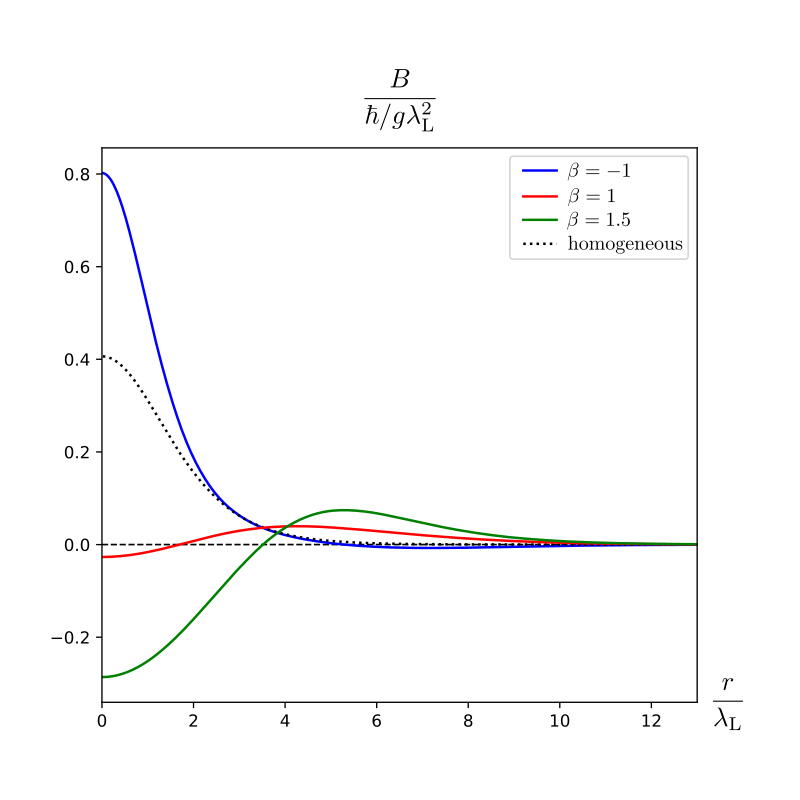}
        }
        \hfill
        \subfigure[]{
                \includegraphics[
                        width=0.45\textwidth
                ]{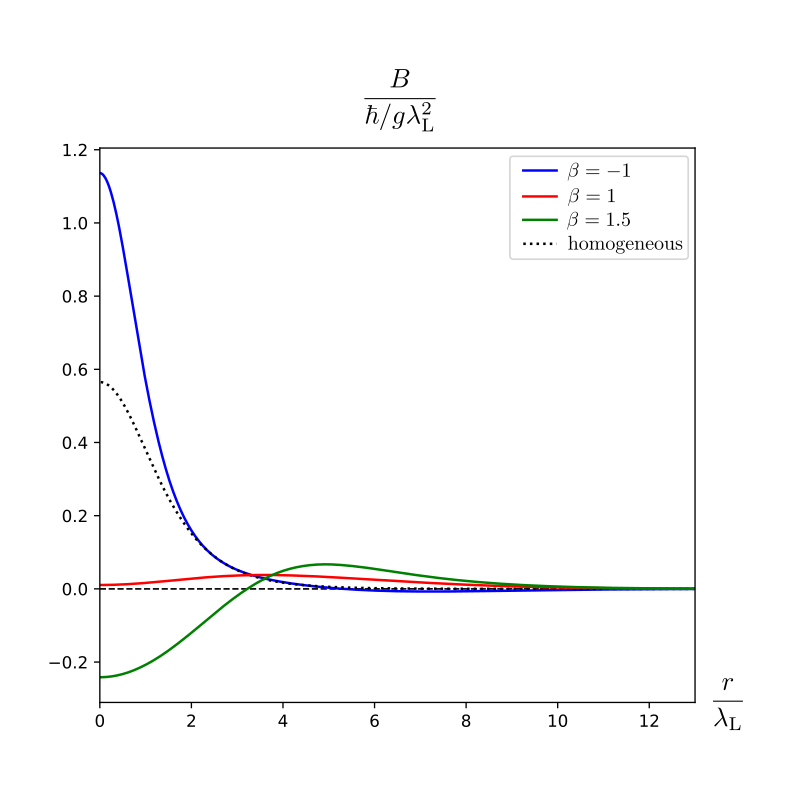}
        }
        \\
        \subfigure[]{
                \includegraphics[
                        width=0.45\textwidth
                ]{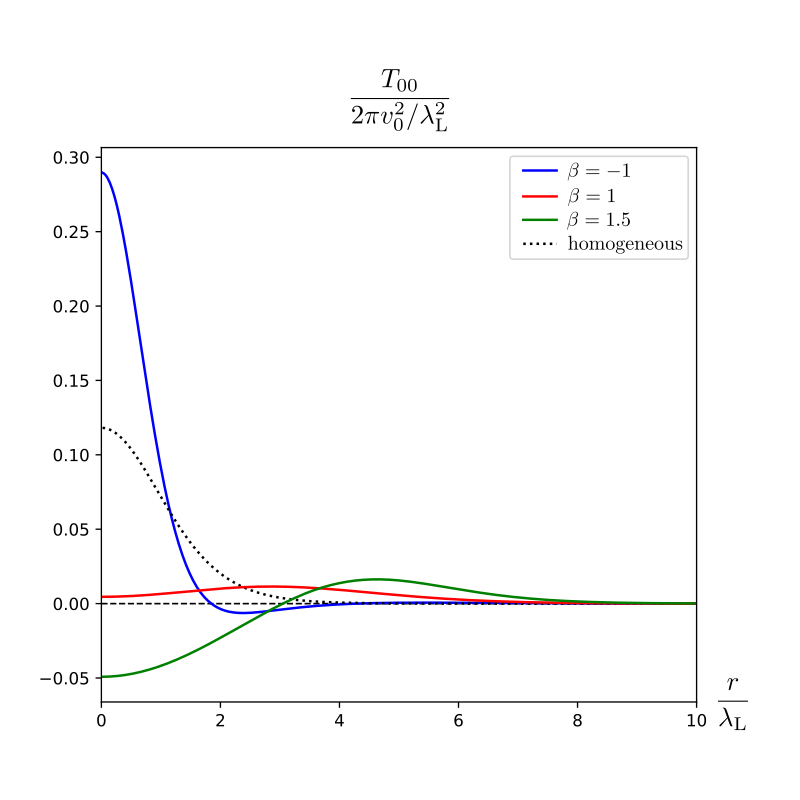}
        }
        \hfill
        \subfigure[]{
                \includegraphics[
                        width=0.45\textwidth
                ]{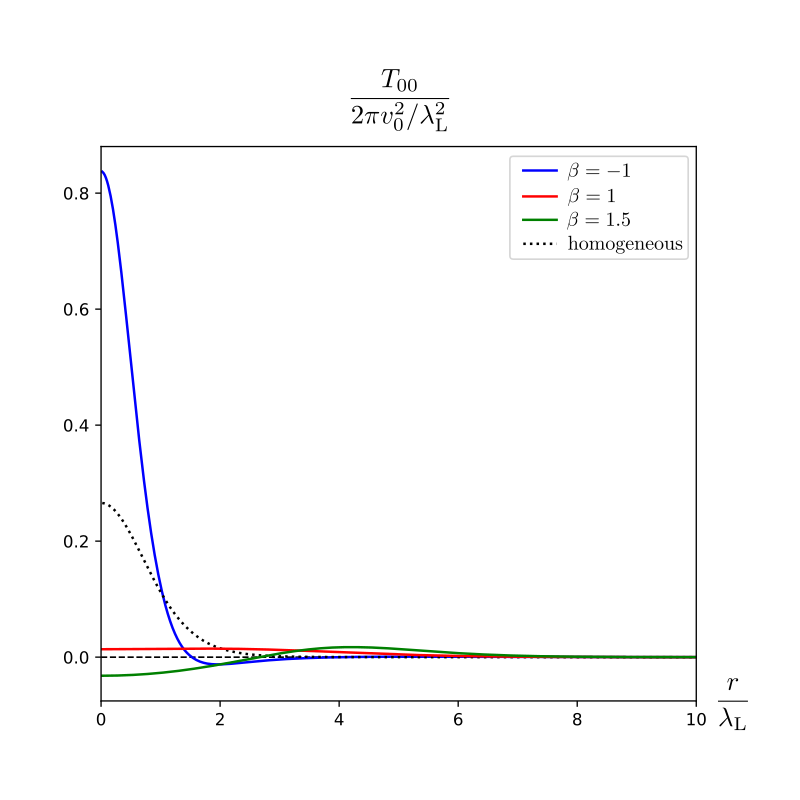}
        }
        \caption{
                Magnetic fields \(B\) for two quartic scalar couplings (a) \(\lambda=0.5\) and (b) \(\lambda=1.5\) and energy density \(-T\indices{^t_t}\) for two quartic scalar couplings (c) \(\lambda=0.5\) and (d) \(\lambda=1.5\) for Gaussian inhomogeneous part with various values of the depth parameter \(\beta\) and fixed size parameter \( \alpha^{-1} = \sqrt{20} \approx 4.47 \).
                The homogeneous counterparts of magnetic field and energy density with zero depth parameter \(\beta=0\) are given by dotted curves for comparison.
        }
\label{fig:408}
\end{figure}

As negative depth parameter \(\beta\) of the Gaussian bump decreases, both the magnetic fields and energy densities of the \(n=1\) vortices become sharply peaked at the origin, decreases to a negative minimum value at finite radius, and then increases to zero at spatial infinity as shown by the blue-colored solid curves in Figure \ref{fig:408}.
Contrarily, as positive depth parameter \(\beta\) of the Gaussian bump increases, value of the magnetic fields and energy densities of the \(n=1\) vortices have the minimum, get smaller and even become negative for sufficiently large \(\beta\) at the origin, and then increases to the positive maximum value at finite radius with further decreasing behavior to zero at spatial infinity as shown by the red- and green-colored solid curves in Figure \ref{fig:408}.
These distribution behaviors of magnetic fields and energy densities occur for arbitrary value of the quartic scalar coupling \(\lambda\).

\begin{figure}[H]
        \centering
        \subfigure[]{
                \includegraphics[
                        width=0.45\textwidth
                ]{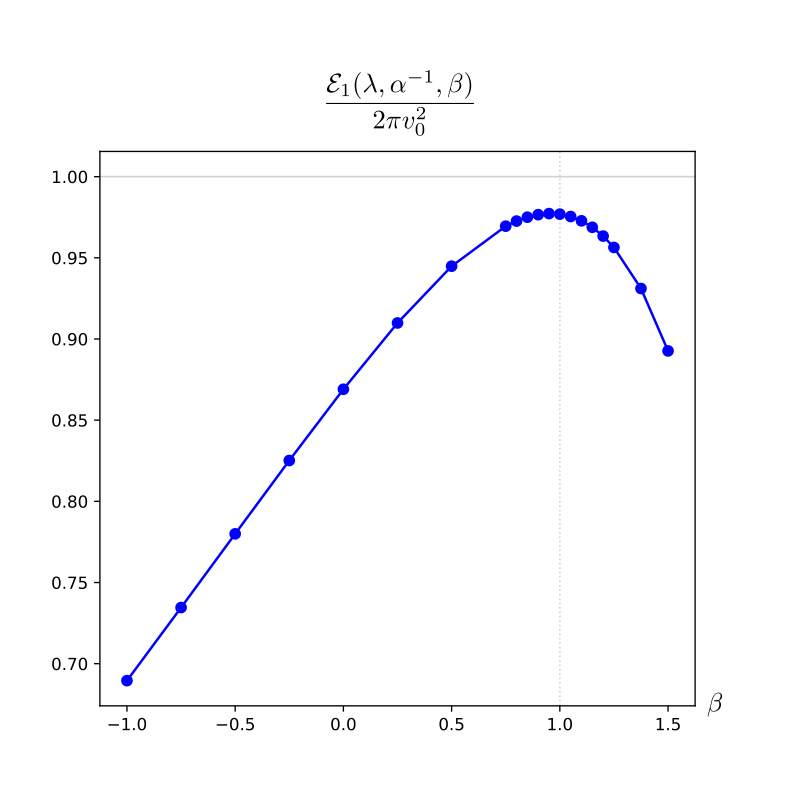}
        }
        \hfill
        \subfigure[]{
                \includegraphics[
                        width=0.45\textwidth
                ]{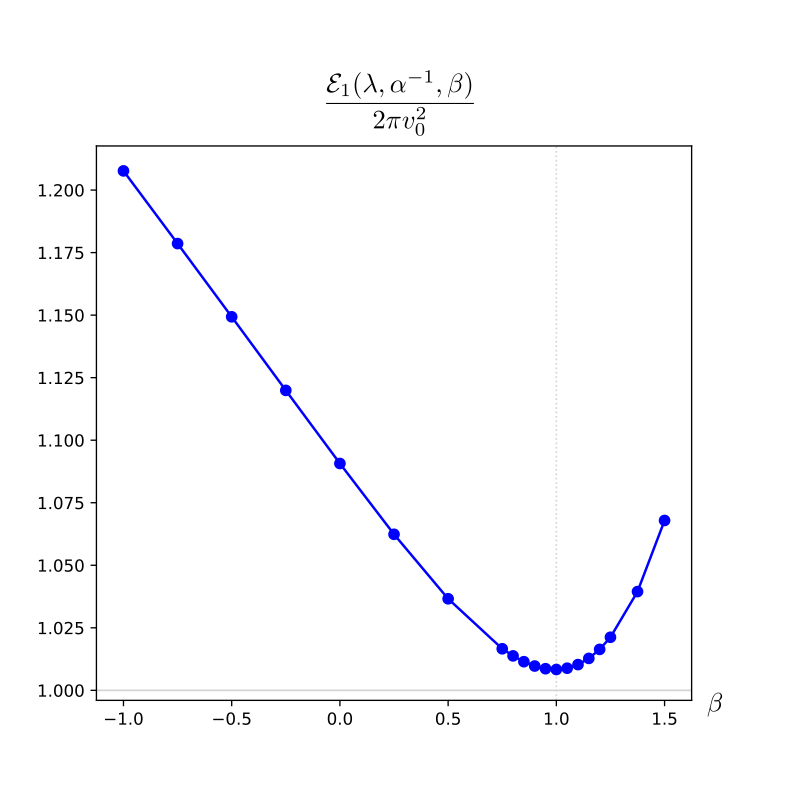}
        }
        \caption{
                The energy of an \(n=1\) vortex \(\mathcal{E}_{1}(\lambda, \alpha^{-1} = \sqrt{20}, \beta)\) for various values of the depth parameter \(\beta\) of Gaussian inhomogeneous part with fixed size parameter \( \alpha^{-1} = \sqrt{20} \approx 4.47 \) for two quartic scalar couplings (a) \(\lambda = 0.5 < 1\) and (b) \(\lambda = 1.5 > 1\).
        }
\label{fig:409}
\end{figure}

Note that the magnetic flux of these \(n=1\) vortices keeps to have a constant value \( \Phi_{B} = \dfrac{2\pi\hbar}{g} \) irrespective of change of the depth (height) of the Gaussian dip (bump).
Similar to the inhomogeneous vacuum energy
\(
        \mathcal{E}_{0}
        ( \lambda, \alpha^{-1}, \beta )
\),
the rest energy of a unit vortex 
\(
        \mathcal{E}_{1}
        ( \lambda, \alpha^{-1}, \beta )
\)
depends on the depth parameter \(\beta\) as long as the quartic scalar coupling is not critical \(\lambda\neq1\).
When it is smaller than the critical value \(\lambda<1\), e.g. \(\lambda=0.5\) in Figure \ref{fig:409}-(a), the rest energy of \(n=1\) non-BPS vortex is always smaller than that of \(n=1\) BPS vortex
\(
        \mathcal{E}_{1}
        ( \lambda < 1 , \alpha^{-1}, \beta )
        <
        \mathcal{E}_{1}
        ( \lambda = 1 , \alpha^{-1}, \beta )
        = 2\pi v_{0}^{2}
\)
for every depth parameter \(\beta\).
When it is larger than the critical value \(\lambda>1\), e.g. \(\lambda=1.5\) in Figure \ref{fig:409}-(b), the rest energy of \(n=1\) non-BPS vortex is always greater than that of \(n=1\) BPS vortex
\(
        \mathcal{E}_{1}
        ( \lambda > 1 , \alpha^{-1}, \beta )
        >
        \mathcal{E}_{1}
        ( \lambda = 1 , \alpha^{-1}, \beta )
        = 2\pi v_{0}^{2}
\)
for every depth parameter \(\beta\).
Numerical data indicate that absolute value of the energy gap
\(
        | \mathcal{E}_{1}
        (\lambda, \alpha^{-1}, \beta)
        - 2\pi v_{0}^{2} |
\)
seems to have minimum value in the vicinity of \(\beta=1\) with numerical validity
\begin{equation}
        | \mathcal{E}_{1}
        (\lambda, \alpha^{-1}, \beta)
        - 2\pi v_{0}^{2} |
        \le
        | \mathcal{E}_{1}
        (\lambda, \alpha^{-1}, \beta=1)
        - 2\pi v_{0}^{2} |
\end{equation}
for which the inhomogeneous part \(\sigma(r)\) \eqref{203} cancels exactly \(v_{0}^{2}\) at the origin,
\(
        v^{2}(r=0)
        = v_{0}^{2} + \sigma(0)
        = 0
\)
\eqref{400}.
Different from the inhomogeneous vacuum solutions in relation with extremum value at \(\beta=0\) in Figure \ref{fig:311}, the extremum values by numerical analysis are observed near \(\beta=1\) for \(n=1\) vortices and still lack reasonable explanation.

\subsection{Interaction between inhomogeneous vortices}
\label{subsec:vortices}

\begin{figure}[H]
        \centering
        \includegraphics[
                width=0.65\textwidth
        ]{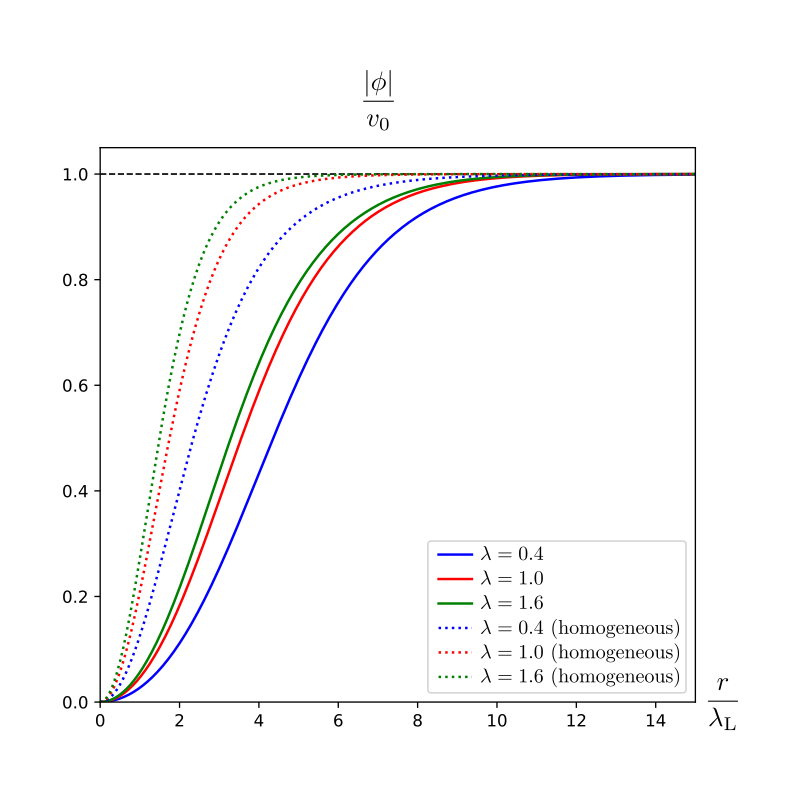}
        \caption{
                Amplitudes of the complex scalar field of \(n=2\) vortices for different values of the quartic scalar coupling \(\lambda = 0.4, 1.0, 1.6 \) and for Gaussian inhomogeneous part with fixed dimensionless size parameter \( \alpha^{-1} = \sqrt{20} \approx 4.47 \) and depth parameter \(\beta=1\).
                The homogeneous counterparts with zero depth parameter \(\beta=0\) are given by three dotted curves for comparison.
        }
\label{fig:410}
\end{figure}

\begin{figure}[H]
        \centering
        \subfigure[]{
                \includegraphics[
                        width=0.45\textwidth
                ]{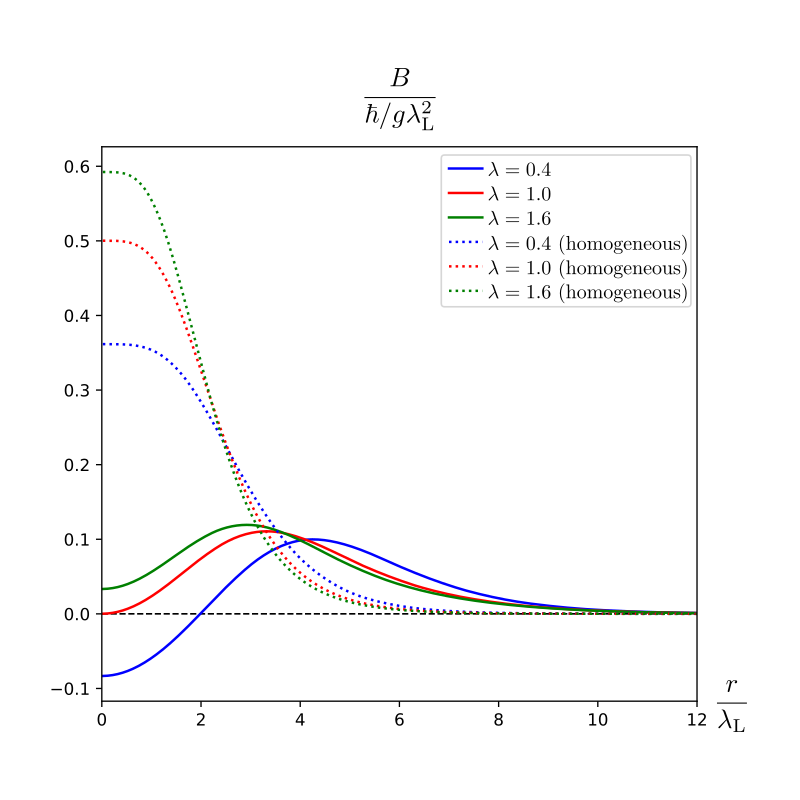}
        }
        \hfill
        \subfigure[]{
                \includegraphics[
                        width=0.45\textwidth
                ]{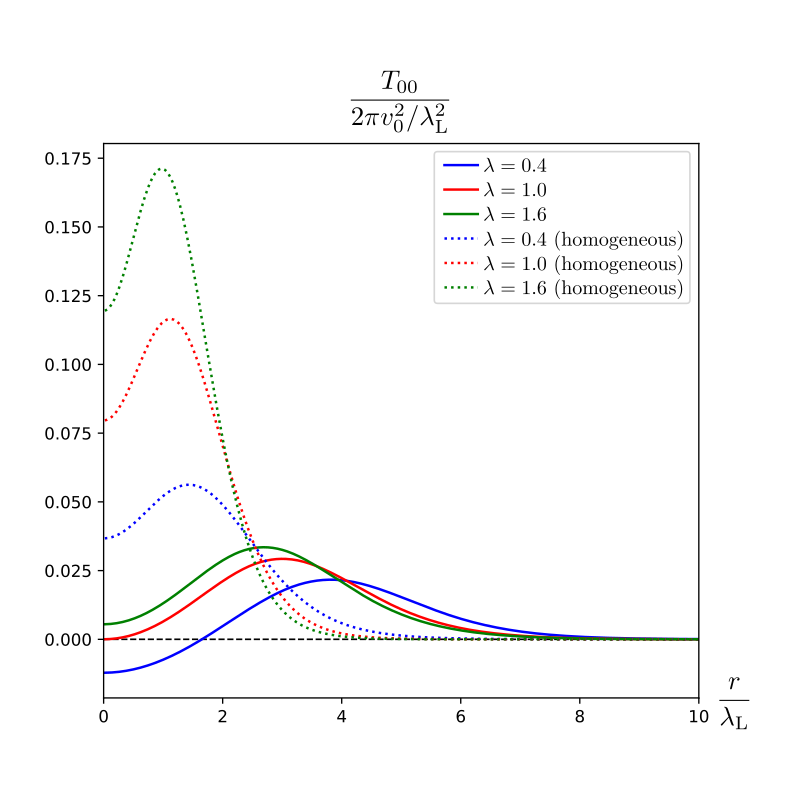}
        }
        \caption{
                (a) Magnetic fields \(B\) and (b) energy densities \(-T\indices{^t_t}\) of \(n=2\) vortices for different values of the quartic scalar coupling \(\lambda = 0.4, 1.0, 1.6 \) and for Gaussian inhomogeneous part with fixed dimensionless size parameter \( \alpha^{-1} = \sqrt{20} \approx 4.47 \) and depth parameter \(\beta=1\).
                The homogeneous counterparts with zero depth parameter \(\beta=0\) are plotted as dotted curves for comparison.
        }
\label{fig:411}
\end{figure}

From now on, let us consider two vortices superimposed at the center of inhomogeneous region.
Numerical works for \(n=2\) vortices in the presence of the Gaussian inhomogeneous part show that the shapes of the scalar amplitude of \(n=1\) vortices and \(n=2\) superimposed vortices are qualitatively the same for equal quartic scalar coupling and Gaussian inhomogeneity as shown in Figure \ref{fig:407} and Figure \ref{fig:410}, respectively.
Only about the origin, the scalar amplitude grows linearly for \(n=1\) but is convex down for \(n=2\) as given in the leading quadratic terms of power series expansion \eqref{402}.
As the quartic scalar coupling \(\lambda\) increases, the correlation length \(\xi\) \eqref{213} decreases and so do the radii of the \(n=2\) vortices for both homogeneous and inhomogeneous cases.
\noindent%
For inhomogeneous vortices, the curves of magnetic field \(B(r)\) and energy density \(-T\indices{^t_t}\) of \(n=2\) vortices in Figure \ref{fig:411}-(a)--(b) are also similar to those of  \(n=1\) vortices in Figure \ref{fig:402}-(a) irrespective of various quartic scalar couplings.
On the other hand, for homogeneous vortices, the curves of energy density \(-T\indices{^t_t}\) of \(n=2\) vortices become ring-shaped as shown in Figure \ref{fig:411}-(b), which are different from those of \(n=1\) vortex with the peak at the origin as shown in Figure \ref{fig:402}-(b).

\begin{figure}[H]
        \centering
        \includegraphics[
                width=0.65\textwidth
        ]{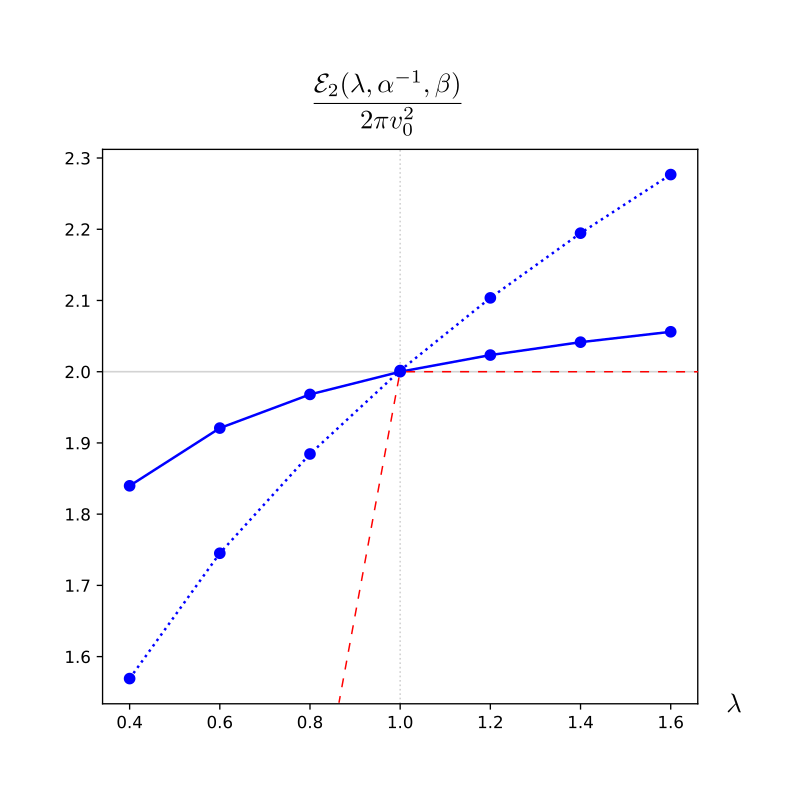}
        \caption{
                The energy of an \(n=2\) vortices for various values of the quartic scalar coupling \(\lambda\) and for Gaussian inhomogeneous part with fixed size parameter \( \alpha^{-1} = \sqrt{20} \approx 4.47 \) and depth parameter \(\beta = 1\) is plotted by a solid curve in the unit of \(n=1\) BPS vortex
                \(
                        \mathcal{E}_{1}
                        ( \lambda=1, \alpha^{-1}, \beta)
                        = 2\pi v_{0}^{2}
                \).
                The homogeneous counterpart with zero depth parameter \(\beta = 0\) is plotted by a dotted curve for comparison.
        }
\label{fig:412}
\end{figure}

\noindent%
If two graphs in Figure \ref{fig:403} and Figure \ref{fig:412} are compared, the curves of the net energy of two superimposed static vortices show almost the same pattern as those of the rest energy of an isolated vortex of unit vorticity except for the values of the horizontal axis, \(1.00\) for \(n=1\) and \(2.00\) for \(n=2\).
Similar to the case of vortex-impurity composite of unit vorticity \(n=1\), the vortex-impurity composite of \(n=2\) is energetically disfavored for \(\lambda<1\) corresponding to type I superconductivity while it is energetically favored for \(\lambda>1\) corresponding to type I$\!$I superconductivity.

The net energy
\(
        \mathcal{E}_{n} =
        \mathcal{E}_{n}
        ( \lambda, \alpha^{-1}, \beta )
\)
of static \(n\) superimposed vortices obtained through numerical works depends on the quartic scalar coupling \(\lambda\) and the shape of inhomogeneity of Gaussian form \eqref{203} centered at the origin.
When every pair of two vortices among the \(n\) vortices are separated sufficiently, the interaction between any pair vanishes completely and hence interaction between \(n\) superimposed inhomogeneous vortices can be measured by the difference \( \mathcal{E}_{n} - n\mathcal{E}_{1} \).
If every separated vortex is located in homogeneous region of \( \sigma(\boldsymbol{x})=0 \) at spatial asymptote, its energy 
\(
        \mathcal{E}_{1} =
        \mathcal{E}_{1}
        ( \lambda, \alpha^{-1}, \beta=0 )
\)
is not contaminated by inhomogeneity of sufficiently localized impurities, e.g. the Gaussian type inhomogeneity \eqref{203}.

\begin{figure}[H]
        \centering
        \includegraphics[
                width=0.65\textwidth
        ]{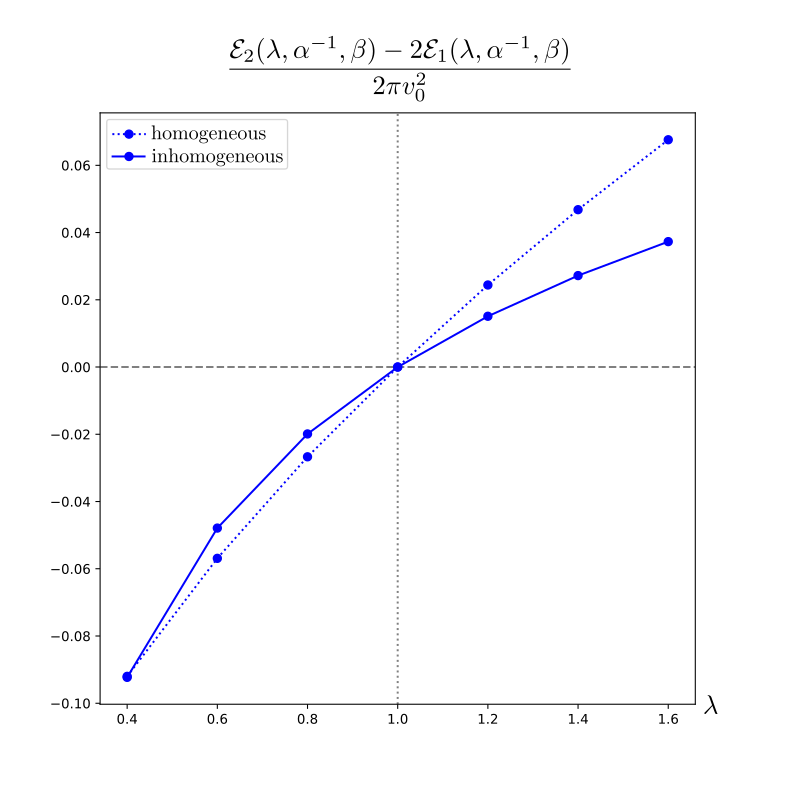}
        \caption{
                A graph of interaction energy \(\mathcal{E}_{2} - 2\mathcal{E}_{1}\) for various quartic scalar coupling \(\lambda\) with fixed size and depth parameters \(\alpha^{-1}=\sqrt{20}\) and \(\beta=1\).
                Dashed curve denotes interaction energy \(\mathcal{E}_{2} - 2\mathcal{E}_{1}\) of the homogeneous counterparts for comparison.
        }
\label{fig:420}
\end{figure}

The simplest form of the interaction is that between two inhomogeneous vortices of unit vorticity \(n=1\).
Thus the obtained interaction energy
\(
        \mathcal{E}_{2}
        ( \lambda, \alpha^{-1}, \beta)
        -
        2 \mathcal{E}_{1}
        ( \lambda, \alpha^{-1}, \beta = 0)
\)
consists of both the interaction between the two vortices in homogeneous limit and the maximized effect of inhomogeneity.
Since it is difficult to separate the inhomogeneity effect and the interaction, we will consider only limited cases of two superimposed vortices on the top of a single cylindrically symmetric Gaussian inhomogeneous part \eqref{203} and discuss the interaction energy
\begin{align}
\mathcal{E}_{2} (\lambda, \alpha^{-1}, \beta) - 2 \mathcal{E}_{1} (\lambda, \alpha^{-1}, \beta) \label{405}
\end{align}
including impurity effect in what follows.
The solid curve in Figure \ref{fig:420} says that the interaction energy \( \mathcal{E}_{2} - 2\mathcal{E}_{1} \) for two superimposed vortices is negative in  weak coupling regime \(\lambda < 1 \), zero in critical coupling $\lambda=1$, and positive in strong coupling regime \( \lambda > 1 \) in the presence of the inhomogeneity \( \sigma(r) \) of Gaussian shape \eqref{203}. These three cases correspond to attractive interaction, vanishing interaction in the BPS limit, and repulsive interaction between the two superimposed inhomogeneous vortices, respectively. In strong coupling regime $(\lambda>1)$,
comparison of the solid and dashed curves implies possible tendency that the amount of interaction energy is decreased by the inhomogeneity. This weaker repulsion implies more probable generation of the vortex-impurity composite of larger vorticity at each impurity site. Therefore, a dirty type I$\!$I superconducting sample allows more penetration of magnetic field and becomes less diamagnetic.%

\begin{figure}[H]
        \centering
        \subfigure[]{
                \includegraphics[
                        width=0.45\textwidth
                ]{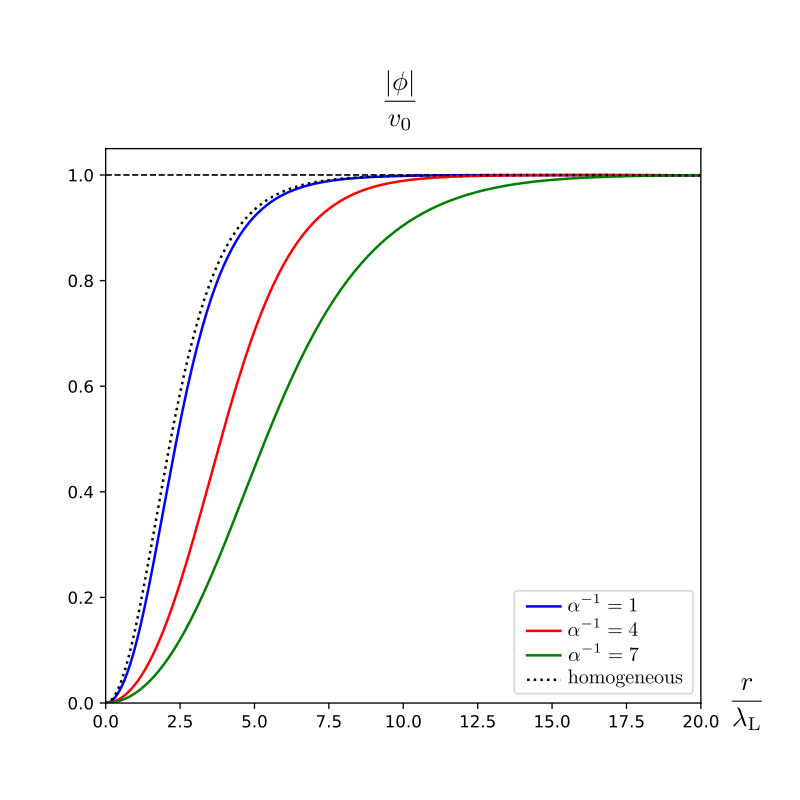}
        }
        \hfill
        \subfigure[]{
                \includegraphics[
                        width=0.45\textwidth
                ]{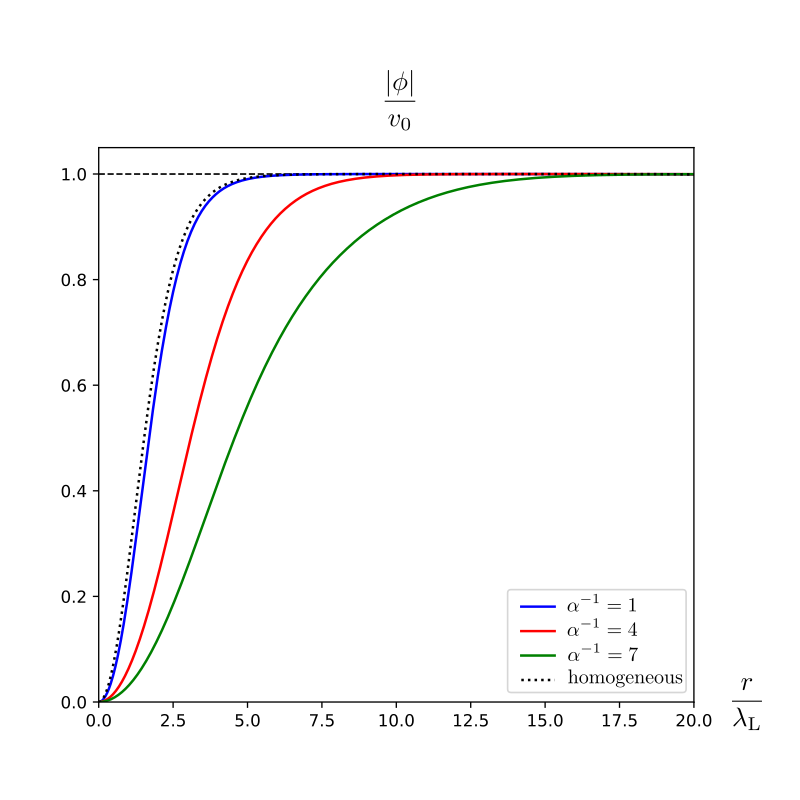}
        }
        \caption{
                Amplitudes of the complex scalar field of \(n=2\) vortices for two quartic scalar couplings (a) \(\lambda=0.5\) and (b) \(\lambda=1.5\) for Gaussian inhomogeneous part with various values of the size parameter \( \alpha^{-1}\) and fixed depth parameter \(\beta = 1\).
                The homogeneous counterpart with zero depth parameter \(\beta=0\) is plotted as dotted curves for comparison.
        }
\label{fig:413}
\end{figure}

\begin{figure}[H]
        \centering
        \subfigure[]{
                \includegraphics[
                        width=0.45\textwidth
                ]{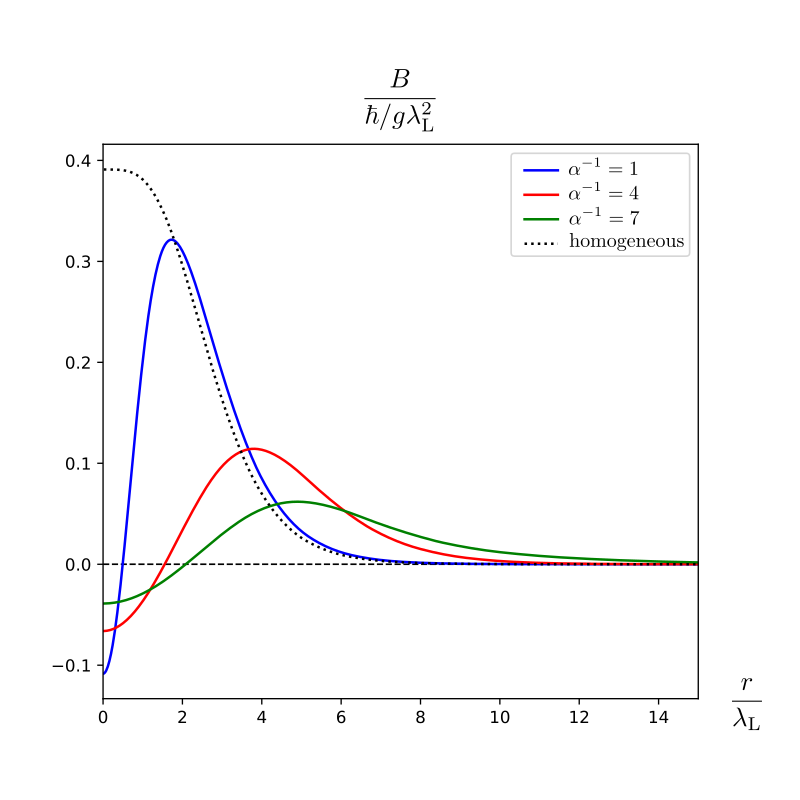}
        }
        \hfill
        \subfigure[]{
                \includegraphics[
                        width=0.45\textwidth
                ]{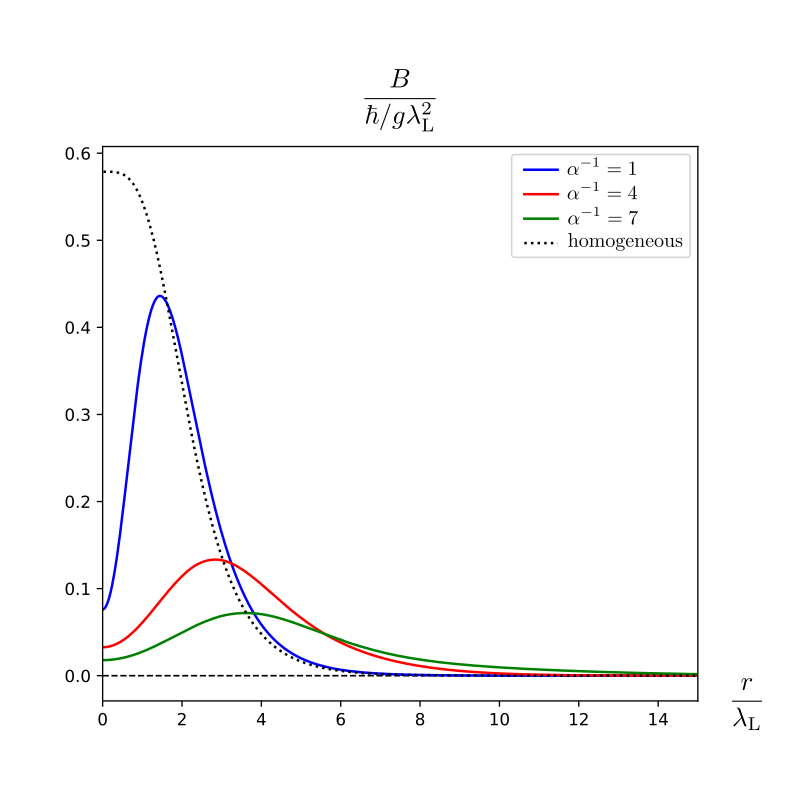}
        }
        \\
        \subfigure[]{
                \includegraphics[
                        width=0.45\textwidth
                ]{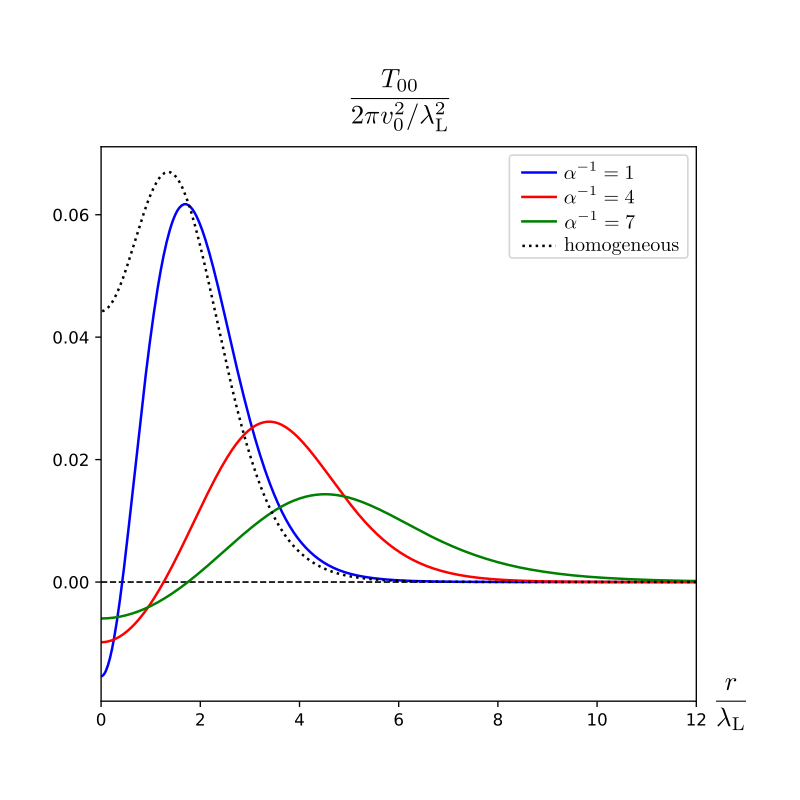}
        }
        \hfill
        \subfigure[]{
                \includegraphics[
                        width=0.45\textwidth
                ]{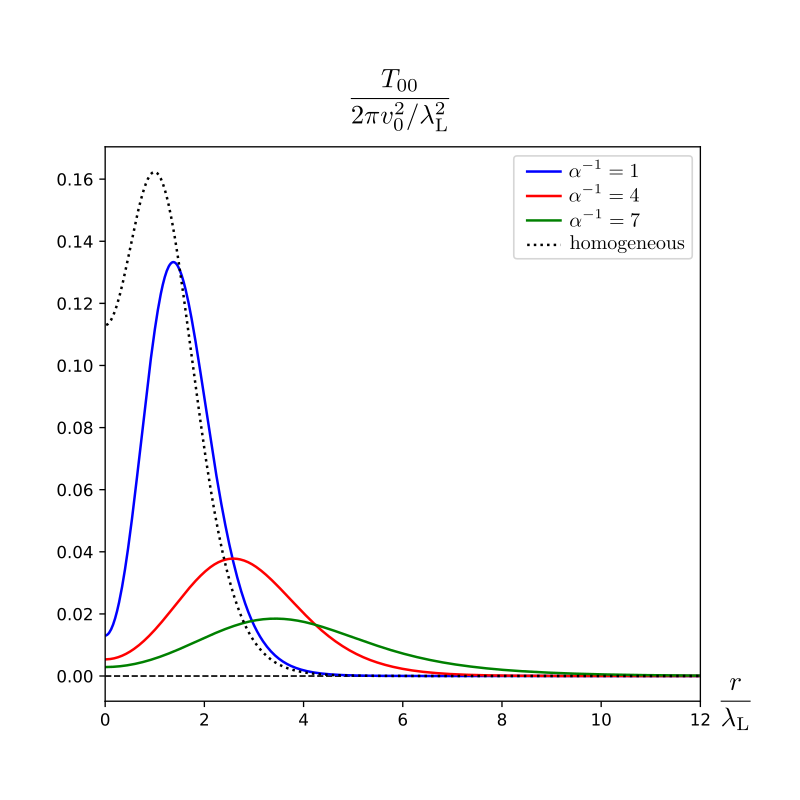}
        }
        \caption{
                Magnetic fields \(B\) of \(n=2\) vortices for two quartic scalar couplings (a) \(\lambda=0.5\) and (b) \(\lambda=1.5\) and energy density \(-T\indices{^t_t}\) of \(n=2\) vortices for two quartic scalar couplings (c) \(\lambda=0.5\) and (d) \(\lambda=1.5\) for Gaussian inhomogeneous part with various values of the size parameter \( \alpha^{-1}\) and fixed depth parameter \(\beta = 1\).
                The homogeneous counterparts of magnetic field and energy density with zero depth parameter \(\beta=0\) are given by dotted curves for comparison.
        }
\label{fig:414}
\end{figure}

When size parameter \(\alpha^{-1}\) varies from small to large with keeping quartic scalar coupling \(\lambda\) and depth parameter \(\beta\), the radii of \(n=2\) superimposed vortices also increase as \(n=1\) vortex.
Comparison of the size parameter \(\alpha^{-1}\) to dependency of the quartic scalar coupling \(\lambda\), \(\lambda=0.5\) for Figure \ref{fig:413}-(a) and \(\lambda=1.5\) for Figure \ref{fig:413}-(b), exhibits competition of two interaction scales. 
\noindent%
Similarity in the profiles of scalar amplitude between \(n=1\) and \(n=2\) inhomogeneous vortices leads to the similar distribution of magnetic field \(B(r)\) and energy density \(-T\indices{^t_t}(r)\) as shown in Figure \ref{fig:414}-(a)--(d).

\begin{figure}[H]
        \centering
        \subfigure[]{
                \includegraphics[
                        width=0.45\textwidth
                ]{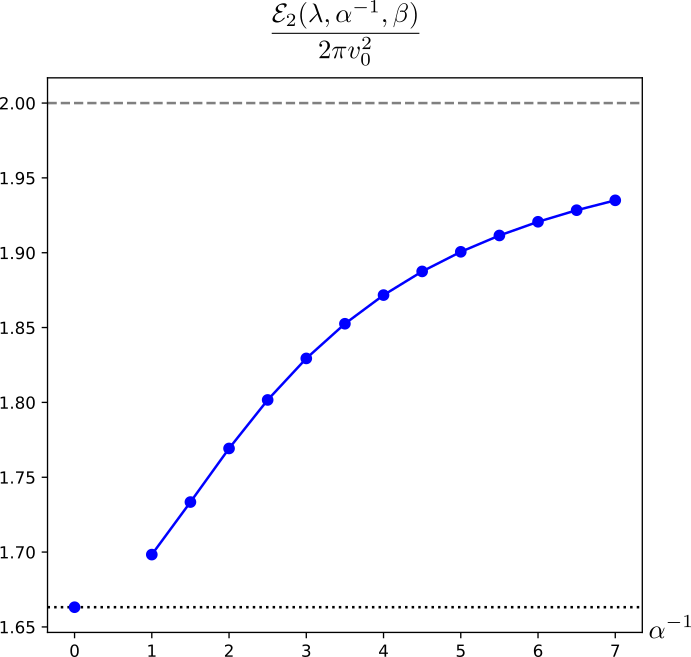}
        }
        \hfill
        \subfigure[]{
                \includegraphics[
                        width=0.45\textwidth
                ]{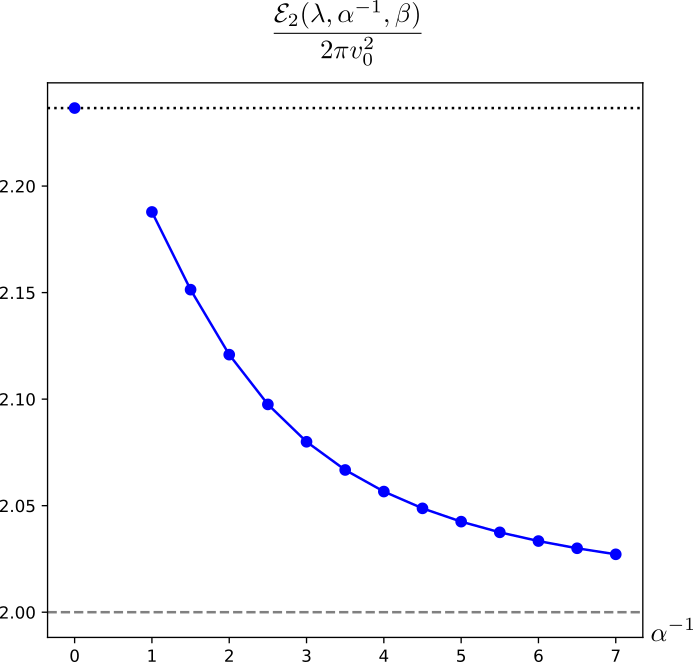}
        }
        \caption{
                The energy of an \(n=2\) vortex \(\mathcal{E}_{2}(\lambda, \alpha^{-1}, 1)\) for various values of the size parameter \(\alpha^{-1}\) of Gaussian inhomogeneous part with fixed depth parameter \(\beta=1\) for two quartic scalar couplings (a) \(\lambda = 0.5 < 1\) and (b) \(\lambda = 1.5 > 1\).
                The homogeneous counterpart with zero depth parameter \(\beta=0\) is plotted by a dotted horizontal line and the energy \( \mathcal{E}_{2} ( \lambda=1, \alpha^{-1}, \beta ) = 4 \pi v_{0}^{2} \) of a BPS vortex is given by a dashed line for comparison.
        }
\label{fig:415}
\end{figure}

\begin{figure}[H]
        \centering
        \includegraphics[
                width=0.65\textwidth
        ]{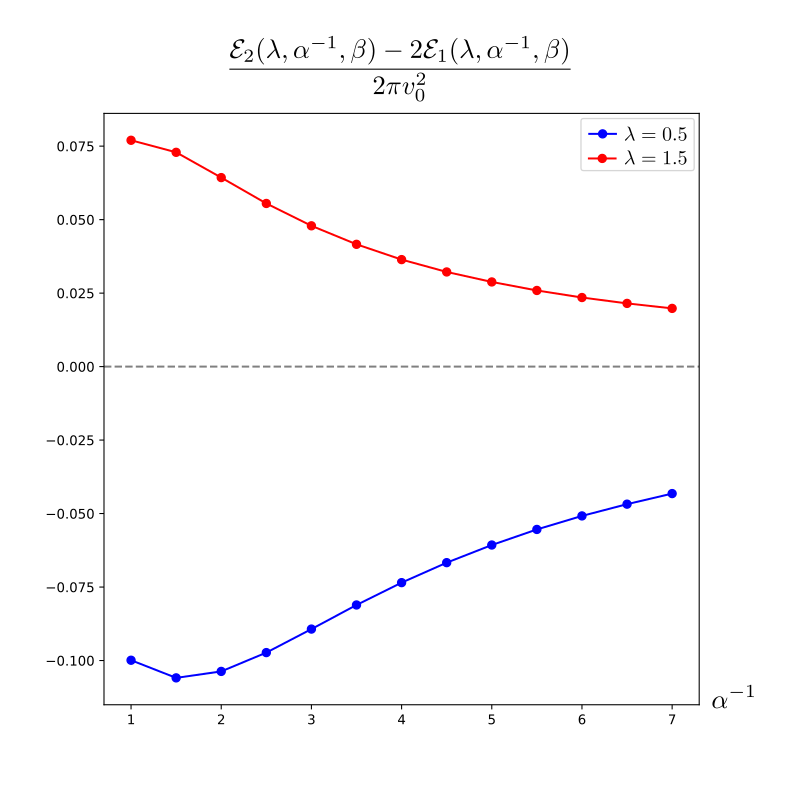}
        \caption{
                Interaction energy \(\mathcal{E}_{2} - 2\mathcal{E}_{1}\) of \(n=2\) inhomogeneous vortices superimposed at the center of inhomogeneity for quartic scalar couplings \(\lambda=0.5\) and \(\lambda=1.5\) and for various size parameter \(\alpha^{-1}\) with fixed depth parameter \(\beta=1\).
                Dashed horizontal line stands for interaction energy of two homogeneous vortices for comparison.
        }\label{fig:421}
\end{figure}

The net energy of \(n=2\) superimposed inhomogeneous vortices is always smaller than that of \(n=2\) BPS vortices for the quartic scalar coupling less than the critical value \(\lambda<1\),
\(
        \mathcal{E}_{2}
        (\lambda<1, \alpha^{-1}, \beta)
        <
        \mathcal{E}_{2}
        (\lambda=1, \alpha^{-1}, \beta)
\)
and is always greater than that for the quartic scalar coupling larger than the critical value \(\lambda>1\).
As the size parameter \(\alpha^{-1}\) increases from zero, the corresponding energy gap
\(
        | \mathcal{E}_{2}
        (\lambda, \alpha^{-1}, \beta)
        - 2 (2\pi v_{0}^{2}) |
\)
decreases from the largest energy gap in the homogeneous limit shown by a blue-colored dots on the dotted horizontal lines in Figure \ref{fig:415}-(a)--(b), irrespective of the values of quartic scalar coupling \( \lambda ~ (\lambda\neq1) \).
\noindent%
The net energy \(\mathcal{E}_{2}\) of \(n=2\) superimposed inhomogeneous vortices obtained through numerical works consists of the rest energy of two \(n=1\) vortices and their 2-body interaction in the presence of common cylindrically symmetric inhomogeneity \(\sigma(r)\) \eqref{203}.
Thus possible interaction energy may be read as energy difference \( \mathcal{E}_{2} ( \lambda, \alpha^{-1}, \beta ) - 2\mathcal{E}_{1} ( \lambda, \alpha^{-1}, \beta ) \) in \eqref{405}
whilst contamination due to inhomogeneity cannot be extracted.
The 2-body interaction is always attractive for \(\lambda<1\), e.g. \(\lambda=0.5\) shown by the blue-colored curve, and more repulsive for \(\lambda>1\), e.g. \(\lambda=1.5\) shown by the red-colored curve, than the homogeneous case shown by the dashed horizontal line in Figure \ref{fig:421}, irrespective of the size of inhomogeneous region.
As size parameter \(\alpha^{-1}\) increases, absolute value of the 2-body interaction energy \( |\mathcal{E}_{2} - 2 \mathcal{E}_{1}| \) \eqref{405} seems to have decreasing tendency for both \(\lambda<1\) and \(\lambda>1\).
Hence the modulation effect to decrease interaction energy increases as the inhomogeneous region grows.

\begin{figure}[H]
        \centering
        \subfigure[]{
                \includegraphics[
                        width=0.45\textwidth
                ]{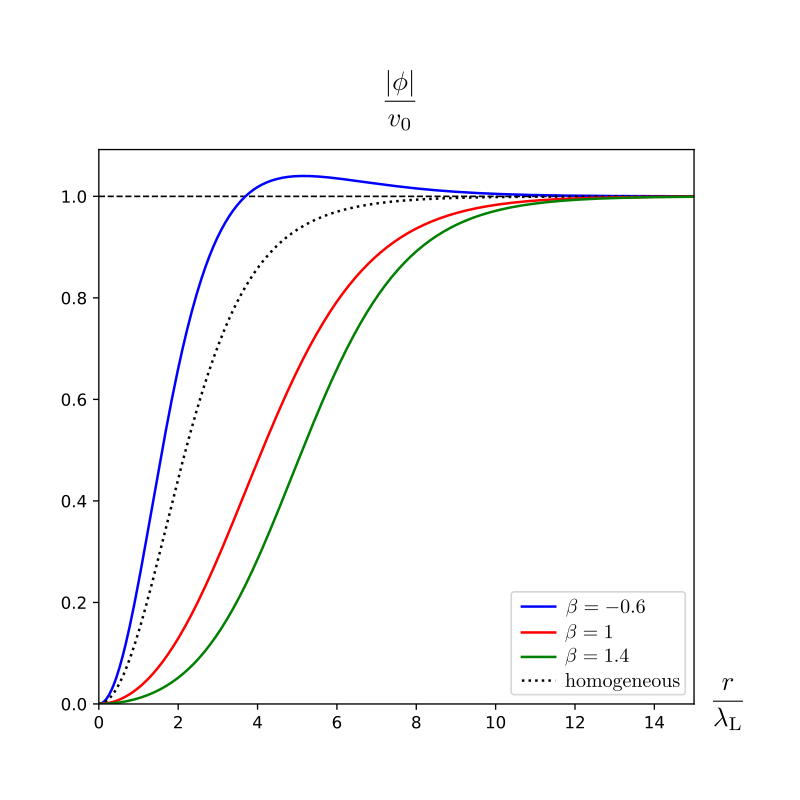}
        }
        \hfill
        \subfigure[]{
                \includegraphics[
                        width=0.45\textwidth
                ]{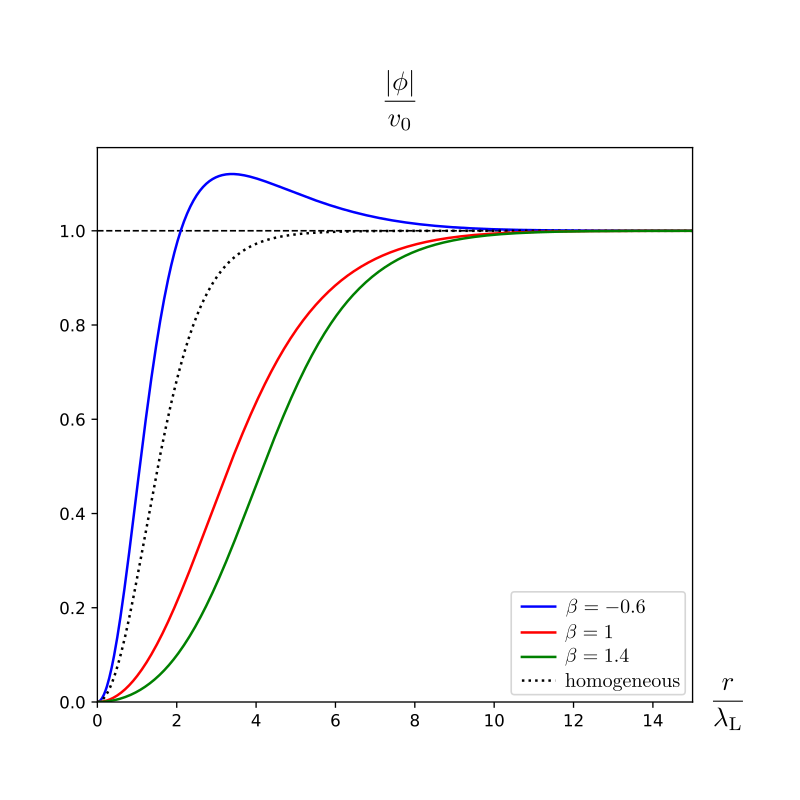}
        }
        \caption{
                Amplitudes of the complex scalar field of \(n=2\) vortices for two quartic scalar couplings, (a) \(\lambda=0.5\) and (b) \(\lambda=1.5\), for Gaussian inhomogeneous part with various values of the depth parameter \(\beta\) and fixed size parameter \( \alpha^{-1} = \sqrt{20} \approx 4.47 \).
                The homogeneous counterpart with zero depth parameter \(\beta=0\) is plotted as a dotted curve for comparison.
        }
\label{fig:416}
\end{figure}

\begin{figure}[H]
        \centering
        \subfigure[]{
                \includegraphics[
                        width=0.45\textwidth
                ]{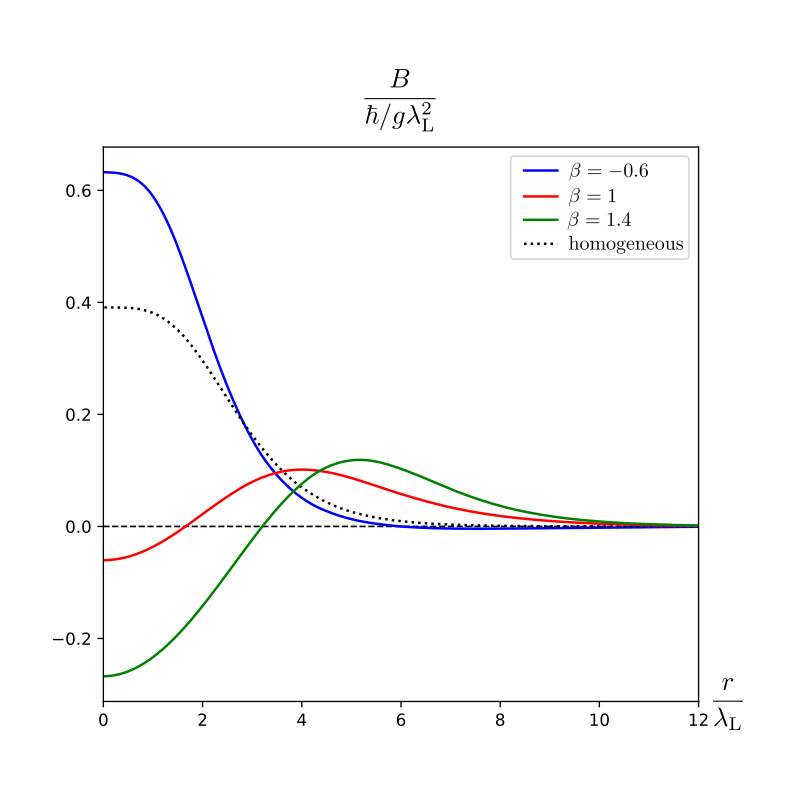}
        }
        \hfill
        \subfigure[]{
                \includegraphics[
                        width=0.45\textwidth
                ]{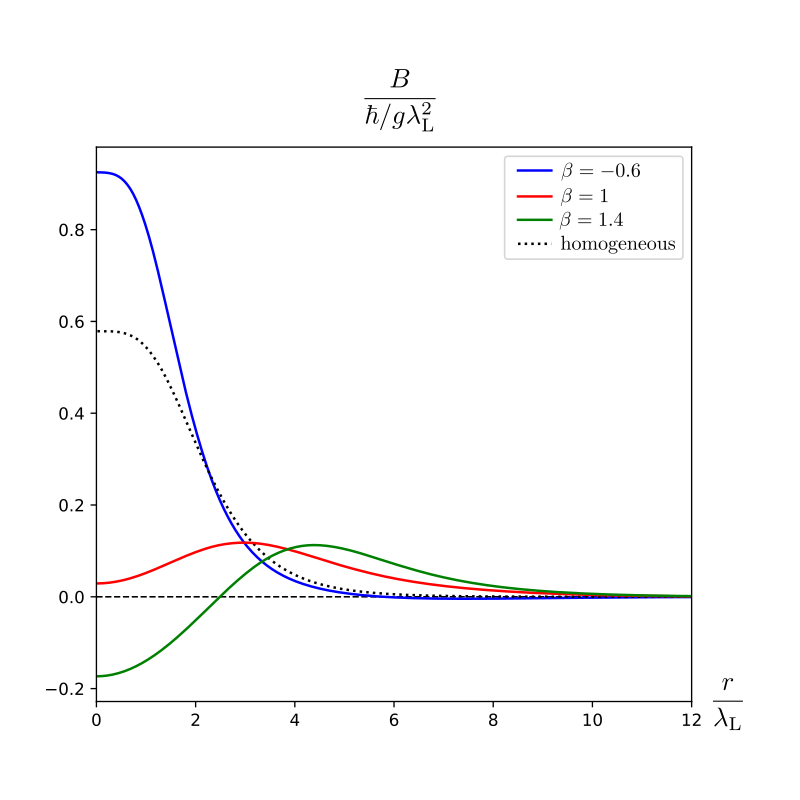}
        }
        \\
        \subfigure[]{
                \includegraphics[
                        width=0.45\textwidth
                ]{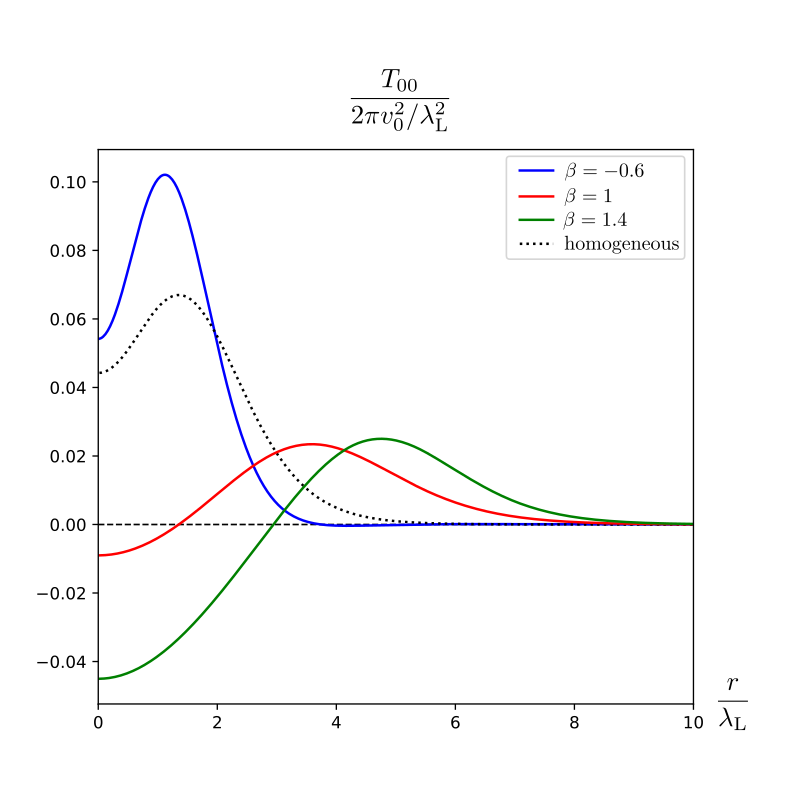}
        }
        \hfill
        \subfigure[]{
                \includegraphics[
                        width=0.45\textwidth
                ]{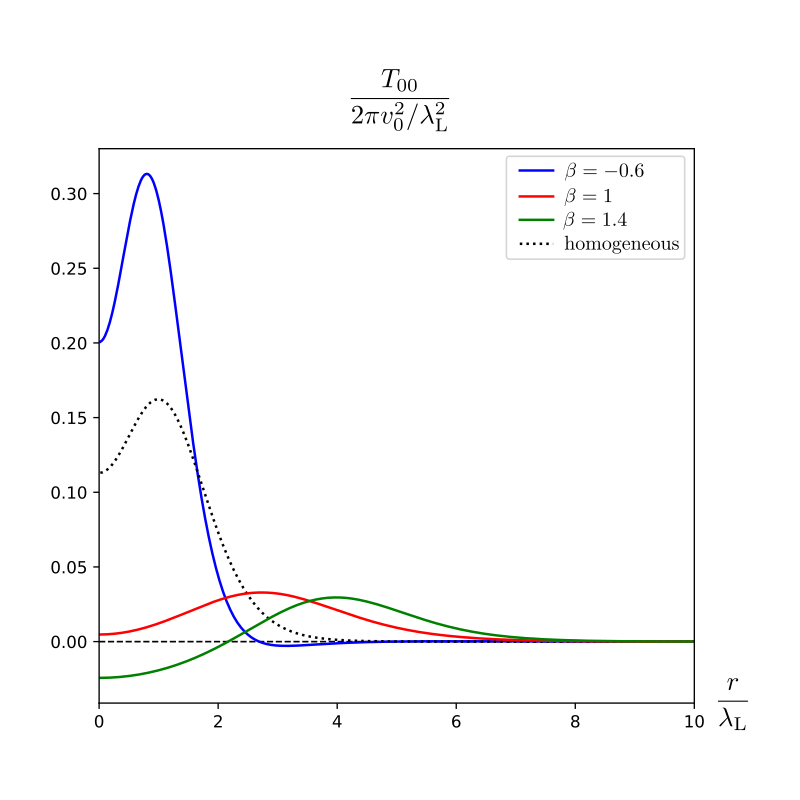}
        }
        \caption{
                Magnetic fields \(B\) of \(n=2\) vortices for two quartic scalar couplings (a) \(\lambda=0.5\) and (b) \(\lambda=1.5\) and energy density \(-T\indices{^t_t}\) of \(n=2\) vortices for two quartic scalar couplings (c) \(\lambda=0.5\) and (d) \(\lambda=1.5\) for Gaussian inhomogeneous part with various values of the depth parameter \(\beta\) and fixed size parameter \( \alpha^{-1} = \sqrt{20} \approx 4.47 \).
                The homogeneous counterparts of magnetic field and energy density with zero depth parameter \(\beta=0\) are given by dotted curves for comparison.
        }
\label{fig:417}
\end{figure}

When depth parameter \(\beta\) varies with keeping quartic scalar coupling \(\lambda\) and size parameter \(\alpha^{-1}\), the profiles of scalar amplitude of \(n=2\) inhomogeneous vortices in Figure \ref{fig:416} show qualitatively the same behavior of those of \(n=1\) inhomogeneous vortices in Figure \ref{fig:407} except for the more convex behavior near the origin, that is consistent with the leading behavior of scalar amplitude \( |\phi| (r) \propto r^{n} \) \eqref{402}.
\noindent%
The aforementioned slight variation of scalar amplitude of \(n=2\) vortices about the origin does not affect the characteristic shapes of magnetic field \(B(r)\) in Figure \ref{fig:417}-(a)--(b) and energy density \(-T\indices{^t_t}(r)\) of the red- and green- colored curves in Figure \ref{fig:417}-(c)--(d) in comparison with those of \(n=1\) vortices in Figure \ref{fig:408}-(a)--(b).
On the other hand, the blue-colored curves of energy density \(-T\indices{^t_t}\) in Figure \ref{fig:417}-(a)--(b) have ring shapes with a positive local minimum at the origin and show qualitative difference from those of \(n=1\) vortices in Figure \ref{fig:408}-(a)--(b), which have a global maximum at the origin.
These two different shapes of magnetic field depending on depth parameter \(\beta\) can be reconciled by the behavior of gauge field near the origin \eqref{403} whose expansion results in the subleading term of magnetic field as
\begin{equation}
        B (r)=
        \frac{\hbar}{g} \frac{1}{r} \frac{dA}{dr}
        \approx
        \frac{\hbar}{g \lambda_{\text{L}}^{2} }
        \Big[
                2 a_{n0}
                + \frac{1}{2}
                ( \alpha^{2} \beta
                  - \phi_{n0}^{2} \delta_{n1} )
                \Big(
                  \frac{r}{\lambda_{\text{L}}}
                \Big)^{2}
                + \cdots
        \Big]
.
\end{equation}
For the vortices of unit vorticity \(n=1\), magnetic field becomes convex when \( \beta > \phi_{10}^{2} / \alpha^{2} \), and, for those of \(n\ge 2\), it becomes convex when \(\beta>0\).

\begin{figure}[H]
        \centering
        \subfigure[]{
                \includegraphics[
                        width=0.45\textwidth
                ]{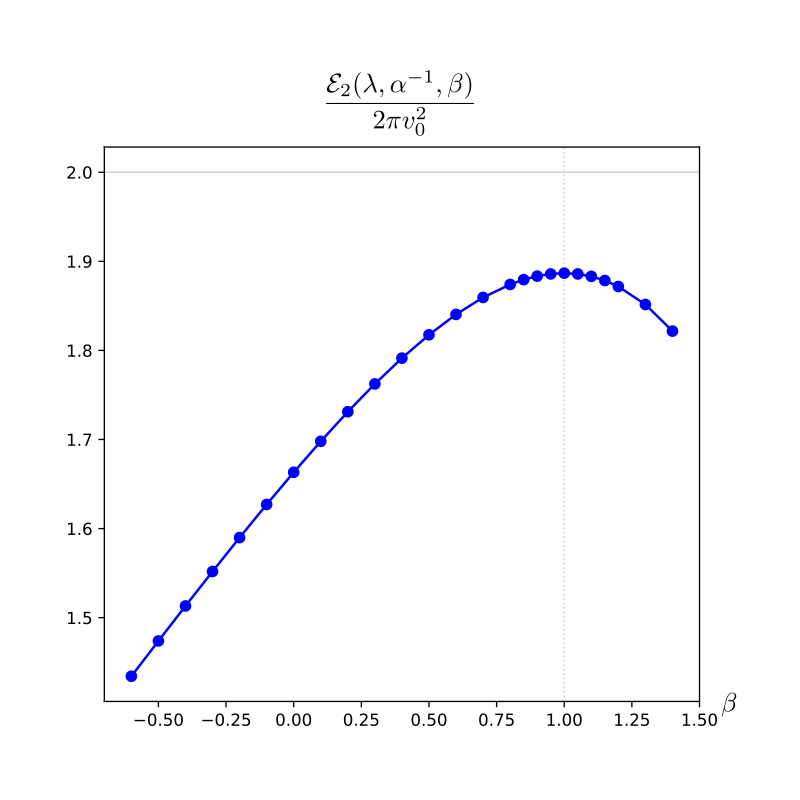}
        }
        \hfill
        \subfigure[]{
                \includegraphics[
                        width=0.45\textwidth
                ]{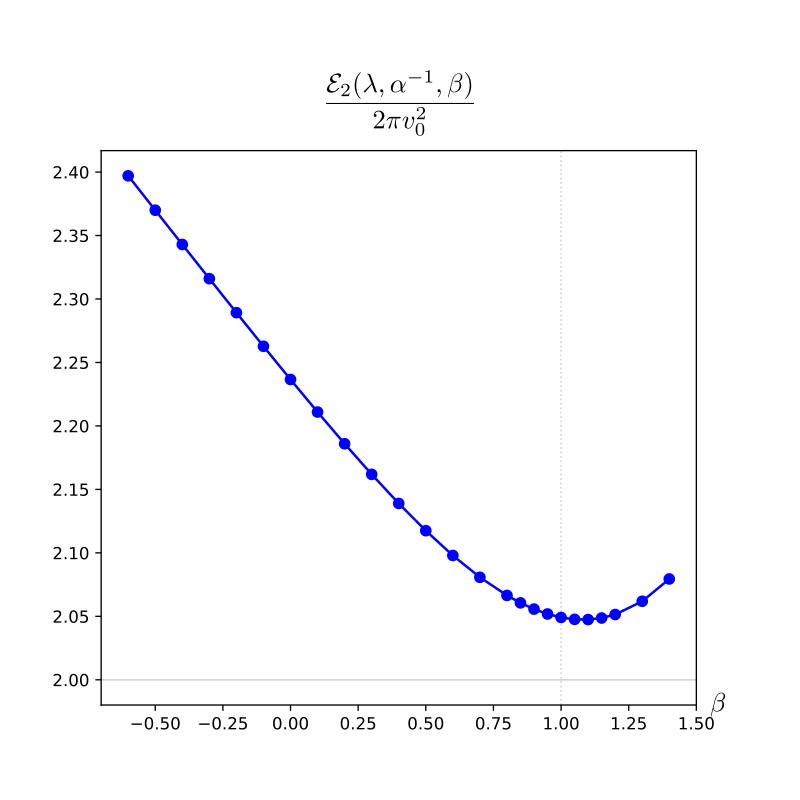}
        }
        \caption{
                The energy of an \(n=2\) vortex \(\mathcal{E}_{2}(\lambda, \sqrt{20}, \beta)\) for various values of the depth parameter \(\beta\) of Gaussian inhomogeneous part with fixed size parameter \( \alpha^{-1} = \sqrt{20} \approx 4.47 \) for two quartic scalar couplings (a) \(\lambda = 0.5 < 1\) and (b) \(\lambda = 1.5 > 1\).
        }
\label{fig:418}
\end{figure}

\begin{figure}[H]
        \centering
        \includegraphics[
                width=0.65\textwidth
        ]{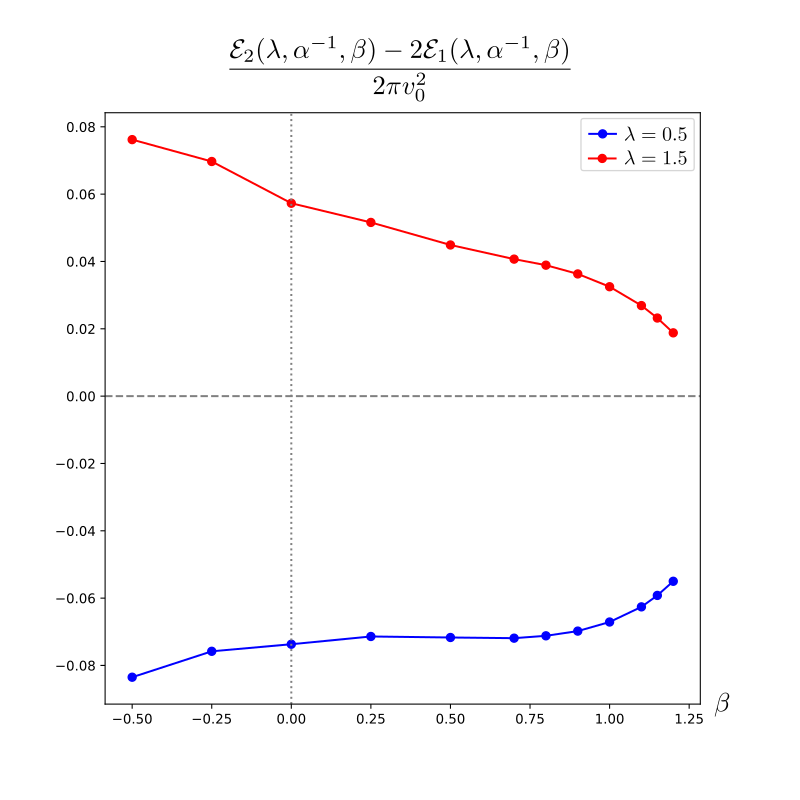}
        \caption{
                Interaction energy \( \mathcal{E}_{2} - 2\mathcal{E}_{1} \) of two inhomogeneous vortices superimposed at the center of inhomogeneity for quartic scalar couplings, \(\lambda=0.5\) and \(\lambda=1.5\), and for various depth parameter \(\beta\) with fixed size parameter \(\alpha^{-1} = \sqrt{20} \approx 4.47\).
                Dashed horizontal line means zero interaction energy of two BPS vortices for comparison.
        }
\label{fig:419}
\end{figure}

The total energy of the two vortices superimposed on top of the Gaussian inhomogeneity also depend on depth parameter \(\beta\) for any quartic scalar coupling \(\lambda\) except for the critical value \(\lambda=1\) of the BPS case.
It is smaller than that of two BPS vortices \( \mathcal{E}_{2} = 4\pi v_{0}^{2} \) for \(\lambda<1\) and greater for \(\lambda>1\) at every value of the depth parameter \(\beta\) as shown in Figure \ref{fig:418}-(a)--(b).
The absolute value of the energy gap
\(
        | \mathcal{E}_{2}
        (\lambda, \alpha^{-1}, \beta)
        - 4\pi v_{0}^{2} |
\)
has the smallest value in the vicinity of \(\beta=1\) similar to the case of a single vortex as in Figure \ref{fig:409}.
As already discussed for \(n=1\) vortices, the extremum value near \(\beta = 1\) for \(n=2\) vortices lacks reasonable explanation.
Nevertheless, if it is found, the explanation is possibly common.
\noindent%
For given quartic scalar coupling, e.g. \(\lambda=0.5<1\) or \(\lambda=1.5>1\), the interaction energy \( \mathcal{E}_{2} - 2\mathcal{E}_{1} \) \eqref{405} is plotted for various depth parameter \(\beta\) with a fixed size parameter \(\alpha^{-1} = \sqrt{20}\) as shown in Figure \ref{fig:419}.
Though extraction of sole effect to \(2\)-body interaction due to inhomogeneity is impossible for these \(n=2\) inhomogeneous vorticees superimposed at the center of inhomogeneity, the interaction of the two superimposed inhomogeneous vortices is always attractive for \(\lambda=0.5<1\) as shown by the blue-colored curve and repulsive for \(\lambda=1.5>1\) as shown by the red-colored curve.
As the depth parameter \(\beta\) increases absolute value of interaction energy \( |\mathcal{E}_{2} - 2\mathcal{E}_{1}| \) decreases for every non-unity quartic scalar coupling \(\lambda\neq1\).

Since the main tool of this paper is numerical computation, it is timely to comment on the accuracy here.
In the BPS limit of critical quartic scalar coupling \(\lambda=1\), exact values of the energy
\(
\mathcal{E}_{n}
(1, \alpha^{-1}, \beta)
= 2\pi v_{0}^{2} |n|
\)
is obtained analytically irrespective of the shape of inhomogeneous part \(\sigma(\boldsymbol{x})\) \eqref{221} if localized \eqref{212}.
Maximum errors
\(      \displaystyle
\bigg|
\frac{ \mathcal{E}_{n} (\lambda=1, \alpha^{-1}, \beta) }
{ 2\pi v_{0}^{2} |n| } - 1
\bigg|
\)
of the numerical data are estimated to be of order \(\mathcal{O}(10^{-4})\) for inhomogeneous vacuum, \(\mathcal{O}(10^{-3})\) for \(n=1\) vortex, and \(\mathcal{O}(10^{-4})\) for \(n=2\) superimposed vortices.

If the values of the net energy of \(n=3\) inhomogeneous vortices superimposed at the origin are evaluated for two typical cases of quartic scalar couplings \(\lambda=0.5\) and \(1.5\), fixed size parameter \(\alpha^{-1}=\sqrt{20}\), and fixed depth parameter \(\beta=1\), numerical data imply that interaction among the three vortices is again attractive \( \mathcal{E}_{3} - 3\mathcal{E}_{1} = -0.187 (2\pi v_{0}^{2}) < 0 \) for \( \lambda=0.5<1 \) and repulsive \( \mathcal{E}_{3} - 3\mathcal{E}_{1} = 0.096 (2\pi v_{0}^{2}) > 0 \) for \( \lambda=1.5>1 \) as expected.
Nevertheless, additional effect of mutual interaction among more than mutual two vortex interaction can possibly be discussed by excess interaction energy of three overlapped vortices \(\mathcal{E}_{3} - 3\mathcal{E}_{1}\) subtracted by the sum of three \(2\)-body interaction energies \( 3(\mathcal{E}_{2} - 2\mathcal{E}_{1})\)
\begin{equation}
        (\mathcal{E}_{3} - 3\mathcal{E}_{1})
        - 3(\mathcal{E}_{2} - 2\mathcal{E}_{1})
        =
        \mathcal{E}_{3}
        - 3\mathcal{E}_{2} + 3\mathcal{E}_{1}
.\label{406}
\end{equation}
Though the sign of the value of the proposed 3-body interaction \eqref{406} are flipped as
\(
        \mathcal{E}_{3}
        - 3\mathcal{E}_{2}
        + 3\mathcal{E}_{1}
        = 0.0139 (2\pi v_{0}^{2}) > 0
\)
for \(\lambda=0.5<1\) and
\(
        \mathcal{E}_{3}
        - 3\mathcal{E}_{2}
        + 3\mathcal{E}_{1}
        = -0.00174 (2\pi v_{0}^{2}) > 0
\)
for \(\lambda=1.5>1\), existence of the net 3-body interaction is not feasible because their absolute values of excess energy are too small to conclude in comparison with numerical error.

In inhomogeneous field theories, interaction between two objects, e.g. two vortices, involves the affection of inhomogeneous part to each object in addition to the interaction between the two objects.
Therefore, the two objects need to be separated and irregularity of inhomogeneous part without cylindrical symmetry is required for systematic study of two inhomogeneous objects \cite{Jacobs:1978ch, Kim:1992yz}.

When Gaussian type inhomogeneous part \eqref{317}--\eqref{318} takes the delta function limit, the field behaviors of the vortices of  vorticity \(n\) near the origin vary
\begin{align}
	|\phi|(r) =&\, f_{0} v_{0} \Big( \frac{r}{\lambda_{ \text{L} }} \Big)^{n+\lambda\eta} \bigg[ 1 - \frac{\lambda}{8(n+\lambda\eta+1)} \Big( \frac{r}{\lambda_{ \text{L} }} \Big)^{2} + \cdots \bigg] ,\\
	A(r) =&\, - \lambda\eta - \frac{f_{0}^{2}}{4(n+\lambda\eta+1)} \Big( \frac{r}{\lambda_{\rm L}} \Big)^{2(n+\lambda\eta+1)} + \cdots ,
\end{align}
where \( n+\lambda\eta\ge 0 \).
%
The U(1) gauge field becomes singular at the origin not because of vorticity \( n \) but because of nonzero effective vorticity $\lambda\eta$.
Then the corresponding magnetic field \( B \) possesses both a delta function type singular configuration and a continuous distribution
\begin{equation}
	B(r) \approx
	\frac{\hbar}{g}
	\bigg[ \lambda \eta 
	\frac{\delta (r)}{r}
	- \frac{f_{0}^{2}}{2\lambda_{ \text{L} }^{2}}
	\Big( \frac{r}{\lambda_{ \text{L} }} \Big)^{2(n+\lambda\eta)}
	+ \cdots
	\bigg]
	.
\end{equation}
The magnetic flux \eqref{220} carried by the vortex-impurity composite consists of two contributions, one from \(n\) vortices and the other from delta function type impurity, e.g. lattice, of continuous strength \(\lambda\eta\),
\begin{equation}
	\Phi_{B} = \frac{2\pi\hbar}{g} (n+\lambda\eta)
	.
\end{equation}
Energy density near the origin also consists of singular delta function pieces and continuous nonsingular part.
Interaction energy of this vortex-impurity composite, e.g. vortex-lattice  composite, in a dirty superconducting sample of nonzero $\lambda\eta$ is attractive but increased in weak coupling regime $(\lambda<1)$ and positive but decreased in strong coupling regime $(\lambda>1)$.
The obtained results with the aforementioned discussions on energetics in relation with Figure \ref{fig:403} may provide a plausible field theoretic elucidation of possible flux-pinning at every lattice site, that is related to imperfect diamagnetism in conventional type \Romtwo\ superconducting samples \cite{tinkham2004introduction, Arovas}. If a delta function bump of negative depth parameter $\beta<0$ is
considered, $n+\lambda\eta$ can become negative. Hence the scalar field diverges at the origin as $|\phi| \sim r^{-|n+\lambda\eta|}$ and no regular solution is allowed.

\section{Conclusions and Discussions}

We have studied the abelian Higgs model with a magnetic impurity term for spatial inhomogeneity. Quartic scalar coupling does not have only the critical value $\lambda=1$ for the BPS limit but also arbitrary value $\lambda\ne1$ for realistic phenomena. The propagation speed $v_{{\rm p}}$ of complex scalar field can be either nonrelativistic or relativistic. Meanwhile we take into account only static configurations, the obtained results are applicable at both regimes. Cylindrically symmetric vacuum and vortex solutions are found by solving numerically the coupled Euler-Lagrange equations in the presence of Gaussian type inhomogeneity controlled by the two dimensionless parameters, the size parameter $\alpha^{-1}$ and depth parameter $\beta$, including the limit of delta function. 

Once inhomogeneity is introduced, the Higgs vacuum of constant scalar field $\phi=v_{0}$ and trivial gauge field can be neither a solution nor the minimum energy configuration. For a given set of the fixed quartic scalar coupling and the fixed size and depth parameters of the Gaussian inhomogeneous part, a cylindrically symmetric nontrivial solution with zero vorticity is identified as the electrically neutral symmetry-broken vacuum since the added positive impurity energy is lowered maximally through the spatial modulation of the scalar and gauge fields by nonlinear interaction. Magnetic field and energy density of the vacuum configuration depend on spatial coordinates but its magnetic flux is left to be zero. As the three parameters consisting of the quartic scalar coupling and the two size and depth parameters vary, profiles of scalar amplitude, magnetic field, and energy density change  accordingly. The energy of the obtained vacuum solution is strictly zero in the BPS limit of the critical quartic scalar coupling $\lambda=1$ for arbitrary localized inhomogeneous part and, away from the BPS limit, is proportional to the quartic scalar coupling. Therefore, in weak coupling regime $0 \le \lambda<1$, the sign of vacuum energy becomes negative due to an over-cancellation of the positive impurity energy, and, in strong coupling regime $\lambda>1$, it becomes positive due to a under-cancellation of the positive impurity energy irrespective of the shape of Gaussian inhomogeneous part. This negativity or positivity of the vacuum energy in inhomogeneous abelian Higgs model of arbitrary quartic scalar coupling relative to the zero energy of constant Higgs vacuum in the homogeneous abelian Higgs model is a measurable energy gap detected by experiments. It suggests that the types I and I$\!$I of conventional superconductors can be distinguished by its vacuum structure through an addition of some impurities in local regions. In the two-dimensional delta function limit from a Gaussian dip, the solution of nonsingular scalar profile possesses a component of infinite magnetic field and energy density concentrated at the site. The corresponding configuration can be interpreted as an inhomogeneous vacuum with an infinitely thin string-like impurity, e.g. lattice, which carries continuous magnetic flux. Its energy is finite while its sign is negative for weak coupling $(\lambda<1)$, zero for critical coupling $(\lambda=1)$, and positive for strong coupling $(\lambda>1)$. Thus this non-zero magnetic flux penetrating the superconducting sample at each lattice site can explain field-theoretically the imperfect diamagnetism of dirty superconductors.      

The impurity-driven vortex configurations are modulated by inhomogeneity through nonlinear interactions, pushing the vortex profiles outward and resulting in ring-shaped magnetic fields and energy densities.
Although the local quantities are modified by the quartic scalar coupling and the inhomogeneity, the electrical neutrality and the quantized magnetic flux remain identical to the homogeneous case.
Numerical results show that vortex energy includes an impurity contribution, matching the rest energy of an isolated vortex at the BPS limit $\lambda = 1$.
It gives insights on diamagnetism in the type I and I$\!$I superconductivity:
When the quartic scalar coupling is weak $(\lambda <1)$, the increased rest energy compared to the homogeneous case energetically disfavors vortex-impurity composite.
It is consistent with the perfect diamagnetism observed in conventional type I superconductivity.
Conversely, for strong coupling $(\lambda >1)$, the decreased rest energy compared to the homogeneous case energetically favors vortex-impurity composite.
This explains consistency of the imperfect diamagnetism in conventional type I$\!$I superconductivity.
As the inhomogeneous part increases its size and depth in strong coupling regime $(\lambda>1)$, the positive interaction energy between two superimposed vortices decreases and this energetics dictates two inhomogeneous vortices less repulsive. This weaker repulsion implies more probable generation of the vortex-impurity composite of larger vorticity at each impurity site. Therefore, a dirty type I$\!$I superconducting sample allows more penetration of magnetic field and becomes less diamagnetic.
 In the delta function limit of Gaussian inhomogeneous part, the vorticity $n$ of the vortices shifts to nonnegative $n+\lambda\eta$ and carries the magnetic flux $\Phi_{B}=2\pi\hbar(n+\lambda\eta)/g$. Interaction  energy of this vortex-impurity composite, e.g. vortex-lattice composite, in a
dirty superconducting sample of nonzero \(\lambda\eta\) is attractive but increased in weak coupling
regime $(\lambda<1)$ and positive but decreased in strong
coupling regime $(\lambda>1)$. This provides a plausible field theoretic elucidation of possible flux-pinning at every lattice site, that is related to imperfect diamagnetism in conventional type \Romtwo\ superconducting samples.
The feasibility of the interaction between two vortices can only be achieved after the study of two separated vortices located at two separated inhomogeneous sites. Three-body interaction remains inconclusive and awaits further accurate numerical studies.

For non-critical scalar couplings \(\lambda\neq1\) in the abelian Higgs model, introduction of inhomogeneous part with magnetic impurity term shows a universal tendency to reduce the net interaction effect irrespective of attractive or repulsive nature.
This universal tendency of decreasing interaction effect is consistent with our daily experiences and is clearly related to usual observations of the blunt signals due to impurities in contaminated condensed matter samples in comparison with the theoretically expected clean signals in the assumed pure samples. To establish the proposed universality of this tendency, this plausible explanation on the basis of energetics in the context of inhomogeneous field theories should be tested in numerous samples in comparison to the theoretical predictions. Along this direction, theoretical studies for the various shapes of inhomogeneous functions, other than the Gaussian type with a limit of delta function, are important, that include the exponential decay, e.g., a hyperbolic tangent, or the power decay, e.g., a Lorentzian type, as the impurity functions in controlled manner. To accumulate the compared data for diversified impurities, defects, disorders, etc., by doping, imperfect growth of samples, junctions of
heterogeneous materials, etc, it may also be necessary to make other couplings, e.g. the quartic scalar coupling, inhomogeneous.

An abelian gauge field of electromagnetism is considered in the current work. It would be beneficial to gain more wisdom by taking into account alternative kinetic terms or by extending gauge symmetry. For the former,  the Chern-Simons term is the first candidate
for planar physics \cite{Hong:1990yh,Jackiw:1990aw,Han:2015tga,Kim:2024gpu} with various shapes of the sextic scalar potential for non-BPS cases,
which is applicable to the fractional quantum Hall effect \cite{Tsui:1982yy} or anyon
superconductivity \cite{Chen:1989xs}. For the latter, extension to $\text{U}(1)^{N}$
or non-abelian gauge symmetry is natural as in the Janus theory \cite{Bak:2003jk, Clark:2005te, DHoker:2006qeo, DHoker:2007zhm, Kim:2008dj, Kim:2009wv} and in the mass-deformed ABJM theory \cite{Kim:2018qle, Kim:2019kns, Arav:2020obl, Kim:2020jrs}. Particularly, the ${\rm U}(1)\times {\rm U}(1)$ gauge theory is intriguing in relation for a possible origin of the inhomogeneity and the magnetic impurity term. 

Two final comments on experiments and tractability are in order. The inhomogeneous abelian Higgs model of our consideration is not enough to be the field theory of conventional superconductors of an \( s \)-wave order parameter, which contain some impurities, but several intriguing nonperturbative properties induced by the localized impurities, e.g., a delta function type for lattice, have been figured out. The achieved nonperturbative properties may shed light toward the field theoretic description of contaminated condensed matter samples, e.g. dirty superconductor. In comparison to the homogeneous field theory, a field theoretic introduction of inhomogeneity through the change of a constant coupling to the corresponding $x$-dependent function  was naively expected to lead to the significant increase of difficulty in the calculations, however it is not the case at least in the tree level except for increase of the amount of tedious numerical computation due to increase of the number of classified cases in order to explain diversified realistic phenomena. This observation suggests a hint on tractability of inhomogeneous field theories.


\section*{Acknowledgement}

The authors would like appreciate Chanju Kim, O-Kab Kwon, and D. D. Tolla for discussions on various topics of inhomogeneous abelian Higgs model, Kwang-Yong Choi for discussions on condensed matter systems including superconductivity, and Beom Jun Kim for crucial discussions on the numerical works.
This work was supported by the National Research Foundation of Korea(NRF) grant with grant number NRF-2022R1F1A1073053 (Y.K.).


\begin{thebibliography}{99}

\bibitem{Bak:2003jk}
D.~Bak, M.~Gutperle and S.~Hirano,
JHEP \textbf{05}, 072 (2003)
doi:10.1088/1126-6708/2003/05/072
[arXiv:hep-th/0304129 [hep-th]].

\bibitem{Clark:2005te}
A.~Clark and A.~Karch,
JHEP \textbf{10}, 094 (2005)
doi:10.1088/1126-6708/2005/10/094
[arXiv:hep-th/0506265 [hep-th]].

\bibitem{DHoker:2007zhm}
E.~D'Hoker, J.~Estes and M.~Gutperle,
JHEP \textbf{06}, 021 (2007)
doi:10.1088/1126-6708/2007/06/021
[arXiv:0705.0022 [hep-th]].

\bibitem{DHoker:2006qeo}
E.~D'Hoker, J.~Estes and M.~Gutperle,
Nucl. Phys. B \textbf{753}, 16-41 (2006)
doi:10.1016/j.nuclphysb.2006.07.001
[arXiv:hep-th/0603013 [hep-th]].

\bibitem{Kim:2008dj}
C.~Kim, E.~Koh and K.~M.~Lee,
JHEP \textbf{06}, 040 (2008)
doi:10.1088/1126-6708/2008/06/040
[arXiv:0802.2143 [hep-th]].

\bibitem{Kim:2009wv}
C.~Kim, E.~Koh and K.~M.~Lee,
Phys. Rev. D \textbf{79}, 126013 (2009)
doi:10.1103/PhysRevD.79.126013
[arXiv:0901.0506 [hep-th]].

\bibitem{Kim:2018qle}
K.~K.~Kim and O.~K.~Kwon,
JHEP \textbf{08}, 082 (2018)
doi:10.1007/JHEP08(2018)082
[arXiv:1806.06963 [hep-th]].

\bibitem{Kim:2019kns}
K.~K.~Kim, Y.~Kim, O.~K.~Kwon and C.~Kim,
JHEP \textbf{12}, 153 (2019)
doi:10.1007/JHEP12(2019)153
[arXiv:1910.05044 [hep-th]].

\bibitem{Arav:2020obl}
I.~Arav, K.~C.~M.~Cheung, J.~P.~Gauntlett, M.~M.~Roberts and C.~Rosen,
JHEP \textbf{11}, 156 (2020)
doi:10.1007/JHEP11(2020)156
[arXiv:2007.15095 [hep-th]].

\bibitem{Kim:2020jrs}
Y.~Kim, O.~K.~Kwon and D.~D.~Tolla,
JHEP \textbf{12}, 060 (2020)
doi:10.1007/JHEP12(2020)060
[arXiv:2008.00868 [hep-th]].

\bibitem{Adam:2019yst}
C.~Adam, J.~M.~Queiruga and A.~Wereszczynski,
JHEP \textbf{07}, 164 (2019)
doi:10.1007/JHEP07(2019)164
[arXiv:1901.04501 [hep-th]].

\bibitem{Adam:2018pvd}
C.~Adam and A.~Wereszczynski,
Phys. Rev. D \textbf{98}, no.11, 116001 (2018)
doi:10.1103/PhysRevD.98.116001
[arXiv:1809.01667 [hep-th]].

\bibitem{Adam:2018tnv}
C.~Adam, T.~Romanczukiewicz and A.~Wereszczynski,
JHEP \textbf{03}, 131 (2019)
doi:10.1007/JHEP03(2019)131
[arXiv:1812.04007 [hep-th]].

\bibitem{Adam:2019djg}
C.~Adam, K.~Oles, J.~M.~Queiruga, T.~Romanczukiewicz and A.~Wereszczynski,
JHEP \textbf{07}, 150 (2019)
doi:10.1007/JHEP07(2019)150
[arXiv:1905.06080 [hep-th]].

\bibitem{Manton:2019xiq}
N.~S.~Manton, K.~Ole{\'s} and A.~Wereszczy{\'n}ski,
JHEP \textbf{10}, 086 (2019)
doi:10.1007/JHEP10(2019)086
[arXiv:1908.05893 [hep-th]].

\bibitem{Adam:2019hef}
C.~Adam, K.~Oles, T.~Romanczukiewicz and A.~Wereszczynski,
[arXiv:1902.07227 [cond-mat.mes-hall]].

\bibitem{Adam:2019xuc}
C.~Adam, K.~Oles, T.~Romanczukiewicz and A.~Wereszczynski,
Phys. Rev. Lett. \textbf{122}, no.24, 241601 (2019)
doi:10.1103/PhysRevLett.122.241601
[arXiv:1903.12100 [hep-th]].

\bibitem{Kwon:2021flc}
O.~K.~Kwon, C.~Kim and Y.~Kim,
JHEP \textbf{01}, 140 (2022)
doi:10.1007/JHEP01(2022)140
[arXiv:2110.13393 [hep-th]].

\bibitem{Hook:2013yda}
A.~Hook, S.~Kachru and G.~Torroba,
JHEP \textbf{11}, 004 (2013)
doi:10.1007/JHEP11(2013)004
[arXiv:1308.4416 [hep-th]].

\bibitem{Tong:2013iqa}
D.~Tong and K.~Wong,
JHEP \textbf{01}, 090 (2014)
doi:10.1007/JHEP01(2014)090
[arXiv:1309.2644 [hep-th]].

\bibitem{Kim:2023abp}
Y.~Kim, O.~K.~Kwon and D.~D.~Tolla,
JHEP \textbf{11}, 074 (2023)
doi:10.1007/JHEP11(2023)074
[arXiv:2308.08136 [hep-th]].

\bibitem{Kim:2024gpu}
Y.~Kim, O.~K.~Kwon, H.~Song and C.~Kim, 
\textit{Inhomogeneous Abelian Chern-Simons Higgs Model with New Inhomogeneous BPS Vacuum and Solitons},
[arXiv:2409.11978 [hep-th]].
  
\bibitem{Kim:2024gfn}
Y.~Kim, S.~Jeon, O.~K.~Kwon, H.~Song and C.~Kim,
\textit{Vacuum and Vortices     in Inhomogeneous Abelian Higgs Model},
[arXiv:2409.12451 [hep-th]].
     
\bibitem{Jeon:2024jbs}
S.~Jeon, C.~Kim and Y.~Kim,
\textit{Existence and Uniqueness of BPS Vacuum and Multi-vortices in Inhomogeneous Abelian Higgs Model},
[arXiv:2409.14054 [math.AP]].

\bibitem{tinkham2004introduction}
M. Tinkham, Introduction to Superconductivity. (Dover Publications,2004).
      
\bibitem{Arovas}
D. Arovas and C. Wu,
\textit{Lecture Notes on Superconductivity (A Work in Progress)}
(2019).

\bibitem{Abrikosov:1956sx}
A.~A.~Abrikosov,
Sov. Phys. JETP \textbf{5}, 1174-1182 (1957).

\bibitem{Jacobs:1978ch}
L.~Jacobs and C.~Rebbi,
Phys. Rev. B \textbf{19}, 4486-4494 (1979)
doi:10.1103/PhysRevB.19.4486.

\bibitem{cohen2016fundamentals}
M. Cohen and S. Louie, Fundamentals of Condensed Matter Physics. (Cambridge University Press,2016).

\bibitem{Bogomolny:1975de}
E.~B.~Bogomolny,
Sov. J. Nucl. Phys. \textbf{24}, 449 (1976)
PRINT-76-0543 (LANDAU-INST.).

\bibitem{Kim:1992mm}
C.~Kim, S.~Kim and Y.~Kim,
Phys. Rev. D \textbf{47}, 5434-5443 (1993)
doi:10.1103/PhysRevD.47.5434

\bibitem{Kim:1992yz}
Y.~Kim and K.~Lee,
Phys. Rev. D \textbf{49}, 2041-2054 (1994)
doi:10.1103/PhysRevD.49.2041
[arXiv:hep-th/9211035 [hep-th]].

\bibitem{Hong:1990yh}
J.~Hong, Y.~Kim and P.~Y.~Pac,
Phys. Rev. Lett. \textbf{64}, 2230 (1990)
doi:10.1103/PhysRevLett.64.2230

\bibitem{Jackiw:1990aw}
R.~Jackiw and E.~J.~Weinberg,
Phys. Rev. Lett. \textbf{64}, 2234 (1990)
doi:10.1103/PhysRevLett.64.2234

\bibitem{Han:2015tga}
X.~Han and Y.~Yang,
JHEP \textbf{02}, 046 (2016)
doi:10.1007/JHEP02(2016)046
[arXiv:1510.07077 [hep-th]].

\bibitem{Tsui:1982yy}
D.~C.~Tsui, H.~L.~Stormer and A.~C.~Gossard,
Phys. Rev. Lett. \textbf{48}, 1559-1562 (1982)
doi:10.1103/PhysRevLett.48.1559

\bibitem{Chen:1989xs}
Y.~H.~Chen, F.~Wilczek, E.~Witten and B.~I.~Halperin,
Int. J. Mod. Phys. B \textbf{3}, 1001 (1989)
doi:10.1142/S0217979289000725

\end{thebibliography}
\end{document}